\documentclass[
reprint,
prx,
longbibliography,
amsmath,
aps
]{revtex4-2}
\usepackage{graphicx}
\usepackage{dcolumn}
\usepackage{bm}

\usepackage{amsmath}
\usepackage{amsthm}
\usepackage{mathtools} 
\usepackage{bm}
\usepackage{bbold}
\usepackage{mathrsfs}

\usepackage{MnSymbol} 

\usepackage{dcolumn}
\usepackage{multirow}
\usepackage{bigstrut}
\usepackage{booktabs}
\usepackage{setspace}

\usepackage{color}
\usepackage{graphicx} 
\usepackage{epstopdf}

\usepackage{chemarrow}
\usepackage{youngtab} 
\usepackage{comment}

\usepackage{algorithm}
\usepackage{algorithmicx}
\usepackage{algpseudocode}

\newtheorem{theorem}{Theorem}[section]

\theoremstyle{definition}

\theoremstyle{remark}

\newcommand{\lr}[1]{\left({#1}\right)}
\newcommand{\ls}[1]{\left[{#1}\right]}
\newcommand{\lc}[1]{\left\{{#1}\right\}}

\newcommand{\x}{\sigma_x}
\newcommand{\y}{\sigma_y}
\newcommand{\z}{\sigma_z}

\begin{document}
	
	\preprint{APS/123-QED}
	
	\title{\textbf{Engineering Precise and Robust Effective Hamiltonians} 
	}%

	\author{Jiahui Chen$^{1,2}$}
	\email{j562chen@uwaterloo.ca}
	\author{David Cory$^{1,3}$}%
	
	\affiliation{%
		$^{1}$Institute for Quantum Computing, Waterloo, Ontario N2L 3G1, Canada\\
		$^{2}$Department of Physics, University of Waterloo, Waterloo, Ontario N2L 3G1, Canada\\
		$^{3}$Department of Chemistry, University of Waterloo, Waterloo, Ontario N2L 3G1, Canada
	}%
	
	
	\date{\today}
	\begin{abstract}
		Engineering effective Hamiltonians is essential for advancing quantum technologies including quantum simulation, sensing, and computing. This paper presents a general framework for effective Hamiltonian engineering,  enabling robust, precise, and efficient quantum control strategies. To achieve efficiency, we focus on creating target zeroth-order effective Hamiltonians while minimizing higher-order contributions and enhancing robustness against systematic errors. The control design identifies the minimal subspace of the toggling-frame Hamiltonian and the full set of achievable, zeroth-order, effective Hamiltonians. The framework also enables robust state transfer, characterization of achievable density matrices, and extension to stochastic parameter fluctuations via a cumulant expansion. Examples are included to illustrate the process flow and resultant precision and robustness.
	\end{abstract}	
	\maketitle
	\section{Introduction}
	Quantum processors have extraordinary potential to transform sensing, secure communication, and enhance computational efficiency, including in the simulation of quantum physics. To realize this potential, three key engineering tasks must be addressed \cite{nielsen2010quantum,divincenzo2000physical,preskill1998reliable}: (1) efficient and robust coherent control \cite{glaser2015training,chow2012universal,pravia2003robust,cappellaro2006principles,koch2016controlling,goerz2014optimal,teske2022qopt,weidner2025robust,doherty1999quantum,dong2010quantum,ansel2024introduction}, (2) the design of complex, interconnected quantum processors \cite{britton2012engineered}, and (3) the implementation of quantum error correction \cite{shor1995scheme}. This paper provides tools for designing coherent control methods to enhance precision and efficiency, using Average Hamiltonian Theory (AHT) as the foundational framework. Introduced by John Waugh over fifty years ago \cite{haeberlen1968coherent}, AHT describes coherent averaging initially developed for NMR \cite{waugh1968approach}. It has since been widely applied to decoupling \cite{mansfield1971symmetrized,rhim1973analysis,burum1979analysis,takegoshi1985magic,cory1990time,cory1991new,boutis2003pulse}, recoupling \cite{raleigh1988rotational,levitt1990theory}, spectroscopy \cite{rose2018high,haas2022nuclear}, sensing \cite{o2019hamiltonian, poggiali2018optimal}, quantum simulation \cite{tseng1999quantum,bookatz2014hamiltonian,cappellaro2007simulations}, multi-body physics \cite{wei2018exploring,niknam2020sensitivity,peng2021floquet,jalabert2001environment}, and quantum computing \cite{souza2011robust}. The formalism Waugh developed for the coherent control of coupled spin-1/2 particles naturally extends to the control of coupled qubits, as the control of qubits and spin-1/2 particles are isomorphic.
	
	AHT is based on the principle that the integrated action of a time-varying total Hamiltonian, $H_\text{tot}(t)$, over a period $[0,T_\text{seq})$, can be represented by a unitary operator or, equivalently, an effective Hamiltonian, $H_\text{eff}$, and its duration, $T_\text{seq}$. 
	The total Hamiltonian of a quantum system is typically partitioned into a user-defined, time-dependent control Hamiltonian and a time-independent internal Hamiltonian
	\begin{equation}
		H_\text{tot}(t)=H_c(t)+H_\text{int}.
		\label{totH}
	\end{equation}
	The total unitary propagator is
	\begin{equation}
		\begin{split}
			U(T_\text{seq})&=\mathcal T\exp\lr{-i\int_0^{T_\text{seq}}H_\text{tot}(t)dt}\\
			&=\exp\lr{-iH_\text{eff}T_\text{seq}},
		\end{split}
		\label{totU}
	\end{equation}
	where $\mathcal T$ is the Dyson time-ordering operator, and $H_\text{eff}$ is the effective Hamiltonian, formally defined as the matrix logarithm of $U(T_\text{seq})$. In practice, AHT uses the Magnus expansion to evaluate the effective Hamiltonian perturbatively.
	
	While control design often targets a unitary propagator or a state transfer \cite{barenco1995elementary,bravyi2005universal,khaneja2005optimal,brif2010control,koch2022quantum,mabuchi2005principles,fortunato2002design,fortunato2002implementation,viola2001experimental,pravia2003robust,palao2008protecting}, there are situations where it is more convenient to design for a target Hamiltonian \cite{haeberlen2012high,britton2012engineered,bookatz2014hamiltonian,sanchez2020perturbation,hafezi2014engineering,petiziol2021quantum,oka2019floquet,peng2021floquet,khromets2025hamiltonian,haas2019engineering,ajoy2019selective,kern2006selective,baum1985multiple,munowitz1987multiple}:
	\begin{enumerate}
		\item When the Hilbert space is large or contains uncertainties that prevent direct computation of a target unitary.
		\item When symmetry-based or scalable solutions are desired that remain valid for any system size.
		\item When one aims to control or tune the relative strength of the dynamics.
	\end{enumerate}
	Moreover, Hamiltonian engineering has been extensively used to realize robust unitaries and state transfers \cite{brown2004arbitrarily,haas2019engineering,wimperis1994broadband,levitt1981composite,shaka1983evaluation,o2013wurst,levitt1984composite,freeman1980radiofrequency,levitt1986composite,levitt1979nmr,low2014optimal,levitt1983composite,ryan2010robust,fung2000improved,waugh1982systematic,shaka1983improved,gordon2008optimal,hayes2014programmable,shaka1988iterative,shaka1983evaluation} where the robustness is achieved by minimizing the effect of an error Hamiltonian.
	
	When designing an effective Hamiltonian using the Magnus expansion, the most efficient solution is often achieved at zeroth order, as higher-order terms typically make increasingly smaller contributions \cite{haeberlen1968coherent,maricq1982application}. However, achieving high precision and robustness in the engineered effective Hamiltonian requires accounting for higher-order corrections, as these terms can accumulate significantly over many cycles.

	Table \ref{obj} lists common use cases of AHT, along with their goals and typical figures of merit.
	\begin{table*}[htbp]
		\centering
		\caption{Overview of applications of AHT with respective targets, goals, and figures of merit (FoM).}
			\begin{tabular}{|c|c|c|c|}\toprule
			\hline
			Use case&	Target & Goal & FoM \\ \hline
			Decoupling&	$H_\text{eff}=s\sum_i\delta_i\vec v_\text{eff}\cdot\vec\sigma_i$ & spectroscopy& line narrowing ($\frac{1}{\pi T_2}$), $s T_{2x}$\\ \hline
			Time suspension&	$H_\text{eff}=0$ & store quantum information& decoherence ($\frac{1}{\pi T_2}$)\\ \hline
			Simulation&	$H_\text{eff}=H_\text{sim}$,  \newline $ U(nT_\text{seq})=e^{-iH_\text{sim}nT_\text{seq}}$ & simulating quantum systems& $\|H_\text{eff}-H_\text{sim}\|$\\ \hline
			Sensing&	$H_\text{eff}=H_\text{sen}$ & parameter estimation& Fisher information\\ \hline
			Gates&	$U(T_\text{seq})=U_\text{target}$ & computation& gate fidelity\\ \hline
			\bottomrule
		\end{tabular}
		\label{obj}
	\end{table*}
One example of a time-independent internal Hamiltonian,
relevant to decoupling spin-spin interactions, refocusing qubit detunings, or simulating
spin systems (Table \ref{obj}), is:
\begin{equation}
	H_\text{int}=\sum_{i\ne j}^NB_{ij}\vec\sigma_i\cdot\mathbf{D}\cdot\vec\sigma_j+\sum_{i=1}^N\delta_i\vec n\cdot\vec\sigma_i
	\label{intexp}
\end{equation}
where $\vec\sigma_i=(\x^i,\y^i,\z^i)$ and 
\begin{equation}
	\sigma_\alpha^i=\mathbb 1^{\otimes(i-1)}\otimes\sigma_\alpha\otimes\mathbb 1^{\otimes(N-i)},
\end{equation}
for $\alpha=\lc{x,y,z}$,
are the Pauli operators acting on the $i$th qubit, $B_{ij}$ are coupling strengths, $\mathbf{D}$ is a rank-2 tensor determining the form of the qubit interactions, $\vec n$ is a unit vector determining the direction, $\delta_i$ the strength of the qubit-dependent detunings, and $N$ is the total number of qubits. 
Note that the algorithms outlined later in the paper do not depend on the particular organization of the Hamiltonians. We will show later that more complex internal Hamiltonians are easily accommodated. 

In Table I, the goal of spectroscopy \cite{tovsner2009optimal} is to measure select contributions to the internal Hamiltonian while suppressing unwanted terms. A common goal is to measure the linear terms while suppressing the bilinear terms in $H_\text{int}$ \cite{haeberlen1968coherent}. The scaling factor $s$ is preferably large and the effective direction $\vec v_\text{eff}$ is typically chosen to be along the quantization axis. Time suspension \cite{cory1990time, viola1999dynamical, zhang2008long} is an extension of decoupling that aims to suppress all of the terms in $H_\text{int}$ so, \( H_\text{eff} = 0 \). It enables qubit storage.
$H_\text{sim}$ and $H_\text{sen}$ are the target effective Hamiltonians for simulation and sensing tasks. 
In quantum simulation, the goal is to transform \( H_\text{int} \) into \( H_\text{sim} \), typically to explore the dynamics of the simulated system. Quantum simulation is often performed stroboscopically with $H_\text{sim}(T_\text{seq})$ being the effective Hamiltonian over a repeated cycle of length $T_\text{seq}$ and $n$ being the index of the stroboscopic measurement. In sensing, the goal is to engineer a Hamiltonian, \( H_\text{sen} \), that offers advantages, such as enhanced Fisher information, for measuring target parameters such as the strength of an external field.
In gate design, the goal is to implement a target unitary \( U_\text{target} \) efficiently and with robustness against errors. The figure of merit (FoM) is the gate fidelity.	

An important step toward fully utilizing AHT in Hamiltonian engineering is to determine which effective Hamiltonians are achievable under a given control setting. This is essential, even under universal control, because the space/set of achievable effective Hamiltonians depends on the chosen interaction frame (toggling frame \cite{haeberlen2012high,mansfield1971symmetrized,mehring2012principles,evans1967time,haas2019engineering}), and there is no guarantee that a desired target Hamiltonian is attainable at a given order. Characterizing controllability is useful for determining whether a desired target is achievable and to identify what additional controls are needed when it is not. Additionally, identifying the space/set of achievable effective Hamiltonians reveals the possible forms/strengths of errors \cite{knill1997theory,knill2000theory,merrill2014progress,kribs2005unified,d2008lie,jiang2017non} and enables more efficient sequence design and numerical optimization by restricting the search to the minimal relevant subspace \cite{larocca2021krylov,chen2017preparing,chen2020subspace,liu2024optimal,larocca2021krylov,heya2023subspace,abdel2008efficient} while suppressing harmful error terms.

There has been extensive study on the controllability of quantum systems \cite{albertini2003notions,albertini2002lie,albertini2001notions,altafini2001controllability,altafini2003controllability,zimboras2015symmetry,chen2017preparing,chen2020subspace,hincks2011equivalent,hodges2008universal,borneman2012parallel,zeier2011symmetry}. These studies typically rely on the symmetry properties of the Lie algebra generated by the total Hamiltonian to determine the set of achievable unitaries or classes of effective Hamiltonians. However, to the best of our knowledge, there is no well-known general framework for characterizing the exact set of achievable effective Hamiltonians when they are engineered through the Magnus expansion.

Recent advances have enhanced the application of AHT. For example, J. Choi et al. \cite{choi2020robust} proposed a symmetry-based method for engineering effective Hamiltonians using control sequences composed of $\pi/2$ pulses and delays, specifically targeting the decoupling of the zeroth-order average Hamiltonian. It enables the design of high-performance sequences for spectroscopy and quantum simulations, demonstrating superior performance compared to existing methods in the presence of strong disorders.
H. Zhou et al. \cite{zhou2023robust} and M. Tyler et al. \cite{tyler2023higher} extended this approach to account for higher-order terms. These methods focus on control with $\pi/2$ rotations of the qubits, anisotropic Heisenberg model, and lower-order corrections. Engineering nonzero many-body effective Hamiltonians using $\pi$ and $\pi/2$ rotations have been considered in \cite{bassler2025general}. 

Reinforcement learning has been utilized to engineer effective Hamiltonians using $\pi/2$ pulses to decouple secular dipolar interactions \cite{peng2022deep}. It has also been used to autonomously design high-fidelity quantum logic gates on superconducting systems \cite{baum2021experimental}.
These methods highlight the potential of advanced numerical algorithms to significantly enhance the design of sequences for Hamiltonian engineering \cite{fortunato2002design}. 

For gate design, H. Haas \textit{et al.} \cite{haas2019engineering} utilized the van Loan equation \cite{golub2013matrix} to enable numerical evaluation of the Magnus expansion for control sequences. The authors further integrated their framework with gradient ascent pulse engineering (GRAPE) \cite{khaneja2005optimal}, facilitating efficient numerical optimization for designing sequences in Hamiltonian engineering. Other numerical methods have also been proposed \cite{puzzuoli2023algorithms}. 

P. Poggi \textit{et al.} \cite{poggi2024universally} proposed a method to achieve leading-order universal robustness to a static Hamiltonian perturbation. \cite{buchanan2025seedless} introduced a package based on GRAPE to design robust operations for NMR experiments. \cite{guo2024engineering,xu2025perturbative} provide a theoretical framework for engineering target Hamiltonians on a periodically driven oscillator using Floquet engineering while accounting for the higher-order corrections.

Several key challenges remain unresolved:
\begin{itemize}
	\item \textbf{Controllability:} A systematic method to find the achievable set of effective Hamiltonians across different systems and controls \cite{d2021introduction,cong2014control} is not well-known.
	\item \textbf{Robustness to control errors:} There is no general methodology for designing robust sequences that mitigate systematic errors across different control systems.
	\item \textbf{Higher-order decoupling:} A generic symmetry-based framework for eliminating higher-order correction terms is still lacking.
\end{itemize}
In addition, many approaches to AHT, and particularly to engineering decoupling sequences, have depended on artisan methods where the researcher intuition is a key step in directing the design. Beyond idealized examples, this is not an effective approach and often fails when robustness to unknown terms is desired.

To address the challenges, we present a unified framework for engineering zeroth-order average Hamiltonians while minimizing higher-order corrections and achieving robustness against systematic errors. The framework also extends to account for stochastic noise and to design robust state transfers.

All achievable zeroth-order average Hamiltonians are shown to lie within a minimal subspace, identified through a simple Lie-algebraic procedure. The achievable set forms a convex region characterized by random sampling and linear programming, revealing attainable targets and efficiency bounds.

Expressing the toggling-frame Hamiltonian in this subspace yields general conditions for Hamiltonian engineering, robustness, and higher-order minimization, all expressed as time-ordered integrals of its coefficients. These integrals admit closed-form solutions for piecewise-constant modulation and naturally generalize to stochastic fluctuations.

Sequence design thus reduces to finding a trajectory in the minimal subspace that satisfies these integral constraints, which define objective functions for numerical optimization. We demonstrate the framework with examples in gate optimization, quantum simulation, and sensing.

The remainder of the paper is organized as follows. Sec. \ref{sectog} introduces the toggling frame and the principles of effective Hamiltonian engineering. Sec. \ref{secset} describes the control setup, parameter dispersion, and the figure of merit for robust control. Secs. \ref{seccon}–\ref{secstate} present the theoretical framework for controllability characterization and robust Hamiltonian engineering. Sec. \ref{secflow} provides a flowchart summarizing the design procedure, followed by examples and comparisons with other approaches in Secs. \ref{secexam} and \ref{seccom}, and concluding with a discussion (Sec. \ref{secdis}) and final remarks (Sec. \ref{secsum}).
\section{Effective Hamiltonian engineering and toggling frame}\label{sectog} 
It is often more convenient to define the effective Hamiltonian in the toggling frame, which is based on partitioning \(H_\text{tot}(t)\) into a primary and a perturbative component:
\begin{equation}
	H_\text{tot}(t)=H_\text{pri}(t)+H_\text{pert}(t).
\end{equation}
The toggling frame is the interaction frame of $H_\text{pri}(t)$ and the toggling-frame Hamiltonian of $H_\text{pert}(t)$ is
\begin{equation}
	H_\text{tog}(t)=U_\text{pri}^\dagger(t)H_\text{pert}(t)U_\text{pri}(t),
\end{equation}
where the primary unitary $U_\text{pri}(t)$ is 
\begin{equation}
	U_\text{pri}(t)=\mathcal T\exp\left(-i\int_0^{t}H_\text{pri}(t')dt'\right).
\end{equation}
The perturbative unitary is defined through an effective Hamiltonian $H_\text{eff}$
\begin{equation}
	\begin{split}
		U_\text{pert}({T_\text{seq}})&=\mathcal T\exp\left(-i\int_0^{T_\text{seq}}H_\text{tog}(t)dt\right)\\
		&=\exp\left(-i H_\text{eff} T_\text{seq}\right).
	\end{split}
\end{equation}
For unitary dynamics, an effective Hamiltonian \(H_\text{eff}\) always exists and describes the system’s evolution over the interval \([0, T_\text{seq})\) in the toggling frame. It is not unique, as adding integer multiples of \(2\pi/T_\text{seq}\) to its eigenvalues yields the same unitary \(U(T_\text{seq})\). 
The total propagator is 
\begin{equation}
	\begin{split}
		U({T_\text{seq}})=U_\text{pri}({T_\text{seq}})U_\text{pert}(T_\text{seq}).
	\end{split}
	\label{ttog}
\end{equation}
Typically, $U_\text{pri}(t)$ is  evaluated through numerical integration while the effective Hamiltonian of $U_\text{pert}(t)$ is evaluated perturbatively through the Magnus expansion:
\begin{equation}
	H_\text{eff}(T_\text{seq})=\sum_{r=1}^\infty\bar H^{(r-1)}(T_\text{seq}),
\end{equation}
where $r-1$ is the perturbation order, and $\bar H^{(r-1)}$ is the $(r-1)$th-order average Hamiltonian. The first few terms of the expansion are
\begin{equation}
	\begin{split}
		&\bar H^{(0)}(T_\text{seq})=\frac{1}{T_\text{seq}}\int_0^{T_\text{seq}}H_\text{tog}(t)dt,\\
		&\bar H^{(1)}(T_\text{seq})=-\frac{i}{2T_\text{seq}}\int_0^{T_\text{seq}}dt_1\int_0^{t_1}dt_2[ H_\text{tog}(t_1),H_\text{tog}(t_2)],\\
		&\bar H^{(2)}(T_\text{seq})=\\
		&-\frac{1}{6T_\text{seq}}\int_0^{T_\text{seq}}dt_1\int_0^{t_1}dt_2\int_0^{t_2}dt_3\\
		&~~~~~~~~~~~~~~\{[H_\text{tog}(t_1),[H_\text{tog}(t_2),H_\text{tog}(t_3)]]\\
		&~~~~~~~~~~~~~~~~+[H_\text{tog}(t_3),[H_\text{tog}(t_2),H_\text{tog}(t_1)]]\},\\
		&\cdots.
		\label{higherterms}
	\end{split}
\end{equation}
In this work, we focus on achieving a target effective Hamiltonian at zeroth order
\begin{equation}
	H_\text{target}=\bar H^{(0)}(T_\text{seq}),
	\label{stra}
\end{equation}
while minimizing higher-order corrections ($\bar H^{(1)},\bar H^{(2)},\ldots$), treating each order independently.
Whenever possible, implementing \(H_\text{target}\) at zeroth order is typically most efficient (e.g., in terms of resources such as the power–bandwidth product). However, for a fixed partitioning of $H_\text{tot}$, not all desired \(H_\text{target}\) are achievable in zeroth order; engineering nonzero higher-order terms will be discussed in a separate work.

In conventional AHT, the control Hamiltonian \(H_c\) is taken as \(H_\text{pri}\) and the internal Hamiltonian \(H_\text{int}\) as \(H_\text{pert}\). This choice is not required; allowing different partitions of \(H_\text{tot}\) expands the set of achievable zeroth-order controls. A more effective strategy is to include control imperfections, parameter dispersions, and components to be engineered or reshaped (e.g., for simulation or sensing) in \(H_\text{pert}\), and assign the remaining well-characterized terms to \(H_\text{pri}\). A discussion on how to choose \(H_\text{pri}\) and \(H_\text{pert}\) is in Appendix \ref{appen0}.
\section{Control Setup}\label{secset}
The control Hamiltonian is most often implemented as a sequence of piecewise-constant intervals of control parameters $\lc{a_k(t)}$. 
For example, it can be
\begin{equation}
	\begin{split}
		&H_c(t)\\
		=&\sum_{i=1}^N [\omega_1^i(t)\cos(\phi_i(t))\x^i+\omega_1^i(t)\sin(\phi_i(t))\y^i\\
		&~~~~~~~~~~~~~~~~~~~~~~~~~~~~~~+(\omega_0-\omega_r^i(t))\z^i],
	\end{split}
	\label{controlexp}
\end{equation}
where the time-dependent amplitude ($\omega_1^i(t)$), frequency ($\omega_r^i(t)$) and phase of the control field ($\phi_i(t)$) on the $i$th qubit are control parameters ($a_k(t)$). 
$a_k(t)$ is typically set to zero at the beginning and the end of each sequence \cite{fortunato2002design}, enabling a simple composition of gates.
\begin{figure*}[htp]
	\centering
	\includegraphics[width=0.8\textwidth]{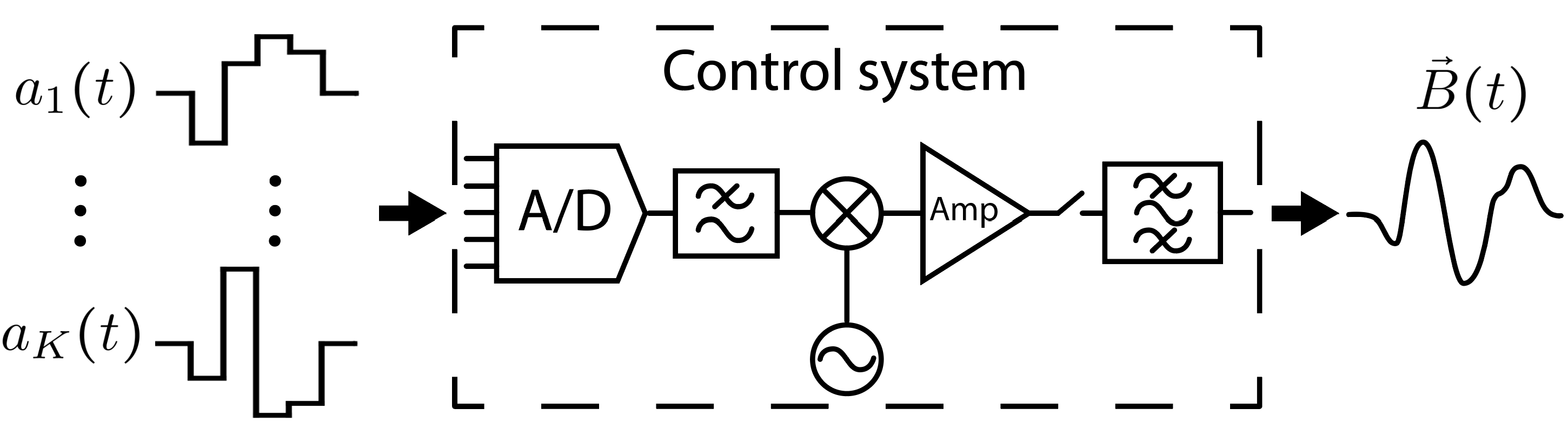}\\
	\caption{Illustration of a very simplified control system. The input signals $\lc{a_k(t)}$ are designed by the user. The control system outputs a continuous control field $\vec B(t)$ acting on the qubit system in the lab frame. The design of $\lc{a_k(t)}$ should account for the model of the system so that it captures the distortion of the hardware along with additive noise. In this case, the control parameters $a_k(t)$ are transformed via convolution with a kernel reflecting the bandwidth of the low- and band-pass filters. More complete mappings are explored in Sec. \ref{nlsec}.}
	\label{condis}
\end{figure*}

The control system (Fig. \ref{condis}) maps $\lc{a_k(t)}$ to a continuous output field $\vec B(t)$ that is conveniently described in the lab frame.
This map is generally a non-invertible, smooth, nonlinear function depending on time-independent parameters of the control system, $\lc{\mu_j}$, such as reactance, spatial and temporal variations of control fields, nonlinearities from kinetic inductance, etc. Variations in $\lc{\mu_j}$ result in a modification of the control Hamiltonian by an error Hamiltonian $\Delta H$. The error Hamiltonian $\Delta H$ can be evaluated by applying the Taylor expansion to the map from $\lc{a_k(t)}$ to $\vec B(t)$ with respect to $\lc{\mu_j}$. Robustness to variations in $\lc{\mu_j}$ can then be achieved by minimizing the effective Hamiltonian of $\Delta H$.
The following methods are capable of finding high-precision control fields that account for $\lc{\mu_j}$ and are robust to their variations. 

The internal Hamiltonian also depends on a set of parameters, $\lc{\eta^i_\text{int}}$, which can include terms such as the Zeeman splitting, chemical shift, coupling strengths, etc. Typically, $\{\eta^i_\text{int}\}$ are governed by a classical probability distribution with a correlation time that is often longer than the total length of a control cycle ($T_\text{seq}$).

Given a sequence $\lc{a_k(t)}$, as well as the internal and model parameters ($\lc{\eta_\text{int}^i}$ and $\lc{\mu_j}$), the density matrix of the system obeys the master equation
\begin{equation}
	\begin{split}
		\frac{d}{dt}\rho(t)=-i[H_\text{tot}(\lc{a_k(t)};\lc{\mu_j},&\lc{\eta_\text{int}^i}),\rho(t)]\\
		&+\mathcal L_\text{relaxation}[\rho(t)]
		\label{master}
	\end{split}
\end{equation}
where $\mathcal L_\text{relaxation}$ is the Lindbladian for relaxation. The solution to Eq. \eqref{master} over the interval $[0,T_\text{seq})$ is 
\begin{equation}
	\rho(T_\text{seq})=\Lambda(\lc{a_k(t)};\lc{\mu_j},\lc{\eta_\text{int}^i})(\rho_0),
	\label{singlesol}
\end{equation}
where $\rho_0$ is the initial density matrix and $\Lambda(\lc{a_k(t)};\lc{\mu_j},\lc{\eta_\text{int}^i})$ is a completely positive trace-preserving (CPTP) map describing the dynamics averaged over a distribution of isochromats. The density matrix of the system for an expectation-valued measurement is the ensemble average of Eq. \eqref{singlesol} over the distributions of $\lc{\mu_j}$ and $\lc{\eta_\text{int}^i}$:
\begin{equation} 
	\bar\rho(T_\text{seq})=\bar\Lambda(\rho_0),
\end{equation}
where the average CPTP map $\bar \Lambda$ is
\begin{equation}
	\begin{split}
		&\bar\Lambda=\\
		&\int d\lc{\mu_j} d\lc{\eta_\text{int}^i} p(\lc{\mu_j})p(\lc{\eta_\text{int}^i})\Lambda(\lc{a_k(t)};\lc{\mu_j},\lc{\eta_\text{int}^i}).
		\label{dan}
	\end{split}
\end{equation}
Here, $p$ denotes the probability densities of the parameters.
The goal is to design $\lc{a_k(t)}$ to achieve the desired FoM with respect to $\bar\Lambda$ or $\bar\rho(T_\text{seq})$ \cite{magesan2013modeling} for tasks in Table \ref{obj}. For example, if the overall unitary target for $U(T_\text{seq})$ is $U_0$, then the average gate fidelity of the sequence can be calculated as
\begin{equation}
	\begin{split}
		\mathcal F_\text{ave}(\lc{a_k(t)},U_0)
		=\int \langle\psi|U_0^\dagger\bar\Lambda(|\psi\rangle\langle\psi|) U_0|\psi\rangle d\psi.
	\end{split}
	\label{avegate}
\end{equation}
When knowing the details of the errors is important (e.g., in implementing quantum error correction, noiseless subsystems, and distinguishing between coherent and stochastic errors), one should report the full average CPTP map $\bar \Lambda$. In general, the coherence of the error can be measured by the orthogonality of $\bar \Lambda$:
\begin{equation}
	\mathcal R(\bar\Lambda)=\frac{\text{Tr}(\bar \Lambda^T\bar \Lambda)}{\text{dim}(\bar \Lambda)}.
	\label{ortho}
\end{equation}
A small orthogonality value indicates limited robustness or an overly long sequence.

Since the FoM and the CPTP map $\bar\Lambda$
are typically estimated via Monte Carlo simulations, they are not suitable for direct use in the design of control sequences. Instead, a combination of fidelity measures for \(U_\text{pri}\) and \(H_\text{target}\) is used in the optimization.  
The fidelity of \(U_\text{pri}(T_\text{seq})\) is evaluated using the overlap  
\begin{equation}
	\mathcal{F}(U_\text{pri}(T_\text{seq}), U_\text{target}) =
	\frac{|\mathrm{Tr}[U_\text{pri}^\dagger(T_\text{seq}) U_\text{target}]|}
	{\mathrm{Tr}[U_\text{target}^\dagger U_\text{target}]}.
	\label{fubi}
\end{equation}
The same fidelity measure, \(\mathcal{F}(U(T_\text{seq}), U_0)\), is also used to visualize control landscapes.  
Next, we show how to determine the set of achievable \(\bar{H}^{(0)}\) and define fidelity measures for \(\bar{H}^{(0)}\) and for minimizing higher-order corrections (Sec. \ref{seceng}). The complete cost function is presented in Sec. \ref{secflow}.
\section{Controllability of $\bar H^{(0)}$}\label{seccon}
This section develops a general method for finding the minimal subspace and set of achievable $\bar H^{(0)}$ for arbitrary $H_\text{pri}$ and $H_\text{pert}$. This needs to make use of the minimal Lie algebra \cite{d2021introduction} containing $H_\text{pri}(t)$ (Appendix \ref{appen1}). For example, $\mathbf{g}_\text{pri}=\text{span}_\text{Lie}\left\{\x,\y,\z\right\}$ for universal control on a single qubit. 
The corresponding Lie group 
$e^{\mathbf{g}_\text{pri}}=\lc{e^{g}|g\in\mathbf{g}_\text{pri}}$
gives the set of achievable $U_\text{pri}(T_\text{seq})$ \cite{d2021introduction}. 

\begin{algorithm}[H]
	\caption{Algorithm for calculating an orthonormal basis of $\mathscr{C}({\mathbf{g}_\text{pri}}, H_\text{pert})$. The function \textsc{FindC}($\mathcal{B}({\mathbf{g}_\text{pri}})$, $H_\text{pert}$) takes as input a basis $\mathcal{B}({\mathbf{g}_\text{pri}}) = \lc{e_1, \ldots, e_{|{\mathbf{g}_\text{pri}}|}}$ of ${\mathbf{g}_\text{pri}}$ and the operator $H_\text{pert}$, and returns an orthonormal basis of $\mathscr{C}({\mathbf{g}_\text{pri}}, H_\text{pert})$. Here `$\leftarrow$' denotes assignment, i.e., the variable on the left is set to the value of the expression on the right.}\label{L2}
	\begin{algorithmic}[1]
		\Function{FindC}{$\mathcal B({\mathbf{g}_\text{pri}})$, $H_\text{pert}$}
		\State $h_{1}\gets H_\text{pert}$
		\State $i_\text{max}\gets1$ \Comment{\textit{Total number of basis vectors.}}
		\State $i_\text{new}\gets1$ \Comment{\textit{Number of new basis vectors.}}
		\While{$i_\text{new}>0$ and $i_\text{max}\le 2^{2N}-1$}
		\State $t\gets0$
		\For{$i_\text{max}-i_\text{new}+1\le i\le i_\text{max}$}
		\For{$1\le j\le |{\mathbf{g}_\text{pri}}|$}
		\If{$h_1,\ldots,h_{i_\text{max}+t},[e_j,h_i]$ are linearly independent}
		\State $t\gets t+1$
		\State $h_{i_\text{max}+t}\gets[e_j,h_i]$
		\EndIf
		\EndFor
		\EndFor
		\State $i_\text{max}\gets i_\text{max}+t$
		\State $i_\text{new}\gets t$
		\EndWhile
		\State Use the Gram–Schmidt process to orthonormalize $(h_1,\ldots,h_{i_\text{max}})$ and \Return the result
		\EndFunction
	\end{algorithmic}
\end{algorithm}

Since $\bar H^{(0)}$ can be any time-weighted average of $H_\text{tog}$ (first Eq. in \eqref{higherterms}), the minimal subspace containing the toggling-frame Hamiltonian is also the subspace of achievable zeroth-order average Hamiltonians. This subspace is
\begin{equation}
	\mathscr{C}({\mathbf{g}_\text{pri}}, H_\text{pert})=\text{span}_\mathbb{R}\left\{U_\text{pri}^\dagger H_\text{pert}U_\text{pri}\middle|U_\text{pri}\in e^{\mathbf{g}_\text{pri}}\right\}.
\end{equation} 
Since $H_\text{tog}(t)$ is determined by the action of $H_\text{pri}(t)$ on $H_\text{pert}$, the subspace can be calculated by taking nested commutators between ${\mathbf{g}_\text{pri}}$ and $H_\text{pert}$ (Appendix \ref{appen2}):
\begin{equation}
	\mathscr{C}({\mathbf{g}_\text{pri}}, H_\text{pert})=\text{span}_{\mathbb R}\left\{[g_1,\cdots[g_L,H_\text{pert}]\cdots]\middle|g_l\in {\mathbf{g}_\text{pri}}, L\ge0\right\}.
\end{equation}
Algorithm \ref{L2} calculates an orthonormal basis of $\mathscr{C}({\mathbf{g}_\text{pri}}, H_\text{pert})$.

Next, we show how to find the set of achievable $\bar H^{(0)}(T_\text{seq})$ in $\mathscr{C}({\mathbf{g}_\text{pri}}, H_\text{pert})$.
This set is 
\begin{equation}
	\begin{split}
		&\mathscr{O'}(\mathbf{g}_\text{pri}, H_\text{pert})\\
		=&\left\{\frac{1}{T_\text{seq}}\int_0^{T_\text{seq}}U_\text{pri}^\dagger(t)H_\text{pert}U_\text{pri}(t)dt\right\}.
	\end{split}
\end{equation}
Assuming \(H_\text{pert}\) is normalized, for any normalized \(H_\text{target} \in \mathscr{C}(\mathbf{g}_\text{pri}, H_\text{pert})\), an achievable scaling factor \(s\) for \(H_\text{target}\) is defined as a real number \(s\) such that there exists a sequence \(\{a_k(t)\}\) satisfying \(\bar{H}^{(0)}(T_\text{seq}) = s H_\text{target}\).
Thus, fully characterizing \(\mathscr{O'}(\mathbf{g}_\text{pri}, H_\text{pert})\) is equivalent to determining the range of achievable scaling factors for each \(H_\text{target} \in \mathscr{C}(\mathbf{g}_\text{pri}, H_\text{pert})\).

Note that \(\mathscr{O'}(\mathbf{g}_\text{pri}, H_\text{pert})\) is the convex hull of $\{U_i^\dagger H_\text{pert} U_i\}$ ($U_i\in e^{\mathbf g_\text{pri}}$), which can be approximated through random sampling \cite{mezzadri2006generate} (Appendix \ref{appen4}). 
\begin{figure}[htp]
	\centering
	\includegraphics[width=0.4\textwidth]{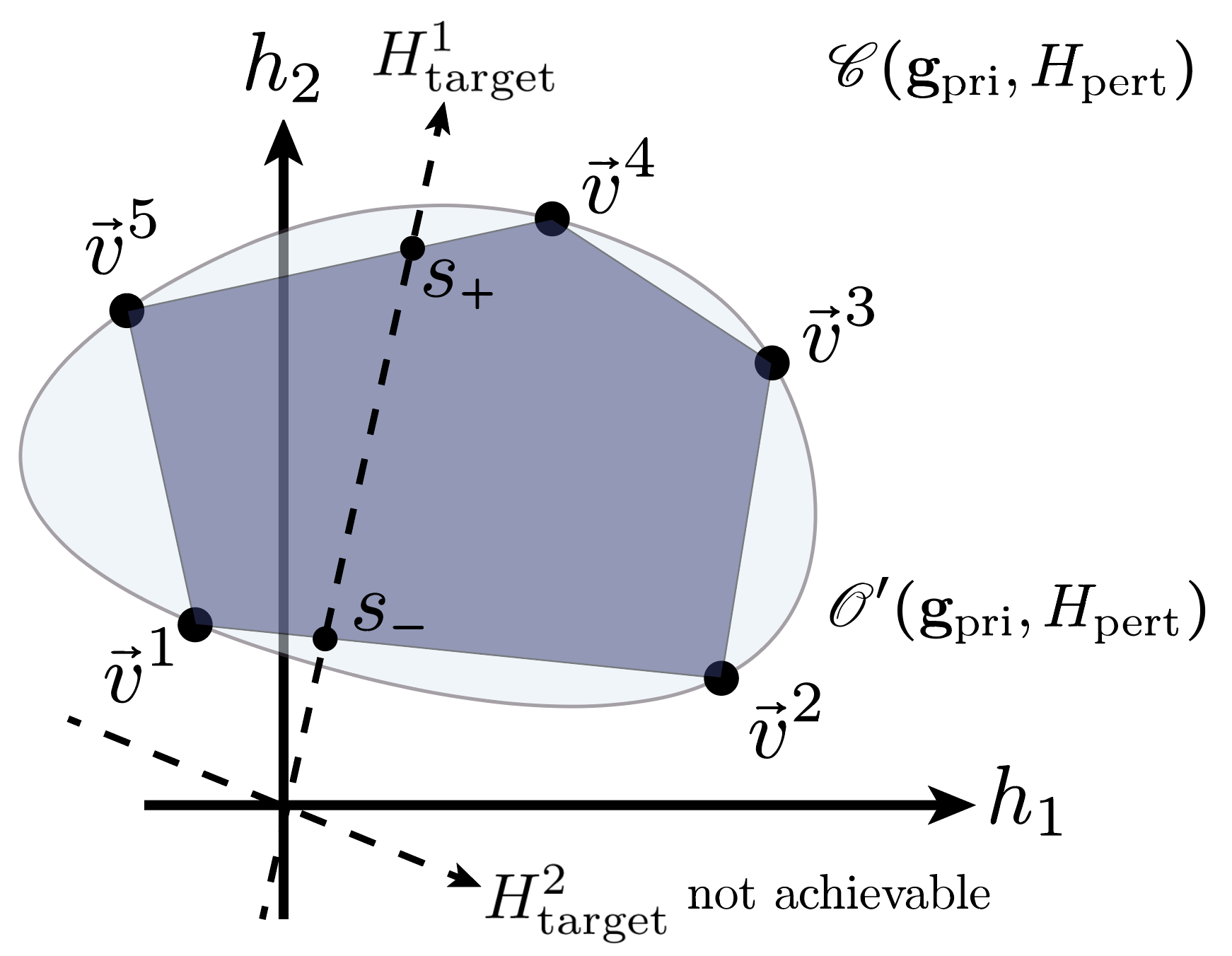}\\
	\caption{Schematic for finding achievable scaling factors. The set of achievable zeroth-order average Hamiltonians $\mathscr{O'}(\mathbf{g}_\text{pri}, H_\text{pert})$ (light blue area) is a convex set in $\mathscr C({\mathbf{g}_\text{pri}},H_\text{pert})$ (spanned by $h_1$ and $h_2$). The convex set is approximated by the polygon (dark blue area) formed by $\lc{\vec v_i=U_i^\dagger H_\text{pert}U_i}$. The achievable scaling factors for $H_\text{target}$ can then be found via linear programming.}
	\label{convex}
\end{figure}
Therefore, along each $H_\text{target}\in\mathscr{C}({\mathbf{g}_\text{pri}},H_\text{pert})$, the range of achievable scaling factors can be found by solving a linear program associated with the sampled set (Fig. \ref{convex}). The procedure is summarized in Algorithm \ref{L3} (formal version provided in Appendix \ref{appen3}).

\begin{algorithm}[H]
	\caption{{\bf{(informal version)}} Algorithm for finding the range of achievable scaling factors for $H_\text{target}$.}\label{L3}
	\begin{algorithmic}[1]
		\State Sample random unitaries $U_i\in e^{{\mathbf{g}_\text{pri}}}$ to form the set $\{U_i^\dagger H_\text{pert} U_i\}$
		\State Solve the linear program that maximizes (minimizes) the projection of the convex combination of $\{U_i^\dagger H_\text{pert} U_i\}$ onto $H_\text{target}$; the solution corresponds to $s_+$ ($s_-$)
		\State \Return $[s_-,s_+]$ as the range of achievable scaling factors
	\end{algorithmic}
\end{algorithm}
\section{Engineering $\bar H^{(0)}$ and minimizing higher-order corrections}\label{seceng}
The representation in the minimal subspace of $H_\text{tog}(t)$ can be used to obtain general conditions for engineering any $H_\text{target}$ while accounting for the higher-order corrections.
Given an orthonormal basis of $\mathscr C(\mathbf g_\text{pri}, H_\text{pert})$:
\begin{equation}
	\mathcal B(\mathscr C)=\lr{h_1,h_2,\dots,h_{|\mathscr C|}},
	\label{cbasis}
\end{equation}
where $|\mathscr C|$ is the dimension of $\mathscr C(\mathbf g_\text{pri},H_\text{pert})$,
the toggling-frame Hamiltonian can be represented as a vector in $\mathscr C(\mathbf g_\text{pri},H_\text{pert})$ (Appendix \ref{appen6}) as
\begin{equation}
	\begin{split}
		|H_\text{tog}(t)\rrangle&=\lr{c_1(t),c_2(t),\ldots,c_{|\mathscr C|}(t)},\\
		&=\mathcal D(U^\dagger_\text{pri}(t))|H_\text{pert}\rrangle.
	\end{split}
\end{equation}
Here, $\mathcal D(U^\dagger_\text{pri}(t))$ is the adjoint representation of $U^\dagger_\text{pri}(t)$ in $\mathscr C(\mathbf g_\text{pri},H_\text{pert})$. The zeroth-order average Hamiltonian is
\begin{equation}
	|\bar H^{(0)}\rrangle=\frac{1}{T_\text{seq}}\int_0^{T_\text{seq}}\mathcal D(U^\dagger_\text{pri}(t))|H_\text{pert}\rrangle.
	\label{pogi}
\end{equation}
The condition for engineering $H_\text{target}$ with a scaling $s$ is then
\begin{equation}
	|\bar H^{(0)}\rrangle=|sH_\text{target}\rrangle.
	\label{0th}
\end{equation}

Now, we consider how to eliminate higher-order corrections for the entire distributions of Hamiltonian parameters. 
Denote the time-ordered integrals of the coefficients $c_i(t)$ ($\mathscr C-$integrals) as
\begin{equation}
	\bar c_{i_1\cdots i_r}=\int_0^{T_\text{seq}}dt_1\cdots\int_0^{t_{r-1}}dt_rc_{i_1}(t_1)\cdots c_{i_r}(t_r),
	\label{Cint}
\end{equation}
the Magnus expansion can be written as
\begin{equation}
	\begin{split}
		&\bar H^{(r-1)}(T_\text{seq})\\
		=&\frac{1}{T_\text{seq}}\int_0^{T_\text{seq}}dt_1\cdots\int_0^{t_{r-1}}dt_rF_r(H_\text{tog}(t_1),\ldots,H_\text{tog}(t_r))\\
		=&\frac{1}{T_\text{seq}}\sum_{i_1,\ldots,i_r=1}^{|\mathscr C|}F_r(h_{i_1},\ldots,h_{i_r})\bar c_{i_1\cdots i_r},
	\end{split}
\end{equation}
where 
\begin{equation}
	\begin{split}
		F_r(h_{i_1},\ldots,h_{i_r})
		=\sum_{\pi}(-i)^{r-1}(-1)^{d_b}\frac{d_a!d_b!}{r!}\pi\lr{h_{i_1}\cdots h_{i_r}},
	\end{split}
\end{equation}
where the sum is taken over all possible permutations $\pi$ of $h_{i_1},h_{i_2},\ldots,h_{i_r}$. $d_a$ and $d_b$ are ascents and descents of $\pi$ \cite{agrachev2017shuffle,arnal2018general}.  This sum over permutations reproduces the nested commutator formula in Eq. \eqref{higherterms}, e.g.,
\begin{equation}
	\begin{split}
		&F_2(h_{i_1},h_{i_2})=-\frac{i}{2}[h_{i_1},h_{i_2}],\\
		&F_3(h_{i_1},h_{i_2},h_{i_3})=-\frac{1}{6}\lr{[h_{i_1},[h_{i_2},h_{i_3}]]+[h_{i_3},[h_{i_2},h_{i_1}]]}.
	\end{split}
\end{equation}
It is easy to show that 
\begin{equation}
	\begin{split}
		&F_r(h_{i_1},\ldots,h_{i_r})=(-1)^{r-1}F_r(h_{i_r},\ldots,h_{i_1}),\\
		&F_r(h,\ldots,h)=0.
	\end{split}
\end{equation}
Therefore,
the condition for $\bar H^{(r-1)}\equiv 0$ for arbitrary $h_1,\ldots,h_r$ is given by
\begin{equation}
	\bar c_{i_1\cdots i_r}(T_\text{seq})-(-1)^{r-1}\bar c_{i_r\cdots i_1}(T_\text{seq})=0,
	\label{higher}
\end{equation}
for all $i_1,\ldots,i_r$ excluding the situations where $i_1=\cdots=i_r$.

Sometimes, it is convenient to decompose $H_\text{pert}$ into several components $H_\text{pert}=\sum_{w=1}^WH_\text{pert}^w$ (e.g., when $H_\text{pert}^w$ transform differently under $e^{\mathbf g_\text{pri}}$ and are subject to different types of variations). Then we can compute $\mathscr C_w(\mathbf g_\text{pri},H_\text{pert}^w)$ for each $w$. Define a composite $\mathscr C$ space as
\begin{equation}
	\mathscr C_\text{comp}=\bigoplus_{w=1}^W \mathscr C_w(\mathbf g_\text{pri},H_\text{pert}^w),
\end{equation}
then the composite toggling-frame Hamiltonian is
\begin{equation}
	\begin{split}
		|H_\text{tog}(t)\rrangle_\text{comp}&=\bigoplus_{w=1}^W|H_\text{tog}(t)^w\rrangle_w\\
		&=(c_1^1(t),\ldots,c_{|\mathscr C_1|}^1,\ldots,c_1^W(t),\ldots,c_{|\mathscr C_W|}^W)\\
		&=(c_1^\text{comp}(t),\ldots,c_{|\mathscr C_\text{comp}|}^\text{comp}(t)).
	\end{split}
\end{equation}
The condition for engineering each $H_\text{pert}^w$ to $H_\text{target}^w$ with a scaling factor $s_w$ is
\begin{equation}
	\begin{split}
		&\frac{1}{T_\text{seq}}\int_0^{T_\text{seq}}dt|H_\text{tog}(t)\rrangle_\text{comp}\\
		=&\frac{1}{T_\text{seq}}(\bar c_1^\text{comp}(t),\ldots,\bar c_{|\mathscr C_\text{comp}|}^\text{comp}(t))\\
		=&\bigoplus_{w=1}^W |s_wH_\text{target}^w\rrangle_w.
	\end{split}
	\label{tot0}
\end{equation}
The joint achievability of $s_wH_\text{target}^w$ is characterized using a generalized form of Algorithm \ref{L3} (Appendices \ref{appen3} and \ref{appenexam}).
The condition for minimizing higher-order corrections, including cross terms between different $H_\text{pert}^w$, is
\begin{equation}
	\bar c_{i_1\cdots i_r}^\text{comp}(T_\text{seq})-(-1)^{r-1}\bar c_{i_r\cdots i_1}^\text{comp}(T_\text{seq})=0,
	\label{totr}
\end{equation}
for all $i_1,\ldots,i_r$ excluding the situations where $i_1=\cdots=i_r$.
The composite $\mathscr C$-integrals admit exact analytic solutions, expressed as explicit functions of the eigenvalues of the primary Hamiltonian, for piecewise-constant modulations (Appendix \ref{appen6}). Consequently, Eqs. \eqref{tot0} and \eqref{totr} define general cost functions that can be directly used in numerical optimization to engineer target effective Hamiltonians and suppress higher-order corrections.
\section{Robust control against imperfections}\label{secro}
As an important application of Eqs. \eqref{tot0} and \eqref{totr}, we show how to achieve robustness against variations in the model parameters, $\lc{\mu_j}$.
When the model parameters $\vec \mu=\lc{\mu_j}$ are subject to variations $\vec\epsilon$, the total Hamiltonian is
\begin{equation}
	H_\text{tot}(t)=H_c(t,\vec \mu)+\Delta H(t,\vec\epsilon)+H_\text{int},
\end{equation}
where
\begin{equation}
	\Delta H(t,\vec\epsilon)=H_c(t,\vec \mu+\vec\epsilon)-H_c(t,\vec \mu),
\end{equation}
and $\Delta H(t,\vec 0)=0$. Denote 
\begin{equation}
	\partial_{j_1}\cdots \partial_{j_i}\Delta H=\left.\frac{\partial^i}{\partial\mu_{j_1}\cdots \partial\mu_{j_i}}\Delta H\right|_{\vec\epsilon=\vec 0},
	\label{derive}
\end{equation}
then
\begin{equation}
	\Delta H(t,\vec\epsilon)=(\vec \epsilon\cdot\vec\partial)\Delta H+\frac{(\vec \epsilon\cdot\vec\partial)^2}{2!}\Delta H+\cdots.
\end{equation}
Appendix \ref{appen10} discusses how to evaluate the derivatives of the error Hamiltonian for a general control system described by a state-space differential equation.

For simplicity, consider only the first-order derivatives, let 
\begin{equation}
	H_\text{pert}^w=\partial_w \Delta H,
\end{equation}
for $w=1,\ldots,J$. Let $H_\text{pert}^{J+1}$ denote the Hamiltonian capturing uncertainties in $H_\text{int}$ (e.g., variations in qubit detuning) and let $H_\text{pert}^{J+2}$ represent the component of the total Hamiltonian to be intentionally reshaped into a nonzero target.
Without loss of generality, assume that $H_\text{pert}^w$ are traceless, set
\begin{equation}
	|H_\text{target}^w\rrangle_w=0,
\end{equation}
for $w=1,\ldots,J+1$, and choose the desired target $H_\text{target}^{J+2}$ and a scaling factor $s_{J+2}$. Eq. \eqref{tot0} gives the condition for engineering the desired target at zeroth order and achieving zeroth-order robustness to $\Delta H$ and $H_\text{pert}^{J+1}$. Eq. \eqref{totr} ($r=2$) gives the condition for achieving first-order robustness including all cross terms and minimizing first-order corrections to $H_\text{target}^{J+2}$. It can be shown that zeroth-order robustness to systematic errors is achievable under assumptions listed in Appendix \ref{appen8}. For higher-order robustness ($r>2$), higher-order derivatives of $\Delta H$ need to be included.

\section{Robust control against stochastic fluctuations}\label{secsto}
Condition \eqref{totr} can be modified to account for random fluctuations in the systems using the cumulant expansion \cite{kubo1962generalized,kubo1963stochastic}. 

First, consider the special case where $H_\text{pert}(t)=\beta(t)H_\text{pert}^1$ where $\beta(t)$ is a random process.
Denote the adjoint representation of $U_\text{pert}$ on the full operator space by $\hat U_\text{pert}$, applying the cumulant expansion yields
\begin{equation}
	\langle \hat U_\text{pert}\rangle=\exp(K^{(0)}+K^{(1)}+\cdots),
\end{equation}
where
\begin{equation}
	\begin{split}
		&K^{(0)}=-i\int_0^{T_\text{seq}}dt_1\langle\beta(t_1)\rangle \text{ad}_{H_\text{tog}(t_1)},\\
		&K^{(1)}=\\
		&-\int_0^{T_\text{seq}}dt_1\int_0^{t_1}dt_2\langle\beta(t_1)\beta(t_2)\rangle\text{ad}_{H_\text{tog}(t_1)}\text{ad}_{H_\text{tog}(t_2)}\\
		&+\frac{1}{2}\int_0^{T_\text{seq}}dt_1\langle\beta(t_1)\rangle \text{ad}_{H_\text{tog}(t_1)}\int_0^{T_\text{seq}}dt_2\langle\beta(t_2)\rangle \text{ad}_{H_\text{tog}(t_2)},\\
		&\cdots.
	\end{split}
\end{equation}
Here, the generator of the adjoint action is defined as
\begin{equation}
	\text{ad}_A(B)=[A,B].
\end{equation}
For simplicity, consider the special case where $\langle \beta(t)\rangle=0$, the leading-order term is
\begin{equation}
	\begin{split}
		&K^{(1)}=\\
		&-\int_0^{T_\text{seq}}dt_1\int_0^{t_1}dt_2\langle\beta(t_1)\beta(t_2)\rangle\text{ad}_{H_\text{tog}(t_1)}\text{ad}_{H_\text{tog}(t_2)}
	\end{split}
	\label{lead}
\end{equation}
We focus on minimizing Eq. \eqref{lead}. Since
\begin{equation}
	\text{ad}_{H_\text{tog}(t_1)}=\sum_{i=1}^{|\mathscr C|} c_i(t)\text{ad}_{h_i},
\end{equation}
The condition for minimizing Eq. \eqref{lead} becomes
\begin{equation}
	\begin{split}
		\sum_{i_1,i_2}\bar c'_{i_1 i_2}(T_\text{seq})\text{ad}_{h_{i_1}}\text{ad}_{h_{i_2}}=0,
	\end{split}
\end{equation}
for all $i_1, i_2$, where $\bar c'_{i_1 i_2}$ are the modified $\mathscr C-$integrals
\begin{equation}
	\bar c'_{i_1 i_2}(T_\text{seq})=\int_0^{T_\text{seq}}dt_1\int_0^{t_1}dt_2\langle\beta(t_1)\beta(t_2)\rangle c_{i_1}(t_1)c_{i_2}(t_2).
\end{equation}
For example, if
\begin{equation}
	\langle\beta(t_1)\beta(t_2)\rangle=\exp(-\theta(t_1-t_2)),t_1>t_2,
\end{equation}
the modified $\mathscr C-$integrals have exact analytic solution for any piecewise-constant modulation (Appendix \ref{appen8}).

In the general case where 
\begin{equation}
	H_\text{pert}=\sum_{w=1}^W \beta_w(t)H_\text{pert}^w,
\end{equation}
the leading-order correction becomes
\begin{equation}
	\begin{split}
		K^{(1)}=
		-\sum_{w_1,w_2}&\int_0^{T_\text{seq}}dt_1\int_0^{t_1}dt_2\\
		&\langle\beta_{w_1}(t_1)\beta_{w_2}(t_2)\rangle\text{ad}_{H_\text{tog}^{w_1}(t_1)}\text{ad}_{H_\text{tog}^{w_2}(t_2)}.
	\end{split}
\end{equation}
Let
\begin{equation}
	\mathcal B(\mathscr C_\text{comp})=(h_1^\text{comp},\ldots,h_{|\mathscr C_\text{comp}|}^\text{comp})
\end{equation}
be an orthonormal basis of the composite $\mathscr C$ space, the condition for minimizing $K^{(1)}$ is
\begin{equation}
	\sum_{i_1,i_2}\bar c^{\text{comp}'}_{i_1 i_2}(T_\text{seq})\text{ad}_{h_{i_1}^\text{comp}}\text{ad}_{h_{i_2}^\text{comp}}=0,
	\label{fluc}
\end{equation}
for all $i_1,i_2$, where
\begin{equation}
	\begin{split}
		&\bar c^{\text{comp}'}_{i_1 i_2}(T_\text{seq})\\
		=&\int_0^{T_\text{seq}}dt_1\int_0^{t_1}dt_2\\
		&~~~~~~~~\langle\beta_{w(i_1)}(t_1)\beta_{w(i_2)}(t_2)\rangle c_{i_1}^\text{comp}(t_1)c_{i_2}^\text{comp}(t_2),
	\end{split}
\end{equation}
with
\begin{equation}
	w(i)= w' \text{, if }\sum_{k=1}^{w'}|\mathscr C_k|<i<\sum_{k=1}^{w'+1}|\mathscr C_k|.
\end{equation}

In the simplest case, 
\begin{equation}
	\langle\beta_{w_1}(t_1)\beta_{w_2}(t_2)\rangle=\Sigma_{w_1w_2}e^{-\theta_{w_1w_2}(t_1-t_2)},t_1>t_2,
\end{equation}
where $\Sigma_{w_1w_2}$ defines the spatial noise covariance matrix. Eq. \eqref{fluc} enables one to incorporate correlations between noise acting on different parts of the control and quantum system, as well as their corresponding cross terms.
\section{Robust state transfer}\label{secstate}
For robust state transfers, conditions Eqs. \eqref{tot0} and \eqref{totr} can be relaxed.
Here, we consider the most common setting where the precision objective is achieved in $U_\text{pri}$ and the robustness objective is achieved in $U_\text{pert}$:
\begin{equation}
	\begin{split}
		&U_\text{target}\rho_0U_\text{target}^\dagger=\rho_\text{target},\\
		&U_\text{pert} \rho_0 U_\text{pert}^\dagger\approx\rho_0,
	\end{split}
\end{equation}
where $\rho_0$ is an initial density matrix and $\rho_\text{target}$ is a desired target density matrix. A method for characterizing the achievable set of $\rho_\text{target}$ based on Algorithm \ref{L3} is discussed in Appendix \ref{appen9}.

Each $\mathscr C_w$ can be decomposed into a subspace that commutes with $\rho_0$ and a complementary subspace. 
To achieve robust state transfer, it is sufficient to suppress errors arising from the complementary subspace. 
Assume that the basis $\{h_i^{\text{comp}}\}_{i\in\mathcal I_{\parallel}}$ spans the entire subspace that commutes with $\rho_0$. 
Then, the zeroth- and higher-order robustness of the state transfer are achieved if
\begin{equation}
	\begin{split}
		&\bar c_{i_1}^{\text{comp}}(T_\text{seq}) = 0,\\
		&\bar c_{i_1\cdots i_r}^{\text{comp}}(T_\text{seq}) 
		- (-1)^{r-1}\bar c_{i_r\cdots i_1}^{\text{comp}}(T_\text{seq}) = 0,
	\end{split}
\end{equation}
excluding cases where $i_1,\ldots,i_r\in \mathcal I_{\parallel}$ for both equations, 
and where $i_1=\cdots=i_r$ for the second equation.

The achievability of zeroth-order robustness can be verified using Algorithm \ref{L3} by constructing the polygon formed by the sampled points $\lc{U_i^\dagger H_\text{pert} U_i}$ projected to the complementary subspace. Zeroth-order robustness is achievable if $H_\text{target}=0$ is reachable in this projected subspace.

\section{Flowchart for effective Hamiltonian Engineering} \label{secflow}
Having developed the general framework for robust Hamiltonian engineering, we now summarize the procedure in a flowchart that integrates all components of the method. The flowchart is the same for all applications in Table \ref{obj}. 
\begin{enumerate}
	\item Describe the quantum system and controls:
	\begin{enumerate}
		\item Provide a description of $H_\text{int}$ including distributions of $\lc{\eta_\text{int}^i}$.
		\item Provide a description of $H_c(t)$ and a model of the control system including finite bandwidth, maximal power, and nonlinearities and distributions of the model parameters $\lc{\mu_j}$. Specify the error Hamiltonian $\Delta H$.
		\item Provide relaxation times when desired.
	\end{enumerate}
	\item Identify the use case from Table \ref{obj} and its objectives. Partition $H_\text{tot}$ into $H_\text{pri}$ and $H_\text{pert}$ accordingly. Identify the components \(\{H_\text{pert}^w\}\) of \(H_\text{pert}\). Specify the target unitary $U_\text{target}$ and/or effective Hamiltonians $H_\text{target}^w$, scaling factors $s_w$, and the required FoM.  
	\item Determine the controllability of $U(T_\text{seq})$:
	\begin{enumerate}
		\item Controllability of $U_\text{pri}$: Check if $U_\text{target}$ is achievable by checking if $U_\text{target}\in e^{\mathbf g_\text{pri}}$. 
		\item Space of achievable $\bar H^{(0)}$: Use Algorithm \ref{L2} to compute $\mathscr C_w(\mathbf g_\text{pri}, H_\text{pert}^w)$, and then check whether $H_\text{target}^w \in \mathscr C_w(\mathbf g_\text{pri}, H_\text{pert}^w)$ for each $H_\text{pert}^w$. 
		\item Range of achievable scaling factors: Use Algorithm \ref{L3} to check the joint achievability of  $s_wH_\text{target}^w$. 
		\item If any of the above objectives is not achievable, go to Step 1 to change $H_c(t)$ if additional control is available in the physical system. Sometimes (see below for an explanation of {\bf Step 3} (d)), one can go to Step 2 to choose a different partitioning to enable a larger set of achievable $\bar H^{(0)}$. 
	\end{enumerate}
	\item Construct the total cost function incorporating the desired \(U_\text{target}\), \(H_\text{target}^w\), and higher-order corrections, and use a numerical optimizer to find the control sequence that minimizes this cost function.
	\item Calculate the control landscape and the FoM, accounting for relaxation and the full model of the control system. Check if the desired FoM and robustness are reached. If yes, return the sequence. Otherwise, if the FoM is low due to relaxation, go to Step 3 (c) to choose a larger $s$. If the robustness is not sufficient, go to Step 4 to include higher-order robustness.
\end{enumerate}

In {\bf Step 1}, the internal Hamiltonian \(H_\text{int}\) is characterized by its operators, amplitudes, and parameter distributions (e.g., \(\mathbf D\), \(\vec n\), \(B_{ij}\), \(\delta_i\)) typically obtained from spectroscopy.  
The control Hamiltonian \(H_c(t)\) is defined by the control inputs \(\{a_k(t)\}\) and the system model (e.g., a first-order state-space equation or transfer function) that maps them to the output field \(\vec{B}(t)\) in the lab frame. Once this mapping is known, \(H_c(t)\) is expressed in the qubit frame, including model parameter distributions \(\{\mu_j\}\) from ``tune-up'' cycles \cite{gerstein1985transient}.  Relevant relaxation times include \(T_1\), \(T_2\), and, when available, \(T_{1\rho}\) \cite{haeberlen1969spin,rhim1976multiple,yan2013rotating}; \(T_2^*\) is incorporated in the qubit energy distribution. Alternatively, the relaxation superoperator can be obtained via quantum process tomography \cite{weinstein2004quantum}.

In {\bf Step 2}, derivatives of \(\Delta H\) with respect to different model parameters are treated as distinct components \(H_\text{pert}^w\). Additional \(H_\text{pert}^w\) terms may represent unwanted uncertainties in \(H_\text{int}\) or parts of the Hamiltonian to be engineered. The remaining well-characterized terms are assigned to \(H_\text{pri}\). For robust control, \(U_\text{target}\) is typically set to the desired total unitary \(U_0\). When the objective is to engineer a nonzero \(H_\text{target}\) (e.g., in spectroscopy, sensing, or simulation), \(U_\text{target}\) is often set to identity. For each \(H_\text{pert}^w\), specify the corresponding \(H_\text{target}^w\): components associated with imperfections or unwanted uncertainties are set to zero, while nonzero \(H_\text{target}^w\) terms are assigned to the parts being engineered. It is also necessary to verify that the FoM is achievable for the given relaxation time and sequence duration \(T_\text{seq}\).

In {\bf Step 3}, to test whether $U_\text{target}$ is achievable under the available controls, we check whether there exists a generator 
$g \in \mathbf g_\text{pri}$ such that 
\(
U_\text{target} = e^{g}.
\)
In practice, this can be implemented by computing $\log U_\text{target}$ and adding integer multiples of $2\pi/T_\text{seq}$ to the eigenvalues to account for branch ambiguities. 
If any such $g$ lies within $\mathbf g_\text{pri}$, the target is achievable. The method for checking the achievability of $H_\text{target}$ and $s$ is in Sec. \ref{seccon}. 
In (d), for gate design, simulation and sensing where $H_\text{pri}=H_c+H_\text{int}$, spectroscopy, and decoupling, if $U_\text{target}$ or $sH_\text{target}$ are not achievable, one needs to introduce additional control fields (e.g., RF, microwave, or laser fields) to modify the controllability. For simulation and sensing where $H_\text{pert}$ includes $H_\text{int}$, one can put part of $H_\text{int}$ in $H_\text{pri}$ to modify controllability of $U_\text{pri}$ and $\bar H^{(0)}$. 

In {\bf Step 4}, the total cost function is
\begin{equation}
	\begin{split}
		f_\text{tot}\lr{\lc{a_k(t)}}=&w_\text{pri}f_\text{pri}\lr{\lc{a_k(t)}}+w_0f^{(0)}\lr{\lc{a_k(t)}}\\
		&+\sum_{r=2}^Rw_{r-1}f^{(r-1)}\lr{\lc{a_k(t)}},
	\end{split}
	\label{totobj}
\end{equation}
where $w_\text{pri}$, $w_\text{0}$ and $w_{r-1}$ are positive weights reflecting relative importance of each objective. The different cost functions are obtained from Eqs. \eqref{fubi}, \eqref{tot0} and \eqref{totr} respectively:
\begin{equation}
	\begin{split}
		&f_\text{pri}\lr{\lc{a_k(t)}}=1-\frac{|\text{Tr}\lr{U_\text{pri}(T_\text{seq})^\dagger U_\text{target}}|}{\text{Tr}\lr{U_\text{target}^\dagger U_\text{target}}},\\
		&f^{(0)}\lr{\lc{a_k(t)}}\\
		&=\left|	\int_0^{T_\text{seq}}dt_1|H_\text{tog}(t_1)\rrangle_\text{comp}-T_\text{seq}\bigoplus_{w=1}^W |s_wH_\text{target}^w\rrangle_w\right|,\\
		&f^{(r-1)}\lr{\lc{a_k(t)}}\\
		&=\ls{\sum_{i_1,\ldots,i_r}'\left|\bar c_{i_1\cdots i_r}^\text{comp}(T_\text{seq})-(-1)^{r-1}\bar c_{i_r\cdots i_1}^\text{comp}(T_\text{seq})\right|^2}^{1/2},
	\end{split}
	\label{subobj}
\end{equation}
where the sum $\sum_{i_1,\ldots,i_r}'$ is taken over all $i_1,\ldots,i_r$ excluding the case where $i_1=\cdots=i_r$. 
Given the relaxation times and desired FoM, determine the maximal $T_\text{seq}$ of a sequence that allows the target FoM. 
Use a numerical optimizer, such as gradient descent \cite{khaneja2005optimal} or simulated annealing \cite{tsallis1996generalized}, to find $\{a_k(t)\}$ that minimizes $f_\text{tot}\lr{\lc{a_k(t)}}$.
A solution should be searched with short $T_\text{seq}$ first and with an increased $T_\text{seq}$ if no solution can be found with the current value of $T_\text{seq}$ so that shorter sequences can be found.

In {\bf Step 5},
the control landscape is obtained by calculating the fidelity between $U(T_\text{seq})$ and target total unitary $U_0$ (Eq. \eqref{fubi}) for each set of values of $\lc{\mu_j}$ and $\lc{\eta^i_\text{int}}$ across their respective distributions.
Simulate the resulting average CPTP map $\bar\Lambda$ and/or the average density matrix $\bar \rho(T_\text{seq})$ of the system (Eq. \eqref{dan}) using, e.g., the Monte Carlo method. Calculate FoM through either the average gate fidelity (Eq. \eqref{avegate}) or the state fidelity Tr$\lr{\sqrt{\sqrt{\rho_\text{target}}\bar\rho(T_\text{seq})\sqrt{\rho_\text{target}}}}^2$. If the FoM is unsatisfactory, check $\bar \Lambda$ to identify the error. If the error is due to a depolarization, check if $T_\text{seq}$ is too long. If not, then the robustness is insufficient, go to Step 4 to include higher-order corrections involving $\Delta H$ and variations of $H_\text{int}$. If the error is coherent, go to Step 4 to increase target fidelity or include higher-order corrections. 
\section{Examples}\label{secexam}
We now present examples to demonstrate the power of the method to achieve robustness to control errors and variations in $H_\text{int}$, characterize controllability, design effective Hamiltonians, and incorporate higher-order corrections.
We explore the application of the method in gate design, quantum simulation, and sensing. 
\subsection{Robust Hadamard}\label{exaro}
We design a Hadamard gate that is robust to variations in the Rabi field strength and qubit detuning. This sequence is useful for high-fidelity quantum operations in NMR, ESR, trapped-ion, quantum dot, Rydberg atom, and superconducting systems.

\underline{ Step 1. System and controls.} The internal and control Hamiltonians are
\begin{equation}
	\begin{split}
		H_\text{int}&=\omega_0\z,\\
		H_c(t)&=\omega_1(t)\cos(\omega_rt+\phi(t))\x,
	\end{split}
	\label{singq}
\end{equation}
where, for this example, we choose $\omega_0=$1 GHz (1 Hz=$2\pi$ rad$\cdot s^{-1}$) as the resonance frequency of the qubit, $\omega_r$ as the transmitter microwave frequency. The control parameters are the amplitude of the applied microwave field ($\omega_1(t)$) and its phase ($\phi(t)$). The maximal value of $\omega_1(t)$ is 20 MHz. The control system has a 500 MHz bandwidth ($\approx$1 ns response time), much shorter than the interval duration (10 ns), so the response can be treated as effectively instantaneous.

It is convenient to move to an interaction frame that rotates at the transmitter frequency $\omega_r$ and take the rotating wave approximation \cite{slichter2013principles}. The internal and control Hamiltonians in this frame are
\begin{equation}
	\begin{split}
		H_\text{int}&=\Delta\omega\z,\\
		H_c(t)&=\omega_1(t)\ls{\cos(\phi(t))\x+\sin(\phi(t))\y},
	\end{split}
	\label{hcon}
\end{equation}
where $\Delta\omega=\omega_0-\omega_r$ is the detuning.

The gate is designed to be robust to variations in the amplitude of the applied microwave field, $(1+\epsilon)\omega_1(t)$, where $\epsilon$ is a random variable sampled from uniform distributions Unif$[-5\%,5\%]$. Thus, the error Hamiltonian is $\Delta H(t)=\epsilon H_c(t)$.
The gate is also designed to be robust to variations in $\Delta\omega$, sampled from Unif$[-2.5\text{ MHz},2.5\text{ MHz}]$. The control, model, and internal parameters are \(a_1(t) = \omega_1(t)\), \(a_2(t) = \phi(t)\), \(\mu_1 = \epsilon\), and \(\eta_\text{int}^1 = \Delta\omega\), respectively.
The qubit decoherence is modeled as a depolarizing channel with a characteristic time of 1 ms. 

\underline{Step 2. Objectives and partitioning of $H_\text{tot}$.} 
Choose the partitioning $H_\text{pri}=H_c(t)$, $H_\text{pert}^1=\Delta H$ and $H_\text{pert}^2=H_\text{int}$. The objectives are
\begin{equation}
	U_\text{target}=\frac{1}{\sqrt 2}\lr{\begin{array}{cc}
			1&1\\
			1&-1
	\end{array}}.
\end{equation}
and $H_\text{target}^1=0$. The target FoM is 0.999.

\underline{ Step 3. Controllability.} The set of generators of $\mathbf g_\text{pri}$ is $\lc{\x,\y}$, and their composition via Algorithm \ref{L1} provides universal control over the qubit space. Therefore, $U_\text{target}$ is achievable. Using Algorithm \ref{L2}, an orthonormal basis of $\mathscr C(\mathbf g_\text{pri},H_\text{int})$ is $\lc{\x,\y,\z}/\sqrt 2$. Choose $H_\text{target}=s\z/\sqrt 2$. Via Algorithm \ref{L3}, the range of achievable scaling factors is [-1,1]. Therefore, $H_\text{target}=0$ is achievable in zeroth order. 

\begin{figure}[htp]
	\centering
	\includegraphics[width=0.45\textwidth]{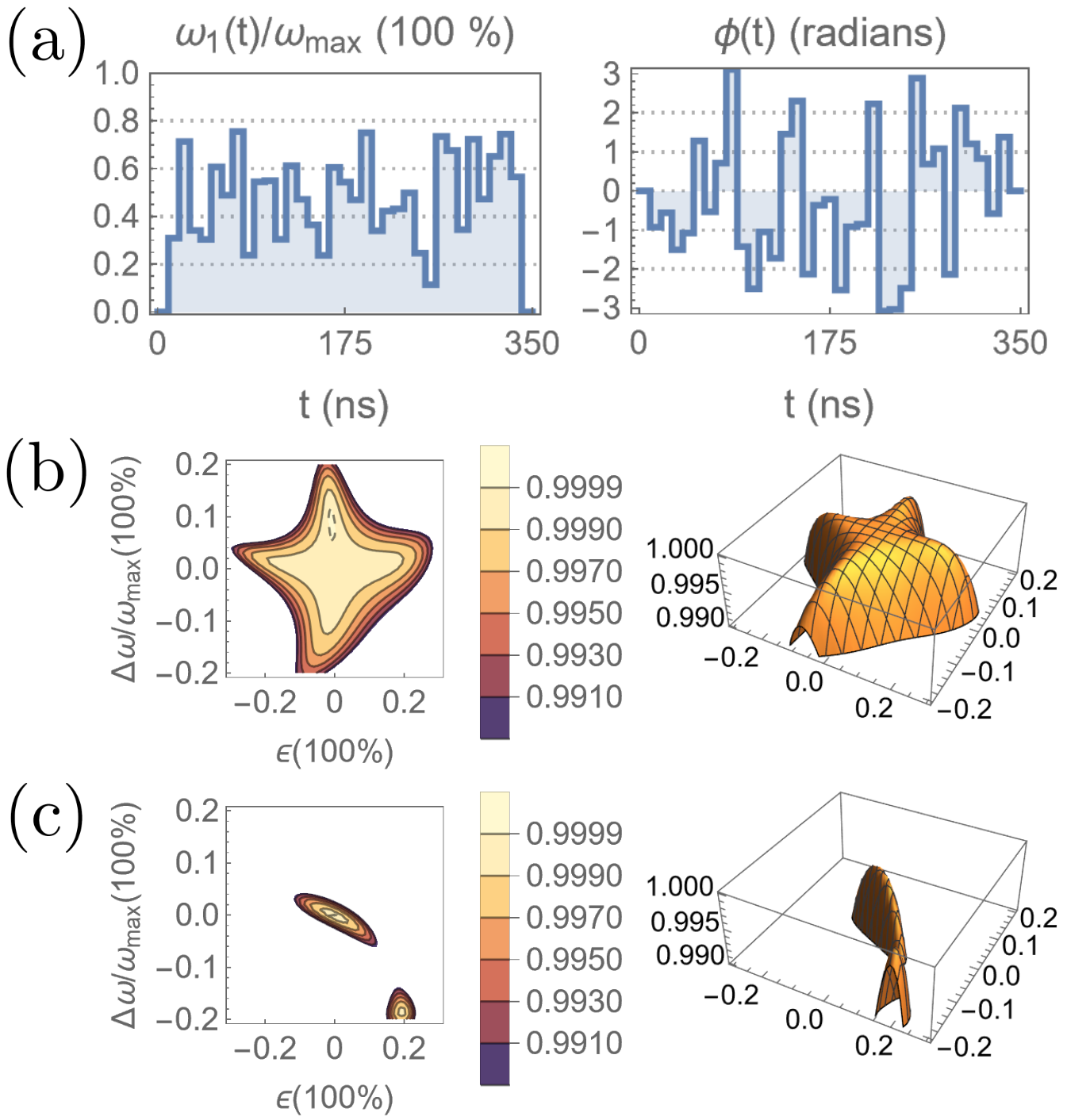}\\
	\caption{(a) Control parameters of a Hadamard gate that is robust to variations in Rabi field strength and detuning. (b) Contour and 3D landscapes of fidelity for the robust sequence. (c) Same as (b), but for a non-robust Hadamard gate of the same $T_\text{seq}$.}
	\label{hada1}
\end{figure}

\underline{Step 4. Design robust sequence.} 
The higher-order terms to minimize in the Magnus expansion include the first-order corrections of $\Delta H$ and $H_\text{int}$, with their cross term. So,
the total cost function includes $f_\text{pri}$, $f_0$ and $f_1$ (Eqs. \eqref{totobj} and \eqref{subobj}).
Based on the target FoM and the strength of the depolarizing channel, the maximum gate time $T_\text{seq}$ is 1 $\mu s$. Within this range, the number of intervals can be varied until the target fidelity is achieved. Here, results are shown for a sequence with 35 intervals and $T_\text{seq}=0.35~\mu s$ (Fig. \ref{hada1}). 
A generalized simulated annealing optimizer \cite{tsallis1996generalized,schanze2006exact} is used to search for an optimized sequence. The sequence and its control landscape are shown in Fig. \ref{hada1}. For comparison, we also show the control landscape of a 
non-robust sequence, obtained by optimizing only over 
$f_\text{pri}$ using the same number of intervals and total duration 
$T_\text{seq}$.

\underline{Step 5. Calculate FoM.}
To simulate the FoM, assume that \( \epsilon \sim \mathcal{N}(0, 0.01^2) \), i.e., a normal distribution with zero mean and variance \( 0.01^2 \) and that $\Delta\omega\sim\mathcal N(0,(0.5 \text {MHz})^2)$. The average gate fidelity (Eqs. (17) and (18)) of the resulting CPTP map of the robust and non-robust sequences are estimated to be $0.99955_{-0.00004}^{+0.00009}$ and $0.9937_{-0.003}^{+0.005}$ where the error bars correspond to the 20th and 80th percentiles. The FoM satisfies the requirement of being at least 0.999. The performance of the robust and non-robust sequences can also be compared by calculating the average CPTP maps using Eq. \eqref{dan}. The representations under the basis $\lc{\mathbb 1,\x,\y,\z}/\sqrt 2$ are
\begin{equation}
	\begin{split}
		\bar\Lambda_\text{robust}&=\lr{\begin{array}{cccc}
				1&&&\\
				&0.0329&0.0036&0.9991\\
				&-0.0076&-0.9996&0.0038\\
				&0.9991&-0.0078&-0.0329
		\end{array}},\\
		\bar\Lambda_\text{non-robust}&=\lr{\begin{array}{cccc}
				1&&&\\
				&0.0104&0.0067&0.9710\\
				&0.0084&-0.9863&0.0238\\
				&0.9748&0.0326&-0.0227
		\end{array}}.
	\end{split}
\end{equation}
Calculating the orthogonality of the two CPTP maps yields $\mathcal R (\Lambda_\text{robust})=0.99945$ and $\mathcal R (\Lambda_\text{non-robust})=0.96710$, indicating that the low FoM of the non-robust sequence primarily arises from poor robustness rather than unitary inaccuracy.

\subsection{Robust Hadamard with finite control bandwidth}
Here, we add distortion to the Hadamard control sequence from having a finite bandwidth (80 MHz) in a linear control system. The gate is designed to be robust to variations in bandwidth and to small anti-symmetric phase transients. For a square pulse, anti-symmetric phase transients refer to quadrature components at the leading and trailing edges with equal amplitude but opposite phase.

\underline{Step 1. System and controls.}
In the rotating frame ($\omega_0=\omega_r=1 \text{ GHz}$), the internal and control Hamiltonians are
\begin{equation}
	\begin{split}
		H_\text{int}&=0,\\
		H_c(t)&=B_x(t)\x+B_y(t)\y,
	\end{split}
	\label{Dcon}
\end{equation}
where $B_x(t)$ and $B_y(t)$ are the time-dependent $x$ and $y$ components of the control field in the rotating frame.  The control parameters are $\omega_1(t)$ (maximal value is 20 MHz) and $\phi(t)$ ($H_c(t)$ in Eqs. \eqref{hcon}), they are mapped to $\vec B(t)$ via a convolution with a kernel $\Phi$
\begin{equation}
	\lr{\begin{array}{c}
			B_x(t)\\B_y(t)
	\end{array}}=\Phi \star \lr{\begin{array}{c}
			\omega_1(t)\cos(\phi(t))\\\omega_1(t)\sin(\phi(t))
	\end{array}},
\end{equation}
with
\begin{equation}
	\Phi=\lr{\begin{array}{cc}
			\phi^A&\phi^P\\
			-\phi^P&\phi^A
	\end{array}}.
\end{equation}
$\phi_A$ and $\phi_P$ represent the amplitude and phase responses of the system respectively:
\begin{equation}
	\phi^A=\begin{cases}
		We^{-Wt}&t>0\\
		0&t\le0
	\end{cases}~~\text{and}~~\phi^P=\begin{cases}
		\delta\lr{1-Wt}e^{-Wt}&t>0\\
		0&t\le0
	\end{cases}.
\end{equation}
Here, the bandwidth $W$ and $\delta =\omega^*-1\text{ GHz}$ are the model parameters, and $\omega^*$ is the free ringing frequency of the system. For a series RLC circuit, $\omega^*=\sqrt{\frac{1}{LC}-\lr{\frac{R}{2L}}^2}$. The distortion given by the kernel $\Phi$ is discussed in detail in \cite{mehring1972phase,barbara1991phase}. $W$ and $\delta$ are random variables that follow $W\sim$Unif$[70,90]\text{ MHz}$ and $\delta\sim$Unif$[-20,20]\text{ MHz}$. The leading-order approximation of the error Hamiltonian is
\begin{equation}
	\begin{split}
		\Delta H&\approx (W-\text{80 MHz})\frac{\partial H_c}{\partial W}(W=\text{80 MHz})+\delta \frac{\partial H_c}{\partial \delta}(\delta=0).\\
	\end{split}
\end{equation}
The control and model parameters are \(a_1(t) = \omega_1(t)\), \(a_2(t) = \phi(t)\), \(\mu_1 = W\), and \(\mu_2 = \delta\); there are no internal parameters.
\underline{Steps 2 and 3} are the same as in the previous example, except
\begin{equation}
	\begin{split}
		&H_\text{pert}^1=\frac{\partial H_c}{\partial W}(W=\text{80 MHz}),\\
		&H_\text{pert}^2=\frac{\partial H_c}{\partial \delta}(\delta=0).
	\end{split}
\end{equation}
Since $H_\text{int}=0$, no $H_\text{target}$ is needed here. The rest of the steps are the same as in the example above.
\begin{figure}[htp]
	\centering
	\includegraphics[width=0.5\textwidth]{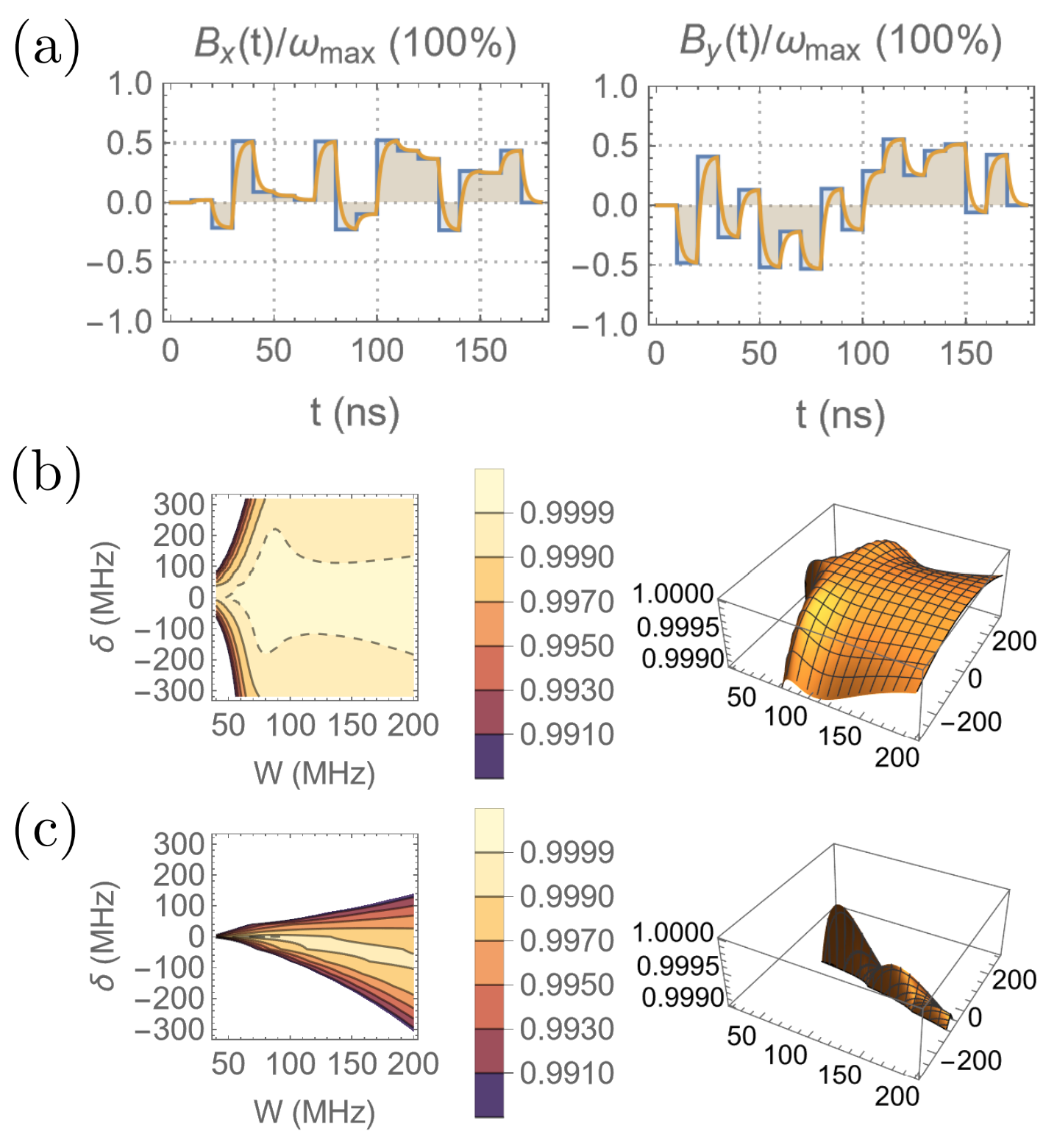}\\
	\caption{(a) Control sequence for Hadamard gate that is robust to $W$ and $\delta$ variations. The blue and orange curves represent the relative amplitudes of $B_x(t)$ and $B_y(t)$ before and after distortion. (b) Contour and 3D landscape of fidelity for the robust sequence. (c) Same as (b) but for a non-robust sequence of the same $T_\text{seq}$.}
	\label{hada2}
\end{figure}

\underline{Step 4. Robustness and optimization.}
Robustness to $\Delta H$ is achieved by minimizing the zeroth-order average Hamiltonians of $\frac{\partial H_c}{\partial W}$ and $\frac{\partial H_c}{\partial \delta}$ simultaneously.
Thus, the total cost function includes $f_\text{pri}$ and $f_0$.
Similar to the previous example, the maximum gate time $T_\text{seq}$ is estimated to be 1 $\mu s$. Fig. \ref{hada2} shows an optimized sequence ($T_\text{seq}=180$ ns) and its control landscape. As a comparison, the control landscape of a non-robust sequence (optimized over $f_\text{pri}$) with the same $T_\text{seq}$ is also shown.

\underline{Step 5. Calculate FoM.}
Assume that $W\sim\mathcal N(80\text{ MHz},(2\text{ MHz})^2)$ ($5\sigma=\text{10 MHz}$) and that $\delta\sim\mathcal N(0,(\text {4 MHz})^2)$ ($5\sigma=20\text{ MHz}$). The FoMs of the robust and non-robust sequences are $0.999818_{-0.000002}^{+0.000003}$ and $0.99976_{-0.00003}^{+0.00004}$, respectively.
These values are limited by $T_1$ relaxation; without accounting for $T_1$, the FoMs are $0.999999_{-0.000002}^{+0.000003}$ and $0.99994_{-0.00003}^{+0.00004}$.
\subsection{Robust control with nonlinear system}\label{nlsec}
The dynamics of the control system are generally nonlinear. Here, we show an important example of a weak nonlinearity for a kinetic inductance.

\begin{figure}[htp]
	\centering
	\includegraphics[width=0.3\textwidth]{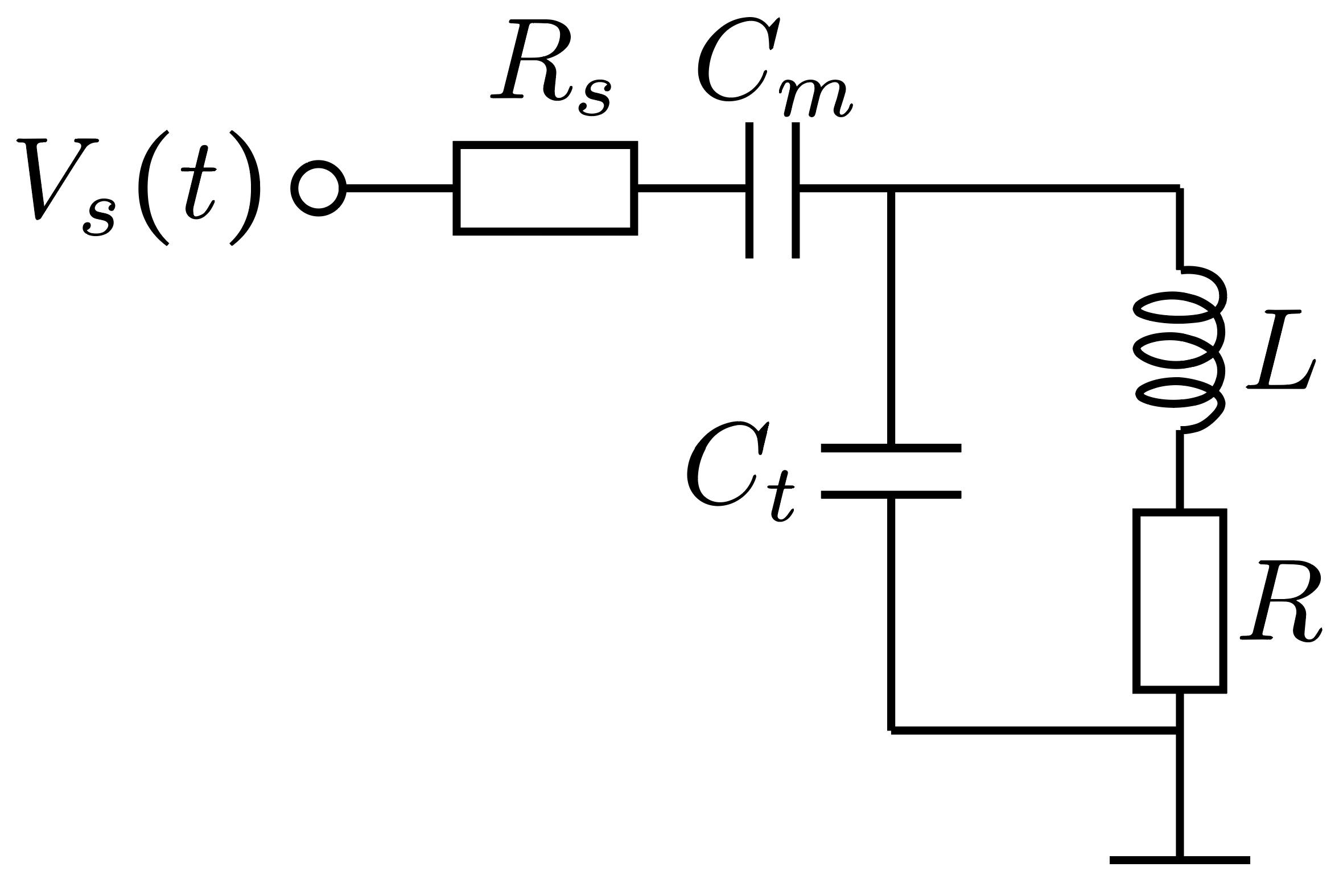}\\
	\caption{A simple resonant circuit with $R_s=50 ~\Omega$, $R=0.01~\Omega$, $C_m=11$ fF, $C_t=1.479$ pF, and $L=170$ pH. The resonance frequency is 10 GHz. }
	\label{concirc}
\end{figure}

Generally, the output field $\vec B(t)$ and the input parameter $\lc{a_k(t)}$ are related through a state-space differential equation (Appendix \ref{appen10})
\begin{equation}
	\frac{d\vec B}{dt}=h(\vec B(t), \lc{a_k(t)},t; \lc{\mu_j}),
\end{equation}
where $h$ is a nonlinear function. We can design control sequences robust to variations in $\lc{\mu_j}$ provided that the equation is not chaotic. As an example, consider the resonant circuit in Fig. \ref{concirc}. We design a Hadamard gate that is robust to small kinetic inductance.
This method can be used to design robust sequences against variations in parameters in general nonlinear control systems.

\underline{Step 1. System and controls.}
Start with the same internal and control Hamiltonians in Eqs. \eqref{Dcon}. The control field $\vec B(t)$ can be obtained by solving the following differential equation in the rotating frame \cite{hincks2015controlling, hincks2020control}
\begin{equation}
	\frac{dx}{dt}=A(x)x+\alpha(t)u,
	\label{e74}
\end{equation}
where
\begin{equation}
	\begin{split}
		x&=\lr{
			\tilde I_L,\tilde V_{C_m},\tilde V_{C_t}
		}^T,\\
		A(x)&=\lr{\begin{array}{ccc}
				-\frac{R}{L}&0&\frac{1}{L}\\0&\frac{-1}{R_LC_m}&\frac{1}{R_LC_m}\\\frac{-1}{C_t}&\frac{-1}{R_LC_t}&\frac{1}{R_LC_t}
		\end{array}}-i\omega_r\mathbb{1},\\
		u&=\lr{
			0,\frac{1}{R_LC_m},\frac{1}{R_LC_t}
		}^T,
	\end{split}
\end{equation}
and the control parameters $\lc{a_k(t)}$ and the components of the output field $\vec B(t)$ are
\begin{equation}
	\begin{split}
		&a_1(t)=\kappa_i\text{Re}(\alpha(t)),~~a_2(t)=\kappa_i\text{Im}(\alpha(t)),\\
		&B_x(t)=\kappa_o\text{Re}(\tilde I_L)\omega_\text{max},~~B_y(t)=\kappa_o\text{Im}(\tilde I_L)\omega_\text{max},
	\end{split}
\end{equation}
where $\omega_\text{max}=$20 MHz, $\kappa_i$=1 V$^{-1}$ and $\kappa_o$=2 A$^{-1}$.

In the presence of kinetic inductance \cite{maas2003nonlinear}, the inductance $L$ depends on the current:
\begin{equation}
	L=L_0(1+\alpha_L|I_L|^2),
\end{equation}
where $\alpha_L$ is a constant, leading to nonlinear behavior in the circuit \cite{maas2003nonlinear, mohebbi2014composite,dahm1997theory}. The model parameters are the electrical component parameters in the circuit where $\alpha_L\sim$Unif$[-0.001,0.001]$ A$^{-2}$. Note, $\alpha_L$ is typically positive, but we can still design around $\alpha_L=0$. The solution is robust to $\alpha_L$ that is either positive or negative. There are no internal parameters in this case. The error Hamiltonian $\Delta H$ can be calculated by solving the differential equation of $\partial x/\partial \alpha_L$ (Appendix \ref{appen10}).

\underline{Steps 2 and 3} are the same as the previous example, except
\begin{equation}
	H_\text{pert}^1=\frac{\partial \Delta H}{\partial \alpha_L}(\alpha_L=0).
\end{equation}

\begin{figure}[htp]
	\centering
	\includegraphics[width=0.45\textwidth]{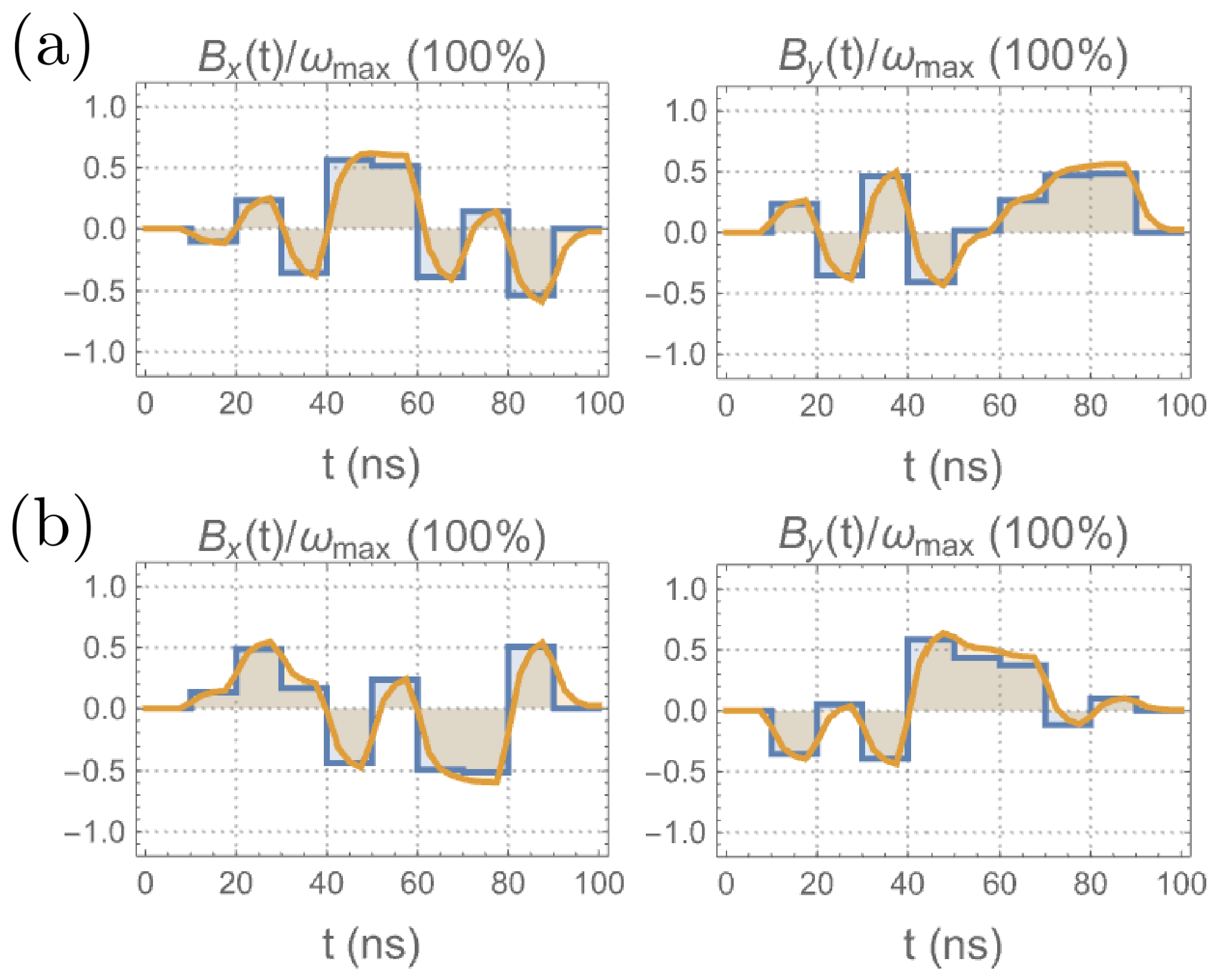}\\
	\caption{The control parameters (blue) and the relative amplitudes (orange) of the $x$ and $y$ components of $\vec B(t)$ at $\alpha_L=0$ for (a) a non-robust sequence and (b) a sequence robust to variations in $\alpha_L$.}
	\label{KIplot}
\end{figure}

\begin{figure}[htp]
	\centering
	\includegraphics[width=0.45\textwidth]{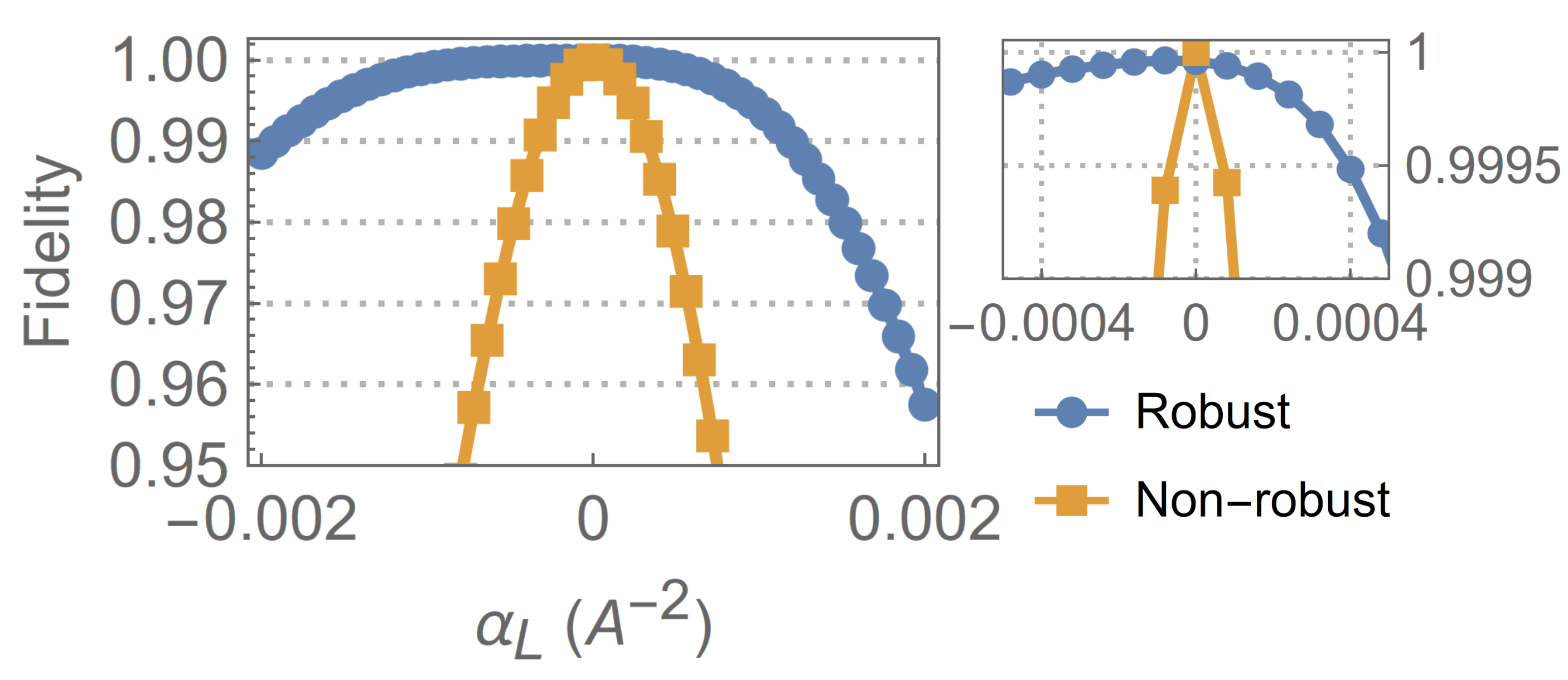}\\
	\caption{Control landscapes of different sequences as a function of the nonlinear coefficient $\alpha_L$. 
		The main panel shows the full fidelity profile; the smaller panel on the right displays a zoomed view around the high-fidelity region near $\alpha_L=0$.
		Although the robustness extends to negative $\alpha_L$, $\alpha_L$ is typically positive.}
	\label{nonlinearro}
\end{figure}

\underline{Step 4. Robustness and optimization.}
The total cost function includes $f_\text{pri}$ and $f_0$.
Design for a FoM of 0.999 or higher.
Fig. \ref{KIplot} shows robust and non-robust (optimized over $f_\text{pri}$) Hadamard gates. Their control landscapes are shown in Fig. \ref{nonlinearro}. 

\underline{Step 5. Calculate FoM.}
To simulate the FoM, we assume that $\alpha_L$ obeys a half Gaussian distribution, $\alpha_L\sim\mathcal {HN}(0,(0.0002~\text{A}^{-2})^2)$ ($5\sigma=\text{0.001 A}^{-2}$). The FoM of the robust and non-robust sequences are $0.99981_{-0.00003}^{0.00008}$ and $0.9969_{-0.002}^{+0.003}$.

\subsection{Robust Control against Stochastic Noise}
The cumulant expansion (Sec. \ref{secsto}) can be used to suppress stochastic noise in an ensemble-average signal. The cost function is defined by the left hand side of Eq. \eqref{fluc}
\begin{equation}
	f_K^{(1)}(\theta)=\left\|\sum_{i_1,i_2}\bar c^{\text{comp}'}_{i_1 i_2}(T_\text{seq})\text{ad}_{h_{i_1}^\text{comp}}\text{ad}_{h_{i_2}^\text{comp}}\right\|.
\end{equation}
Higher-order cost functions can be defined similarly by considering $K^{(2)},K^{(3)},\ldots$. These cost functions can be used alongside \eqref{subobj} or replace them when the sole objective is to suppress noise in an ensemble-averaged signal by incorporating its statistical properties.

\begin{figure*}[htp]
	\centering
	\includegraphics[width=0.9\textwidth]{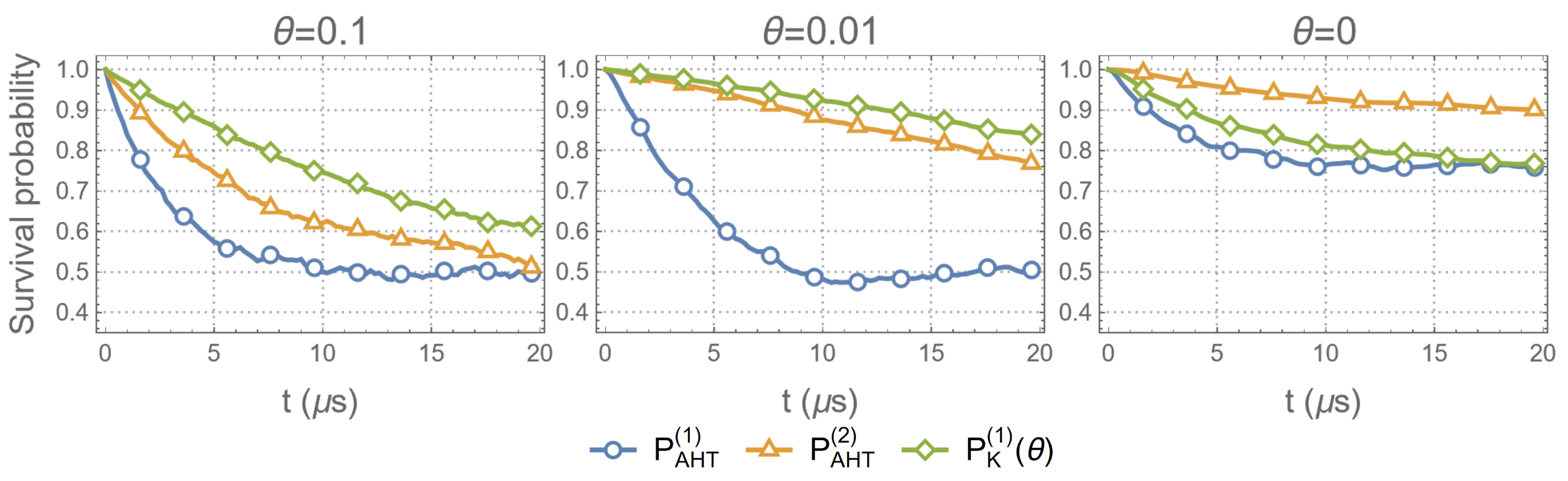}\\
	\caption{Simulated survival probability of $|+\rangle$ under different sequences for different $\theta$. Each $P_K^{(1)}(\theta)$ minimizes $f_K^{(1)}(\theta)$ for the corresponding value of $\theta$.}
	\label{Kplot}
\end{figure*}

We compare single-qubit decoupling sequences ($\omega_\text{max}= 20$ MHz, $T_\text{seq}=200$ ns, $U_\text{target}=\mathbb 1$ and $ H_\text{target}=0$) designed by minimizing different cost functions (Table \ref{pseq}). The control (primary) and internal (perturbative) Hamiltonians are \eqref{singq} with $\omega_0$ replaced by an Ornstein-Uhlenbeck process \cite{uhlenbeck1930theory}, $\beta(t)$, with a mean reversion speed $\theta$ ($10^8\text{ rad}\cdot s^{-1}$) and a stationary standard deviation $\sigma=1.6$ MHz ($\beta(t)\sim\mathcal N(0,\sigma^2), \forall t\ge0$). The simulated survival probability (averaged over 500 random trajectories) of $|+\rangle$ is shown in Fig. \ref{Kplot}. 
\begin{table*}[htbp]
	\centering
	\caption{Sequences and their corresponding minimized cost functions.}
		\begin{tabular}{|c|c|c|c|}\toprule \hline
		Sequences &	$P_\text{AHT}^{(1)}$ & $P_\text{AHT}^{(2)}$ & $P_K^{(1)}(\theta)$  \\ \hline
		Cost functions &$f_\text{pri}, f^{(0)}, \newline f^{(1)}$& $f_\text{pri}, f^{(0)}, \newline f^{(1)}, f^{(2)}$ & $f_\text{pri}, \newline f_K^{(1)}(\theta)$ \\ \hline
		\bottomrule
	\end{tabular}
	\label{pseq}
\end{table*}
The result shows effective suppression of stochastic noise through minimizing $f_K^{(1)}(\theta)$ for nonzero $\theta$, which can lead to even better performance than those of the sequences (e.g., $P_\text{AHT}^{(2)}$ in Fig. \ref{Kplot}) that correct higher-order errors in the Magnus expansion. When $\theta=0$, however, $P_\text{AHT}^{(2)}$ outperforms $P_K^{(1)}(\theta)$ (which is essentially a $P_K^{(2)}(\theta)$ since $f_K^{(2)}(\theta=0)\equiv0$). This indicates that, when coherent errors dominate, AHT can enable design of more robust sequences.

\subsection{Two-qubit gates}
\begin{figure}[htp]
	\centering
	\includegraphics[width=0.5\textwidth]{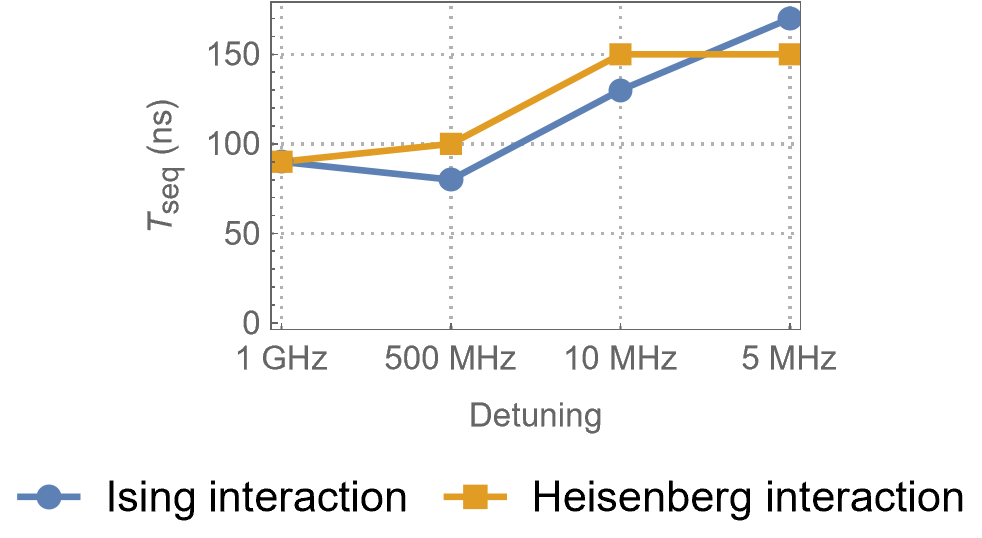}
	\caption{Minimal length of sequence for a CNOT gate with different detunings and interaction forms.}
	\label{CNOTt}
\end{figure}

The methods for designing robust two-qubit gates are the same as those discussed above. There are additional terms such as the leakage between controls on different qubits which can reduce the efficiency of a sequence. Additionally, different forms of qubit interactions can result in different efficiencies. Fig. \ref{CNOTt} shows the minimal lengths of the sequences we found for a CNOT. We compare the efficiency of sequence design under Ising and Heisenberg interactions, using the same interaction strength $J$:
\begin{equation}
	\begin{split}
		&H_\text{Ising}=J \z\otimes\z,\\
		&H_\text{Heisenberg}=J(\x\otimes\x+\y\otimes\y+\z\otimes\z)/\sqrt 3.
	\end{split}
\end{equation}
The minimal sequence length increases as the qubit detuning $\Delta$ decreases (i.e., as leakage increases). The results also reveal efficiency differences arising from the form of the interaction. This indicates that even when controllability remains unchanged, different system modalities can impact the efficiency of sequence design.	
\subsection{Simulating a Tunable Quantum Harmonic Oscillator}\label{exaQHO}
We show an example of simulating a truncated quantum harmonic oscillator (QHO) using two qubits. The purpose of this simple example is to illustrate how effective Hamiltonian engineering can be used in quantum simulation to achieve desired dynamics in a controlled manner.
The Hamiltonian of a QHO is $H_\text{QHO}=\hbar \Omega (\hat N+1/2)$. Using the code \cite{somaroo1999quantum}
\begin{equation}
	\begin{split}
		|n=0\rangle=|00\rangle,\\	|n=1\rangle=|01\rangle,\\
		|n=2\rangle=|11\rangle,\\
		|n=3\rangle=|10\rangle,\\
	\end{split}
\end{equation}
an evolution under $H_\text{QHO}$ can be simulated by implementing
\begin{equation}
	U(T_\text{seq})=\exp[i\lc{\z^2(1+\z^1/2)-2}\Omega T_\text{seq}].
\end{equation}
With universal control over the two qubits, a solution to realize $U(T_\text{seq})$ can be easily found using algorithms such as GRAPE \cite{khaneja2005optimal, borneman2010application} or chopped random basis (CRAB) optimal control \cite{caneva2011chopped,muller2022one}. We consider a different approach using Hamiltonian engineering that enables the direct control of the scaling of the simulated dynamics with an input parameter. 

\underline{Step 1. System and controls.}
Assume that the bandwidth of the control system is very large and variations in control and model parameters are negligible. In the rotating frame of the qubits, the control and internal Hamiltonians are
\begin{equation}
	\begin{split}
		H_c(t)=&\omega_1^1(t)\ls{\cos(\phi_1(t))\x^1+\sin(\phi_1(t))\y^1}+\Delta\omega_1(t)\z^1\\
		&+\omega_1^2(t)\ls{\cos(\phi_2(t))\x^2+\sin(\phi_2(t))\y^2}+\Delta\omega_2(t)\z^2,\\
		H_\text{int}=&J(\x^1\x^2+\y^1\y^2+\z^1\z^2).
	\end{split}
\end{equation}
The control parameters are $\omega_1^i(t), \theta_i(t)$ and $\Delta\omega_i$ for $i=1,2$. $J=1.4$ MHz and $\Delta\omega\sim$Unif[-1,1] MHz are the internal parameters. There are no model parameters. $T_1=100$ ms.

\underline{Step 2. Objectives and partitioning of $H_\text{tot}$.} 
Set $\Delta\omega_1(t)\equiv \Delta\omega$ and $\Delta\omega_2(t)\equiv 0$, and define the primary and perturbative Hamiltonians as
\begin{equation}
	\begin{split}
		H_\text{pri}=&\omega_1^1(t)(\cos(\phi_1(t))\x^1+\sin(\phi_1(t))\y^1)\\
		&+\omega_1^2(t)(\cos(\phi_2(t))\x^2+\sin(\phi_2(t))\y^2)\\
		&+J(\x^1\x^2+\y^1\y^2+\z^1\z^2),\\
		H_\text{pert}^1=&\Delta\omega\z^1.
	\end{split}
\end{equation}
$H_\text{pert}$  depends on the detuning $\Delta\omega$ which can be adjusted at the start of each experiment to modify the rate of the dynamics. The objectives are $U_\text{pri}=\mathbb 1$ and $H_\text{target}^1=s\Delta\omega (\z^2(1+\z^1/2))$.

\underline{ Step 3. Controllability.}
Using Algorithm \ref{L1}, $\mathbf g_\text{pri}$ generates universal control on the two qubits. Thus, $U_\text{pri}$ is achievable.
Using Algorithm 2, $\mathscr C(\mathbf g_\text{pri},H_\text{pert})$ is fifteen-dimensional and includes $H_\text{target}$. 
Using Algorithm 3, we find that the achievable range of $s$ is $[-0.6(2),0.6(8)]$. Since the strength of $H_\text{target}$ can be controlled by adjusting $\Delta\omega$, we design for $s>0.1$.

\begin{figure}[htp]
	\centering
	\includegraphics[width=0.4\textwidth]{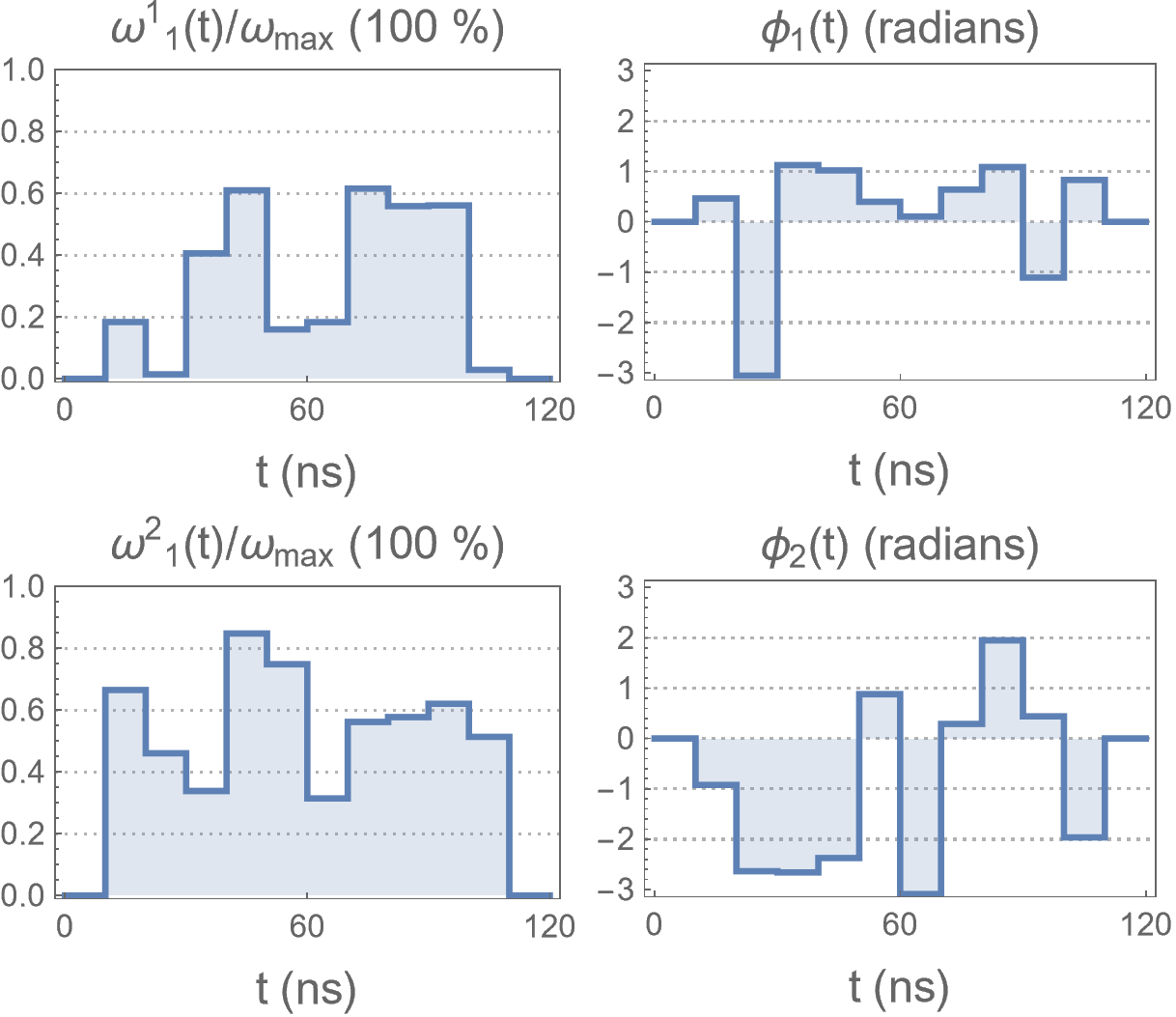}\\
	\caption{Time-dependent control parameters of an optimized sequence that engineers a tunable QHO.}
	\label{pQHO}
\end{figure}

\begin{figure}[htp]
	\centering
	\includegraphics[width=0.5\textwidth]{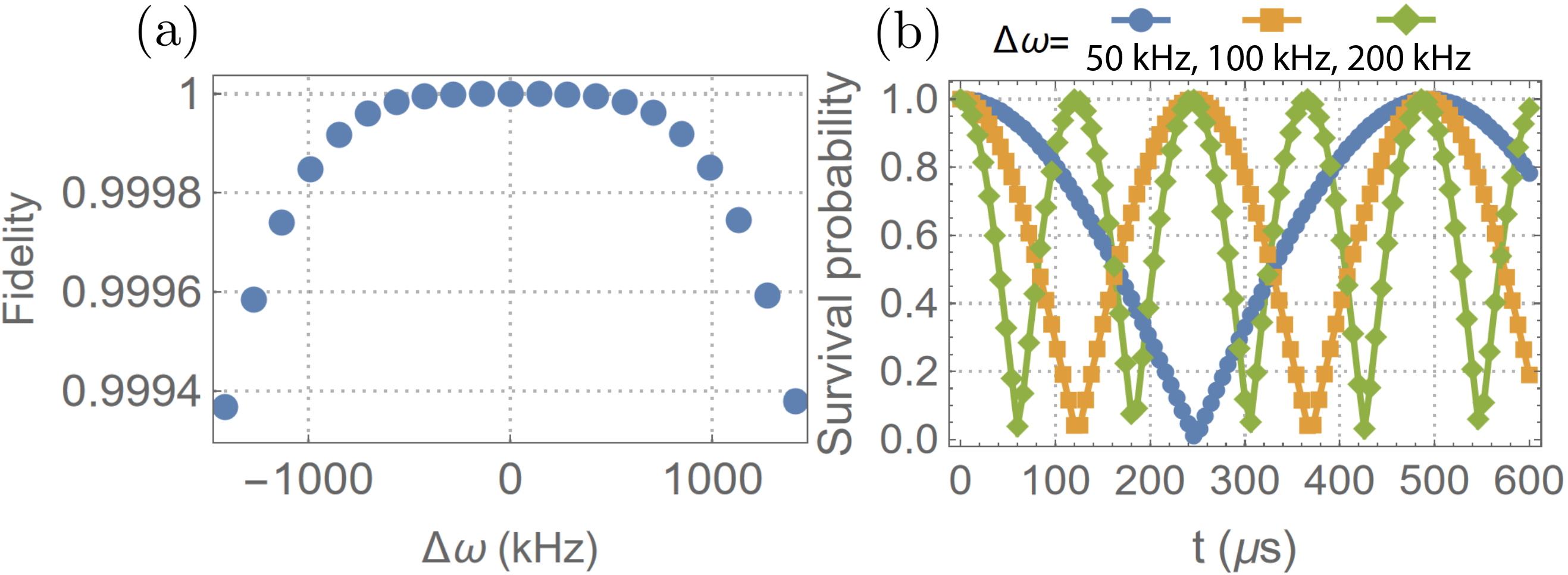}\\
	\caption{(a) Control landscape of the sequence (Fig. \ref{pQHO}) for simulating a QHO as a function of $\Delta\omega$. (b) Simulations of the evolution of the state $(|n=1\rangle+|n=3\rangle)/\sqrt 2$ under the sequence.}
	\label{qhosim}
\end{figure}

\underline{Step 4. Robustness and optimization.} To achieve precision in the engineered effective Hamiltonian, we choose to minimize $\bar H^{(1)}$. So, the total cost function includes $f_\text{pri}$, $f_0$ and $f_1$. Fig. \ref{pQHO} shows an optimized sequence ($T_\text{seq}=120$ ns), the control landscape is shown on the left of Fig. \ref{qhosim}. This sequence achieves a scaling factor of $0.221$.

\underline{Step 5. Calculate FoM.} Assume that $\Delta\omega\sim \mathcal N(0,(200\text{ kHz})^2)$, the FoM of the sequence is 0.999997$_{-0.000005}^{+0.000006}$.

As mentioned before, the sequence allows simulation of a tunable QHO whose frequency can be adjusted by directly changing $\Delta\omega_1(t)\equiv\Delta\omega$. The right-hand side of Fig. \ref{qhosim} shows the survival probability of $(|n=1\rangle+|n=3\rangle)/\sqrt 2$ under the sequence with different values of $\Delta\omega$.
\subsection{Sensing with a Scalable Spin Amplifier}\label{exaSA}
Consider designing a scalable solution for realizing a spin amplifier for measuring the state of a qubit. The quantum system of the target qubit and the spin amplifier is initially prepared in the state $|\phi_T\rangle\otimes|0_10_2\cdots 0_N\rangle$. The goal is to design a sequence such that at the end of the sequence we can learn the state of the target qubit $|\phi_T\rangle$ by measuring the collective magnetization $M_z$ of qubits 1 to $N$ in the spin amplifier. 

A scalable solution can be obtained by engineering conditioned dynamics of the spin amplifier controlled by the state of the target qubit \cite{cappellaro2005entanglement}. This can be achieved using internal interactions of the system and collective rotations of qubits in the spin amplifier. 

We design a conditional decoupling of the internal dynamics of the spin amplifier. 
Under the applied sequence, the internal interaction is suppressed when $|\phi_T\rangle = |0\rangle$ and active when $|\phi_T\rangle = |1\rangle$. 
In the experiment, a collective $\pi/2)_x$ pulse first rotates the amplifier polarization to the transverse plane, after which the system evolves under the engineered dynamics, followed by a $-\pi/2)_x$ readout pulse. 
When the interaction is active ($|\phi_T\rangle = |1\rangle$), the amplifier evolves into an entangled state with an average longitudinal magnetization $M_z \approx 0$ for large $N$, whereas when it is decoupled ($|\phi_T\rangle = |0\rangle$), $M_z$ remains close to its initial value. 
Measuring $M_z$ at the end of the sequence thus reveals the state of $|\phi_T\rangle$.

\underline{Step 1. System and controls.}
The internal and control Hamiltonians are
\begin{equation}
	\begin{split}
		H_c(t)&=\omega_1(t)\lr{\cos(\phi (t))\sum_{i=1}^N\x^i+\sin(\phi (t))\sum_{i=1}^N\y^i},\\
		H_\text{int}&= \z\otimes\sum_{i=1}^N\omega_i\z^i\\
		& + \mathbb{1}\otimes\sum_{i<j}^N\omega_{ij}(2\z^i\otimes\z^j - \x^i\otimes\x^j - \y^i\otimes\y^j).
	\end{split}
\end{equation}
The control parameters are $\omega(t)$ ($\omega_\text{max}=$ 20 MHz) and $\phi (t)$. The internal parameters are the coupling strength: $\omega_i\sim$Unif$[441,459]$ kHz and $\omega_{ij}\sim$Unif$[-22.5,22.5]$ kHz. No model parameters are considered and $T_1=$100 ms.

\underline{Step 2. Objectives and partitioning of $H_\text{tot}$.}
Define the primary and perturbative Hamiltonians as
\begin{equation}
	\begin{split}
		H_\text{pri}(t)&= H_c(t) + \z\otimes\sum_{i=1}^N\omega_0\z^i,\\
		H_\text{pert}^1&=\z\otimes\sum_{i=1}^N(\omega_i-\omega_0)\z^i,\\
		H_\text{pert}^2&=\mathbb{1}\otimes\sum_{i<j}^N\omega_{ij}(2\z^i\otimes\z^j - \x^i\otimes\x^j - \y^i\otimes\y^j),
	\end{split}
\end{equation}
where $\omega_0=450$ kHz. The objectives are
\begin{equation}
	\begin{split}
		&U_\text{pri}=\mathbb{1},\\
		&H_\text{target}^2 =\\ &s\sqrt{2}\ls{|1\rangle\langle 1|\otimes\sum_{i<j}^N\omega_{ij}(2\z^i\otimes\z^j - \x^i\otimes\x^j - \y^i\otimes\y^j)}.
	\end{split}
\end{equation}

\underline{Step 3. Controllability.}
To check the controllability, it is sufficient to consider two spins from the spin amplifier. $\mathbf g_\text{pri}$ is three-dimensional and $U_\text{pri}=\mathbb 1$ is achievable. Using Algorithm \ref{L2}, $\mathscr C_2(\mathbf g_\text{pri},H_\text{pert}^2)$ is ten-dimensional and includes $H_\text{target}^2$. 
Using Algorithm \ref{L3}, we can find the achievable range of the scaling factor $s$ (Table \ref{achi3}). For comparison, Table \ref{achi3} also lists some other achievable $H_\text{target}^2$ and their ranges of scaling factors.

\begin{table*}[htbp]
	\centering
	\caption{Achievable ranges of scaling factors for different target effective Hamiltonians. They correspond to conditioned decoupling,  double/zero quantum dynamics, and double/single quantum dynamics. Note, normalization constants have been omitted for simplicity.}
	\begin{tabular}{|c|c|}\toprule
		\hline
		Target Hamiltonian & Achievable range \\ \hline
		$|1\rangle\langle 1|\otimes\lr{2\z\otimes\z-\x\otimes\x-\y\otimes\y}$ &[-0.353(5),0.706(8)] \\
		$|0\rangle\langle 0|\otimes\lr{\x\otimes\x-\y\otimes\y}+|1\rangle\langle 1|\otimes\lr{2\z\otimes\z-\x\otimes\x-\y\otimes\y}$ & [-0.408(2),0.811(4)] \\
		$|0\rangle\langle 0|\otimes\lr{\x\otimes\y+\y\otimes\x}+|1\rangle\langle 1|\otimes\lr{\x\otimes\z+\z\otimes\x}$ & [-0.576(4),0.576(8)] \\ 
		\hline
		\bottomrule
	\end{tabular}
	\label{achi3}
\end{table*}

\begin{figure}[htp]
	\centering
	\includegraphics[width=0.4\textwidth]{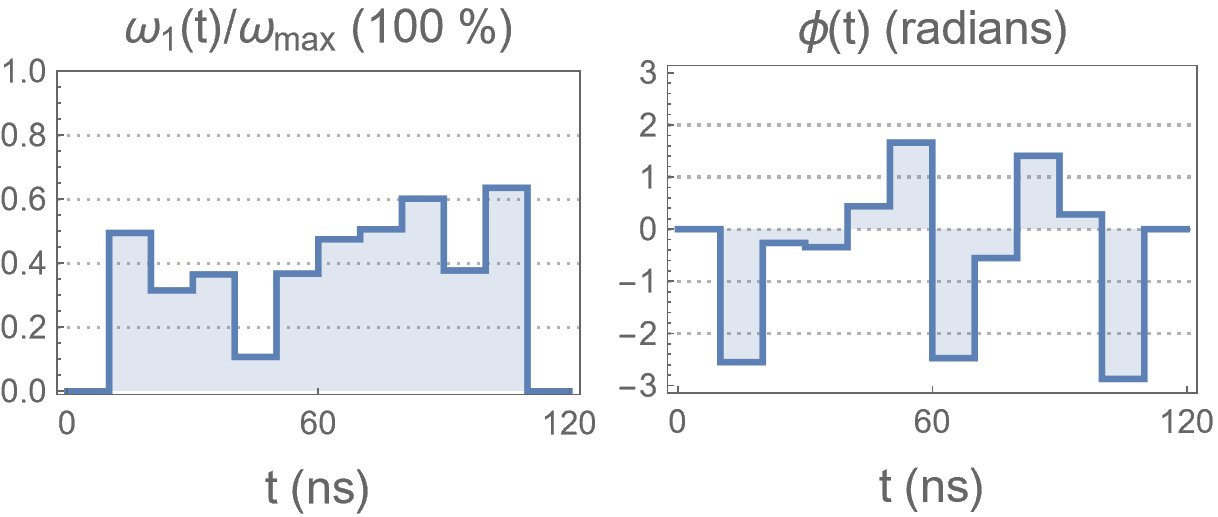}\\
	\caption{ Time-dependent control parameters of the sequence that achieves a scalable spin amplifier through engineering a conditioned decoupling.}
	\label{sacon}
\end{figure}

\begin{figure}[htp]
	\centering
	\includegraphics[width=0.4\textwidth]{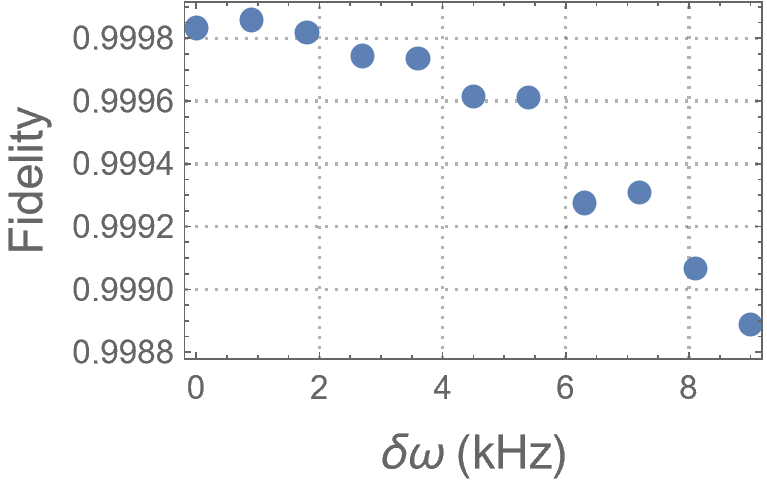}\\
	\caption{Control landscape of the sequence in Fig. \ref{sacon}. The parameter $\Delta\omega$ determines the width of the distribution of the coupling strength between the target qubit and the spin amplifier. The oscillations in fidelity arise from averaging over randomly sampled coupling strengths.}
	\label{fidsa}
\end{figure}

\underline{Step 4. Robustness and optimization.} Design for a FoM of 0.9999 or higher. Fig. \ref{sacon} shows an optimized sequence ($T_\text{seq}=120$ ns). It achieves a scaling factor of $s=0.327$.
To obtain its control landscape, consider a spin amplifier of 6 qubits with the coupling strengths generated from $\omega_i\sim$Unif$[450-\delta\omega,450+\delta\omega]$ kHz and $\omega_{ij}\sim$Unif$[-22.5,22.5]$ kHz. The mean fidelity for different values of $\delta\omega$ is plotted in Fig. \ref{fidsa}.

\underline{Step 5. Calculate FoM.} To calculate the FoM of the sequence, use a spin amplifier of 6 qubits with the coupling strengths generated from $\omega_i\sim$Unif$[441,459]$ kHz and $\omega_{ij}\sim$Unif$[-22.5,22.5]$ kHz. The FoM, calculated as the average gate fidelity of the sequence, is $0.9997_{-0.0001}^{+0.0001}$.

\begin{figure}[htp]
	\centering
	\includegraphics[width=0.45\textwidth]{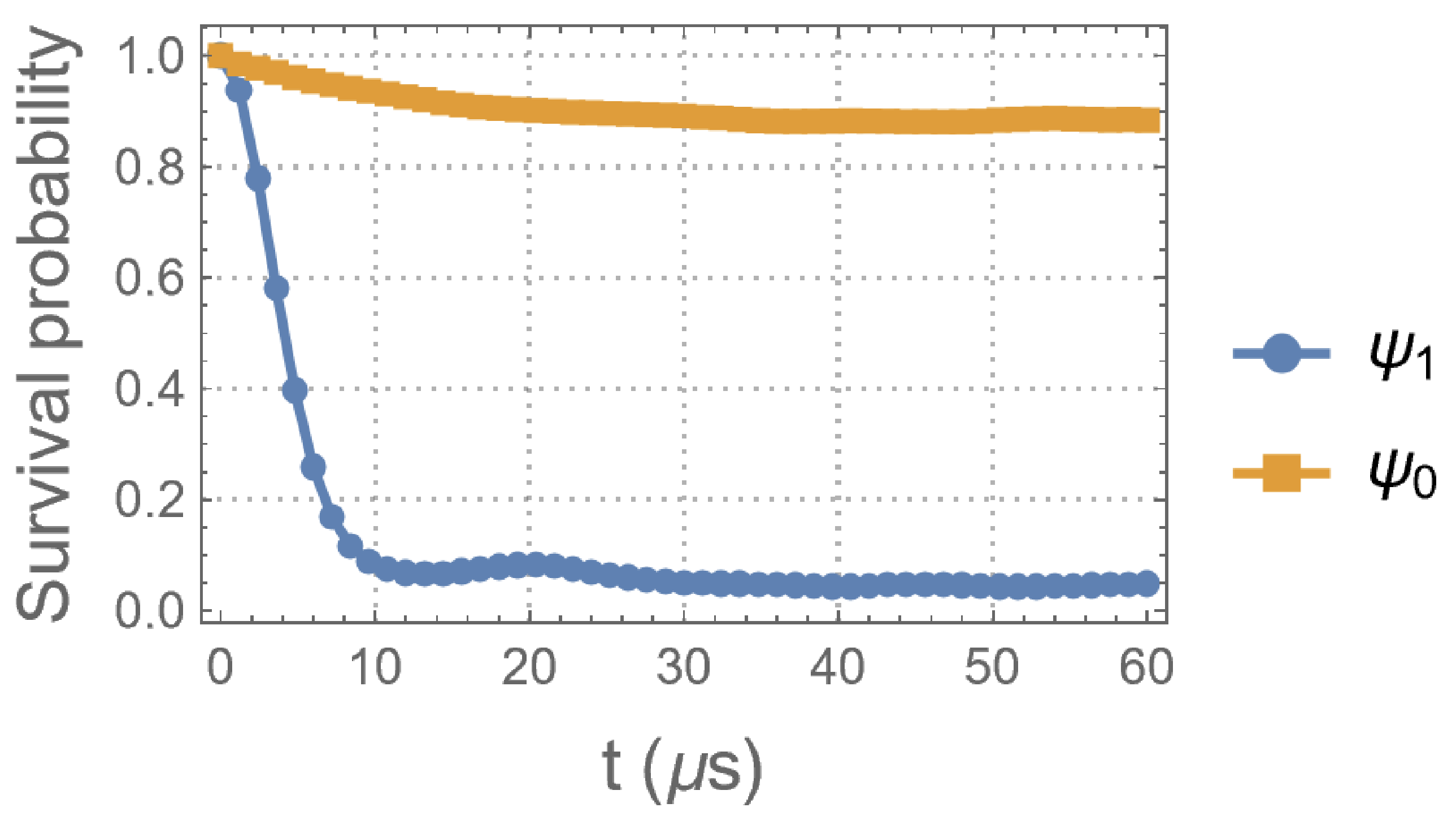}\\
	\caption{Simulated survival probabilities of the states $|\psi_{0/1}\rangle$ under the sequence in Fig. \ref{sacon}. }
	\label{CUsig}
\end{figure}

To test the sequence, we simulate the same spin amplifier used in the calculation of the FoM. 
The initial state of the system evolves under the internal dynamics modulated by the designed sequence. 
Before and after each simulated sequence, a $\pi/2)_x$ and a $-\pi/2)_x$ pulse are applied, respectively. 
The survival probability of the initial state is then measured at the end of the simulation. 
Repeating this procedure for different numbers of applied sequences yields the time evolution of the survival probability. The survival probabilities of the states
\begin{equation}
	|\psi_{0/1}\rangle = |\phi_T=0/1\rangle\otimes|00\cdots0\rangle
\end{equation}
under the sequence are plotted in Fig. \ref{CUsig}. The survival probability is proportional to $M_z$ and can be used to learn the state $|\phi_T\rangle$. 
As can be seen from the result, after 20 $\mu s$, the survival probability of $|\psi_0\rangle$ is above 0.8 while that of $|\psi_1\rangle$ is below 0.1. So the states $|\psi_{0/1}\rangle$ can be distinguished by measuring the magnetization of the spin amplifier, which is proportional to the survival probability.

In principle, one can include robustness by requiring $H_\text{target}^1=0$ and including higher-order corrections \footnote{To minimize the cross term between $H_\text{pert}^1$ and $H_\text{pert}^2$, consider $N=2$, the total Hamiltonian is
	\begin{equation}
		H_\text{tot}(t)=H_\text{pri}(t)+(\omega_1-\omega_0)\z\otimes\z\otimes\mathbb 1+(\omega_2-\omega_0)\z\otimes\mathbb 1\otimes\z+H_\text{pert}^2.
	\end{equation}
	Since both $H_\text{pri}(t)$ and $H_\text{pert}^2$ are symmetric with respect to a permutation of the two qubits in the spin amplifier, we only need to consider $(\omega_1-\omega_0)\z\otimes\z\otimes\mathbb 1$ for leading-order robustness.}. In practice, however, we could not find a sequence that fulfills these requirements simultaneously while maintaining a high scaling factor without significantly increasing the length of the sequence. In turn, decreased scaling factor or fidelity of $U_\text{pri}$ results in decreased FoM and inferior overall performance in sensing. This is because $\|H_\text{pert}^1\|\ll \|H_\text{pert}^2\|$, thus a larger scaling factor on $H_\text{pert}^2$ is much more important. Large scaling factors are crucial when engineering nonzero effective Hamiltonians, as higher-order corrections are typically much smaller and less significant than the dominant low-order scaling.
\section{Comparison with other methods}\label{seccom}
In this section, we compare our framework with selected representative works on robust control and effective Hamiltonian engineering. The goal is not to provide a comprehensive literature review, but rather to clarify the conceptual distinctions and advantages of our approach, highlight its novelty, and illustrate how it can complement and improve existing control strategies.
\subsection{Time-ordered integrals}
Time-ordered integrals of time-dependent Hamiltonians, and of their components under a basis, form the mathematical foundation of a wide range of theories in physics and control.
They underlie the Dyson expansion \cite{dyson1949radiation} and time-dependent perturbation theory in quantum mechanics \cite{cohen1986quantum,sakurai2020modern}, the calculation of the S-matrix in scattering theory \cite{dyson1949s,schwinger1951green,feynman1950mathematical}, stochastic Liouville \cite{kubo1963stochastic} and generalized cumulant expansions \cite{kubo1962generalized} for open systems, path-integral formulations of both quantum \cite{feynman1979path} and stochastic dynamics \cite{chaichian2018path}, the Magnus expansion \cite{magnus1954exponential,blanes2009magnus}, and the Peano–Baker series in control theory \cite{baake2011peano,dollard1979product} and geometric numerical integration \cite{hairer2006geometric}.
Accordingly, it is natural that robust quantum control and effective Hamiltonian engineering methods \cite{haeberlen1968coherent,cheng2004stochastic,paz2016dynamical,bylander2011noise,viola1998dynamical,viola1999universal,bylander2011noise,green2012high,green2013arbitrary,paz2014general,haas2019engineering}, which build upon one or more of these theoretical frameworks, make use of similar time-ordered integral constructions. In particular, the representation in $\mathscr C(\mathbf g_\text{pri},H_\text{pert})$ is analogous to the control matrix \cite{green2013arbitrary,clausen2010bath} and superoperator cumulant coefficients \cite{hangleiter2021filter,ball2021software} in filter-function formalism, the frame matrix in Hamiltonian engineering of multi-spin systems \cite{choi2020robust,tyler2023higher,zhou2023robust}, and expansion coefficients in Floquet engineering \cite{sameti2019floquet,guo2024engineering} and Dyson series \cite{abrams2025efficient,kalev2021integral,chen2021quantum}. Our contribution lies in identifying the minimal subspace $\mathscr C(\mathbf g_\text{pri},H_\text{pert})$ that fully captures the evolution of the toggling-frame Hamiltonian, demonstrating that key properties of a control sequence (i.e., the engineered effective Hamiltonian and robustness) can be expressed entirely through the $\mathscr C$-integrals, and in deriving an exact solution to the corresponding 
$\mathscr C$-integrals for piecewise-constant modulation. This enables a unified and efficient framework for both error analysis and the systematic design of robust control sequences.
\subsection{Universal robust control}
Ref. \cite{poggi2024universally} proposed a useful method based on minimizing the quantum Fisher information to achieve leading-order universal robustness against static Hamiltonian perturbations by minimizing the nontrivial part of the superoperator 
\begin{equation}
	M_0=\frac{1}{T_\text{seq}}\int_0^{T_\text{seq}}dt (U_\text{pri}(t)\otimes U_\text{pri}^*(t))^\dagger,
\end{equation}
and proved that such robustness is always achievable under universal control. 

The condition in~\cite{poggi2024universally} corresponds to a special case of Eq.~\eqref{pogi}, obtained when $\mathscr C(\mathbf g_{\text{pri}}, H_{\text{pert}})$ is the full operator space and the requirement $|\bar H^{(0)}\rrangle = 0$ is imposed for every $H_{\text{pert}}$ except the identity.

Unlike universal robustness, which discards all information about the perturbation, our framework leverages the symmetry and systematic time dependence of the Hamiltonians to target specific error mechanisms, enabling the design of more efficient and adaptable robust sequences based on the available information about the system. In addition to leading-order robustness, our framework provides a 
general condition for achieving higher-order robustness. 
In particular, when $\lc{H_\text{pert}^w}$ is a basis of the full operator 
space, Eq.~\eqref{totr} provides the condition for achieving 
higher-order universal robustness.
Moreover, our method applies to any choice of $H_\text{pri}$ and $H_\text{pert}$, and we prove that zeroth-order robustness to systematic control errors under assumptions listed in Appendix \ref{appen8} is always achievable while offering a systematic procedure to test the achievability of robustness to other error types (e.g., errors outside of $\mathbf g_\text{pri}$) under partial control (Algorithm \ref{L2} and \ref{L3}).
\subsection{Hamiltonian engineering in multi-spin systems}
Ref. \cite{choi2020robust} presents an algebraic rule set that provides human-readable symmetry conditions for constructing decoupling sequences from collective $\pi/2$ rotations and equal delays for general secular many-body Hamiltonians with on-site disorder; it treats finite-pulse effects, rotation-angle errors, and phase (axis) errors to leading order, and outlines an extension to three-body interactions. Refs. \cite{zhou2023robust,tyler2023higher} build on \cite{choi2020robust} to develop a clear, rule-based approach to higher-order Hamiltonian engineering with $\pi/2$ pulses: they derive toggling-frame symmetry motifs (time-reversal, row/block balance) that are sufficient to cancel low-order Magnus terms, use them to design both decoupling sequences and nonzero engineered models (e.g., tunable XXZ), and handle finite-pulse effects via exact first-order overlap corrections together with a leading-order second-order approximation. 

Using our framework, their formulation corresponds to choosing $H_\text{pri}$ and $H_\text{pert}$ as the collective control and internal Hamiltonians of the system, respectively. In the limit of infinitely short pulses, the symmetry conditions they derive can be reproduced by evaluating the 
$\mathscr C$-integrals when the toggling-frame Hamiltonians are restricted to a discrete set of candidate operators. Their conditions for eliminating higher-order terms under finite pulse widths, such as sequence symmetrization, are generally more restrictive than Eq. \eqref{totr}.

By contrast, our work replaces motif rules with a general formulation in terms of 
$\mathscr C$-integrals for all orders for any system and control and choices of $H_\text{pri}$ and $H_\text{pert}$, and turns design into a unified optimization. We derive exact conditions for engineering desired nonzero $\bar H^{(0)}$ and for minimizing higher-order corrections under any modulation. In addition, our framework provides an exact characterization of the set of achievable $H_\text{target}$ at zeroth order, including their relative scaling, and allows one to incorporate  systematic and stochastic errors in both the control and the quantum system.
\subsection{Filter-function formalism}
Dynamical decoupling and the filter-function formalism \cite{gullion1990new,maudsley1986modified,khodjasteh2007performance,ng2011combining,zhang2007dynamical,souza2012effects,khodjasteh2005fault,kofman2001universal,martinis2003decoherence,uhrig2007keeping,uhrig2008exact,cywinski2008enhance,biercuk2011dynamical,alvarez2011measuring,bylander2011noise,paz2017multiqubit,uys2009optimized,biercuk2009optimized,soare2014experimental,green2012high,ajoy2011optimal,alvarez2010performance,ryan2010robust,souza2011robust} are widely used for analyzing and suppressing errors arising from random fluctuations in quantum systems. The leading-order robustness is achieved by minimizing an overlap between the noise power spectrum $S(\omega)$ and the filter function $F(\omega)$ of the decoupling sequence in the frequency domain:
\begin{equation}
	\chi(T_\text{seq})=\frac{1}{\pi}\int_{-\infty}^\infty\frac{d\omega}{\omega^2}S(\omega)F(\omega)d\omega.
\end{equation}
The generalized cumulant expansion formalism has also been widely studied and has recently been applied to design robust control to quantum crosstalk \cite{zhou2023quantum}. By defining an operator-valued filter function \cite{paz2014general,hangleiter2021filter}
\begin{equation}
	F(\omega)=\int_0^{T_\text{seq}} dt_1\int_0^{t_1} dt_2 e^{i\omega(t_1-t_2)}\text{ad}_{H_\text{tog}(t_1)}\text{ad}_{H_\text{tog}(t_2)},
\end{equation}
the condition Eq. \eqref{lead} becomes formally equivalent to the conventional filter-function formalism. Representing the filter function in the minimal error space $\mathscr C_\text{comp}$ enables more efficient and systematic sequence design as well as more transparent error analysis.

Note that, however, the filter-function formalism and AHT are not hierarchical; minimization of $K^{(1)}$ (Eq. \eqref{lead}) does not imply complete decoupling of $H_\text{pert}$ in general.
Our unified formalism, which encompasses both AHT and the cumulant expansion, allows coherent systematic errors and random stochastic fluctuations to be treated on the same footing. This extends the applicability of both approaches and provides a general framework for analyzing and mitigating combinations of coherent and stochastic error mechanisms.
\section{Discussion and outlook}\label{secdis}
The design process developed in this work is compatible with a wide range of internal and control Hamiltonians and applies across the full range of quantum platforms.
Typical control terms for two-level system qubit transitions (e.g., spin-based systems \cite{cory1997ensemble, borneman2012bandwidth, balasubramanian2008nanoscale}) include Rabi strength, phase, and detuning of on-resonance or near-resonance drive. These controls act through microwave or laser fields and provide universal SU(2) control for each qubit.
Similar control terms are used for effective qubits realized by selective driving of two energy levels in multi-level systems (e.g., superconducting qubits \cite{krantz2019quantum}, trapped ions \cite{biercuk2010ultrasensitive,blatt2012quantum,haffner2008quantum}, and Rydberg atom arrays \cite{saffman2010quantum}), where leakage must be effectively managed \cite{motzoi2009simple,gambetta2011analytic}.
Other control building blocks include tuning of local potentials and tunneling rates in quantum dots \cite{loss1998quantum,philips2022universal}, setting amplitudes and phases of photonic modes in linear optics \cite{knill2001scheme}, and applying local modulations and mode couplings in cavity arrays \cite{houck2012chip}. These control terms define the control Hamiltonian $ H_c(t) $ and the time-dependent control parameters $\lc{a_k(t)}$, while the electronic settings of the corresponding hardware specify the model parameters $\lc{\mu_j}$. Uncertainties in $\lc{\mu_j}$ give the error Hamiltonians $\lc{\Delta H_j}$ (Sec. \ref{secro}).	

Multi-qubit gates often rely on coupling between qubits, such as scalar or Fermi contact interactions, dipolar spin-spin coupling, coupling via shared resonators or direct capacitive or inductive elements in superconducting qubits, motional coupling in trapped ions, tunneling-enabled exchange interactions in quantum dots and neutral atoms, and effective interactions via measurement and post-selection in photonic systems. These couplings determine the internal Hamiltonians $H_\text{int}$ and the internal parameters $\lc{\eta_\text{int}^i}$. Typically, single-qubit effects like static detunings and inhomogeneities are also in $H_\text{int}$ since they are not under direct control. 

A partitioning of $H_\text{tot}$ into $H_\text{pri}$ and $H_\text{pert}$ can then be chosen based on the control and internal Hamiltonians. The method described here identifies the set of achievable primary unitaries and perturbative effective Hamiltonians and designs control sequences that meet precision and robustness goals.

As illustrated in Examples \ref{exaro}, Hamiltonian components subject to uncertainties or parameter dispersions (whether in the model or internal parameters) are typically included in \(H_\text{pert}\) to facilitate robust design against such variations.	In Example \ref{exaSA}, the internal dynamics of the spin amplifier are included in $H_\text{pert}$ to enable the design of a scalable solution.
When selecting a well-characterized, finite-dimensional component of \(H_\text{tot}\) to include in \(H_\text{pert}\) for engineering (Example \ref{exaQHO}), the partitioning is guided by three considerations: (1) optimizer efficiency, as different splits affect the search complexity; (2) controllability, since reachable targets depend on the chosen partition; and (3) parameter tunability, as only parameters in \(H_\text{pert}\) appear in the engineered Hamiltonian. For high-fidelity implementations, fixed parameters are best placed in \( H_\text{pri} \), as \( U_\text{pri} \) is generated by numerical integration and avoids errors due to truncation of the Magnus expansion. If a sequence is hard to find, the partitioning of the total Hamiltonian can be adjusted. A practical strategy is to allow the optimizer to adaptively modify the partitioning using auxiliary parameters that control the weight of terms in \( H_\text{pri} \) and \( H_\text{pert} \), provided controllability is preserved. This can lead to shorter and more easily discovered sequences.

Characterizing controllability involves two key steps: identifying the minimal subspace of achievable zeroth-order average Hamiltonians for each $H_\text{pert}^w$, and determining the range of achievable scaling factors. These steps are crucial because the controllability of both nonzero effective Hamiltonians and robustness depends on the choice of $\mathbf{g}_\text{pri}$ and on how $H_\text{tot}$ is partitioned. For instance, an Ising-coupled qubit system cannot be fully decoupled using only collective rotations, while including part of the internal Hamiltonian in $H_\text{pri}$ can enable engineering of multibody interactions at zeroth order, something unattainable with control Hamiltonians alone.  

The computed $\mathscr{C}_w(\mathbf{g}_\text{pri}, H_\text{pert}^w)$ defines the minimal error subspace for each perturbative term and provides the foundation for the general conditions of precise and robust Hamiltonian engineering (Eqs.~\eqref{tot0} and \eqref{totr}). The achievable scaling factors bound the efficiency of the engineered dynamics and determine the joint achievability of $\lc{s_w H_\text{pert}^w}$, which together form the engineered effective Hamiltonian. This joint controllability also quantifies the relative strength of errors from different $H_\text{pert}^w$ and enables more targeted error-suppression strategies.

While controllability characterization determines whether a target is reachable, the time and quality of a resulting sequence can depend on the choice of optimizer and partitioning. Although generalized simulated annealing \cite{tsallis1996generalized} was employed in the examples for its ability to escape local minima, the focus of this work is not on the specific choice of optimizer. The cost functions in Eqs.~\eqref{totobj} and~\eqref{subobj} are compatible with a wide range of numerical optimization algorithms, such as Nelder-Mead \cite{nelder1965simplex, lagarias1998convergence}, nonlinear conjugate gradient \cite{hager2005new}, Powell’s method \cite{powell1964efficient}, particle swarm \cite{kennedy1995particle}, sequential convex programming \cite{kosut2013robust}, reinforcement learning \cite{sutton1998reinforcement}, or differential evolution \cite{storn1997differential}. These algorithms are readily available in standard numerical libraries, such as those implemented in Python \cite{virtanen2020scipy,harris2020array,paszke2019pytorch}.

The control workflow (Sec. \ref{secflow}) supports optimization for the largest possible scaling factor, which is key to achieving high efficiency. When designing a sequence for nonzero $\bar H^{(0)}$, achieving large scaling factors is typically more important than accounting for higher-order corrections. This is due to two reasons: first, higher-order terms typically scale as $O((\|H_\text{pert}\|/\omega_\text{mod})^n)$ so when the strength of the modulation $\omega_\text{mod}$ is larger than $\|H_\text{pert}\|$, $\bar H^{(0)}$ dominates higher-order terms; and second, higher-order terms are modulated by $\bar H^{(0)}$ due to a second averaging effect \cite{haeberlen2012high,cory1990multiple}. Take $\bar H^{(1)}$ as an example,
\begin{equation}
	U(T_\text{seq})\approx\exp\ls{-i\lr{s\bar H^{(0)}+\bar H^{(1)}}T_\text{seq}}.
\end{equation}
In the interaction frame of $s\bar H^{(0)}$, $\bar H^{(1)}$ becomes $e^{i s\bar H^{(0)}T_\text{seq}}\bar H^{(1)}e^{-i s\bar H^{(0)}T_\text{seq}}$. If $[\bar H^{(0)},\bar H^{(1)}]\ne0$, $\bar H^{(1)}$ is modulated by $s\bar H^{(0)}$ and its average scales as $1/s$ over repeated applications of the sequence.
However, when $s$ is small or $[\bar H^{(0)},\bar H^{(1)}]=0$, it is still important to account for $\bar H^{(1)}$ and higher-order terms.	

The method naturally includes parameter dispersions with moderate to long correlation times. Control landscapes (fidelity maps over internal and model parameters) provide insight into precision and robustness. A broad, flat high-fidelity region generally indicates greater robustness to parameter variations. Sequences can be designed for high fidelity across parameter distributions, with performance quantified by an average gate fidelity FoM.
The computation of the FoM allows the inclusion of Markovian noise and relaxation processes, either via Lindbladian master equations or by composing with noise channels when they commute with the target unitary (e.g., depolarizing channels).
Finally, by outputting the average CPTP map $\bar\Lambda$ or a corresponding Kraus representation, we gain diagnostic access to residual errors. The orthogonality $\mathcal R(\bar \Lambda)$ captures robustness, while its fidelity reflects both precision and robustness. These diagnostics help assess whether a sequence meets requirements or needs further refinement. Further diagnosis of possible error mechanisms can be achieved by expressing $\bar \Lambda$ in its Kraus sum representation \cite{kraus1971general}, which provides the set of possible errors that is important for the design of quantum error correction codes \cite{knill1997theory}. 

The conditions for minimizing higher-order corrections (Eq. \eqref{totr}) are the most general and apply to all \(h_1^\text{comp}, h_2^\text{comp}, \ldots\). Conventional approaches, such as sequence symmetrization to eliminate the first-order term, can fail when $H_{\text{pri}}$ contains part of $H_{\text{int}}$ or when $H_{\text{pert}}$ is time-dependent.
In such cases, the general conditions presented here must be used. When the symmetry or topological properties of \(h_1^\text{comp}, h_2^\text{comp}, \ldots\) are taken into account, Eq. \eqref{totr} can be further relaxed, enabling characterization of the achievable space of higher-order terms and the engineering of interactions, such as multi-body couplings, that are inaccessible at zeroth order. This will be reported separately.

The generalization in Sec.~\ref{secsto} can be applied to suppress errors arising from random fluctuations in any model or internal parameter without relying on standard approximations (e.g., Gauss-Markov) of the stochastic process. When combined with the method in Sec. \ref{secro}, it further enables robustness to variations in the noise model itself by minimizing the effect of a stochastic error Hamiltonian \(\Delta H'\), for instance, achieving robustness to variations in the correlation time of the noise. The \(\mathscr{C}\)-space representation also allows selective suppression or enhancement of specific components, which can be exploited for noise spectroscopy to isolate, for example, particular correlations \(\langle \beta_{w_1}(t)\beta_{w_2}(t)\rangle\) while suppressing others \cite{alvarez2011measuring,norris2016qubit,sung2019non,paz2017multiqubit}. By appropriately choosing \(H_\text{pri}\) and \(H_\text{pert}\) (e.g., including the coherent component of the noise in \(H_\text{pri}\)), one can even utilize the noise as a control resource while mitigating decoherence from stochastic fluctuations. The result will be published separately.

\section{Conclusion}\label{secsum}	
This work presents a general framework for precise and robust quantum control via effective Hamiltonian engineering, addressing three key challenges: (1) characterizing the set of achievable zeroth-order effective Hamiltonians and demonstrating the advantage of identifying the corresponding subspace when designing control sequences; (2) establishing general conditions for engineering target zeroth-order average Hamiltonians and eliminating higher-order corrections; and (3) achieving robustness to parameter variations by minimizing the effective contribution of the error Hamiltonian. The framework supports robust state transfer and characterizes the set of reachable density matrices. Combined with a cumulant expansion, it enables the design of sequences robust to stochastic parameter fluctuations, and it is integrated with numerical optimization to produce high-performance control sequences.

Importantly, the process does not depend on the researcher’s intuition to direct the algorithm to a solution. It systematically engineers precise zeroth-order effective Hamiltonians while minimizing higher-order corrections, and establishes the foundation for extending control design to nonzero higher-order terms. This unified approach enables the autonomous design of coherent control methods applicable to Hamiltonian systems beyond coupled spins.

The process flow incorporates controllability characterization to ensure that the desired unitary or effective Hamiltonian is achievable. The minimal subspace of the toggling-frame Hamiltonian identifies possible error forms, enabling efficient and targeted suppression of errors and systematic error analysis of any control sequence. The explicit objective functions (Eqs. (60) and (61)) for achieving precision and robustness allow automated sequence design with a high success rate. Overall, the method highlights the power of AHT at zeroth order and combines it with explicit control of errors and represents a useful step toward the automated design of high-performance sequences for quantum control in different systems.
\section*{Online Content}
Mathematica implementations of the algorithms presented in this paper are openly available at https://github.com/Jiahui-Chen-quantum/Hamiltonian-engineering.
The package contains notebooks for controllability analysis, 
$\mathscr C$-integral computation, and a runnable example.
Licensed under MIT.
\section*{Acknowledgments}
We thank Cheng Zheng for helpful comments on the manuscript.
This work was supported by the Canadian Excellence Research Chairs (CERC) program, the Natural Sciences and Engineering Research Council of Canada (NSERC) Discovery program, the Canada First Research Excellence Fund (CFREF), and the Innovation for Defence Excellence and Security (IDEaS) Program of the Canadian Department of National Defence.
\appendix
\section{Choosing $H_\text{pri}$ and $H_\text{pert}$}\label{appen0}
Achieving the target effective Hamiltonian at zeroth order has several advantages. First, when \(\|H_{\text{pert}}\| T_{\text{seq}} < 1\), the zeroth-order term dominates the higher-order corrections, so implementing $H_targe$t at zeroth order is more efficient. This is typically, though not necessarily, the regime in which the Magnus expansion converges \cite{haeberlen1968coherent,blanes2009magnus}. Second, higher-order terms live in a larger operator space and require suppressing lower-order terms, which makes them substantially harder to engineer.
Third, for any finite-dimensional quantum system, one can show that any effective Hamiltonian that is achievable for the given control system can be represented within $\mathscr C(\mathbf g_\text{pri},H_\text{pert})$ and treated at zeroth order, provided the partitioning of \(H_{\text{tot}}\) is chosen correctly.
To see this, let $\lc{H_j}$ be an Hermitian basis of the minimal operator subspace containing $H_\text{tot}(t)$. Thus, the total Hamiltonian can be written as
\begin{equation}
	H_\text{tot}(t)=\sum_{j}e_j(t)H_j.
\end{equation}
The total Lie algebra $\mathbf g_\text{tot}$ is spanned by the right-normed commutators
\begin{equation}
	\lc{[iH_{j_1},[iH_{j_2},\ldots,[iH_{j_{l-1}},iH_{j_l}]\ldots]]|\forall\text{ $j_1,\ldots,j_l$ and $l$}}.
	\label{bas1}
\end{equation}
$\mathbf g_\text{tot}$ determines all achievable effective Hamiltonians. Consider the partitioning
\begin{equation}
	\begin{split}
		&H_\text{pri}(t)=\sum_{j}(1-p_j(t))e_j(t)H_j,\\
		&H_\text{pert}(t)=\sum_{j}p_j(t)e_j(t)H_j,
	\end{split}
\end{equation}
where $0\le p_j(t)<1$ are weights that control the partitioning. Thus, the primary Lie algebra is identical to the total Lie algebra. Given that $\lc{p_j(t)}$ can be freely adjusted
$\mathscr C(\mathbf g_\text{pri},H_\text{pert}(t))$ is spanned by
\begin{equation}
	\lc{[g_{j_1},[g_{j_2},\ldots,[g_{j_{l-1}},iH_{j_l}]\ldots]]|\forall\text{ $j_1,\ldots,j_l$ and $l$}},
	\label{bas2}
\end{equation}
where $g_{j_1},\ldots,g_{j_{l-1}}\in\mathbf g_\text{pri}$. \eqref{bas1} is a subset of \eqref{bas2}. Therefore, any achievable effective Hamiltonian is included in $\mathscr C(\mathbf g_\text{pri},H_\text{pert}(t))$ and achievable at zeroth order.

However, because the partitioning determines how different components of \(H_{\text{tot}}\) are treated, not every choice of \(H_{\text{pri}}\) and \(H_{\text{pert}}\) is suitable and for a fixed partitioning of $H_\text{tot}$, not all target effective Hamiltonians are achievable at zeroth order. The primary unitary \(U_{\text{pri}}\), obtained by numerically integrating \(H_{\text{pri}}\), is well-suited for implementing precise target unitaries since it avoids any perturbative truncation.
On the other hand, effective Hamiltonian engineering of \(H_{\text{pert}}\) is well-suited for designing scalable, symmetry-based, and tunable Hamiltonian targets. To preserve the advantages of both approaches, \(H_{\text{pri}}\) is chosen so that \(\mathbf g_{\text{pri}}\) remains finite-dimensional and does not contain uncertain parameters. In addition, (asymptotically) infinite-dimensional spaces \(\mathscr C(\mathbf g_{\text{pri}}, H_{\text{pert}}(t))\) needs to be avoided, as truncation can compromise the precision of the engineered effective Hamiltonians. Based on these considerations, we propose the following guidelines for partitioning \(H_{\text{tot}}\):
\begin{enumerate}
	\item The partitioning does not lead to (asymptotically) infinite-dimensional $\mathbf g_\text{pri}$ or $\mathscr C(\mathbf g_\text{pri},H_\text{pert}(t))$
	\item Uncertain parameters are included in $H_\text{pert}$
	\item Finite-dimensional and accurate terms are included in $H_\text{pri}$ unless one wishes to design scalable or tunable solutions. 
\end{enumerate}

There is another important degree of freedom in choosing a partition that is not addressed by the above rules: when a Hamiltonian term is distributed between \(H_{\text{pri}}\) and \(H_{\text{pert}}\) (so that controllability is unchanged), how should one choose the relative allocation? 
For example, one may wish to place a Hamiltonian term associated with an uncertain parameter in \(H_{\text{pri}}\). According to the second rule above, this requires selecting a primary value for that parameter (typically the mean of its distribution) and including the corresponding Hamiltonian contribution in \(H_{\text{pri}}\). By placing the dominant part of a noisy Hamiltonian in \(H_{\text{pri}}\), the robustness region is centered around the chosen primary parameter value. This also enables robustness beyond the perturbative regime: by connecting multiple such robustness regions, one can construct broadband robust sequences that maintain high fidelity across a wide parameter range. This also provides a general mechanism for using noise as a resource to enhance controllability while mitigating its detrimental effects.

A finite-dimensional and accurately known term may also be split between \(H_{\text{pri}}\) and \(H_{\text{pert}}\) to engineer efficient and tunable dynamics. In this case, one must balance how much of the term is retained in \(H_{\text{pri}}\) to ensure controllability and how much is placed in \(H_{\text{pert}}\) to provide tunability.
If too little of the term is included in \(H_{\text{pri}}\), the sequence may become inefficient, as a longer evolution time is required to steer \(H_{\text{tog}}(t)\) in the desired direction. Conversely, if too little is placed in \(H_{\text{pert}}\), the engineered dynamics become weak, again reducing the efficiency of the sequence.
Therefore, a good partitioning should place only the necessary part of the Hamiltonian in $H_{\text{pri}}$ to ensure controllability, with the remainder assigned to $H_{\text{pert}}$ to maximize its strength.
\section{Algorithm for Calculating Minimal Lie Algebra}\label{appen1}
Here, we present an algorithm adapted from Ref. \cite{d2021introduction} to determine the primary Lie algebra \(\mathbf{g}_\text{pri}\).
For a given primary Hamiltonian \(H_\text{pri}(t)\), let \(\{H_j\}\) be a basis of the minimal operator subspace that contains all possible realizations of \(H_\text{pri}(t)\) for arbitrary trajectories of the control parameters \(\{a_k(t)\}\).
$\lc{iH_j}$ is the set of generators of the Lie algebra, $\mathbf g_\text{pri}$, which can be calculated using
Algorithm \ref{L1}.
\begin{algorithm}[H]
	\caption{Algorithm for calculating the Lie algebra generated by $\lc{iH_j}$. The function \textsc{FindLieAlgebra}($\lc{iH_j}$) takes a set of generators $\lc{iH_j}$ and returns an orthonormal basis of the generated Lie algebra.}\label{L1}
	\begin{algorithmic}[1]
		\Function{FindLieAlgebra}{$\lc{iH_j}$}
		\State $\mathcal A_0\gets \lc{iH_j}$
		\State $v\gets 0$
		\While{$\mathcal A_v$ is not empty and $\left| \bigcup_{i=1}^{v}\mathcal A_i\right|<4^{N}-1$}
		\State $\mathcal A_{v+1}\gets\lc{}$
		\For{$g\in \mathcal A_0$ and $h\in \mathcal A_v$}
		\If {$[h,g]$ is linearly independent with $\bigcup_{i=1}^{v+1}\mathcal A_i$}\Comment{\textit{Linear dependence can be checked by calculating the rank of the matrix of the vectorizations of the operators.}}
		\State Append $[h,g]$ to $\mathcal A_{v+1}$
		\EndIf
		\EndFor
		\State $v\gets v+1$
		\EndWhile
		\State Use the Gram–Schmidt process to orthonormalize $\bigcup_{i=1}^{v}\mathcal A_i$ and \Return the result.
		\EndFunction
	\end{algorithmic}
\end{algorithm}
\section{Calculating $\mathscr C({\mathbf{g}_\text{pri}},H_\text{pert})$}\label{appen2}
Here,  we prove a theorem that enables the calculation of $\mathscr C({\mathbf{g}_\text{pri}},H_\text{pert})$ using Algorithm \ref{L2} in the main text. Next, we discuss the representation for calculating $\mathscr C({\mathbf{g}_\text{pri}},H_\text{pert})$ and how to engineer multiple $H_\text{pert}^w$ in parallel.

\begin{theorem}
	For a Hamiltonian $H_\text{pert}$, define 
	\begin{equation}
		\begin{split}
			&\mathscr{C}({\mathbf{g}_\text{pri}}, H_\text{pert})\\
			&=\text{span}_{\mathbb R}\lc{[g_1,\cdots[g_L,H_\text{pert}]\cdots]|g_l\in {\mathbf{g}_\text{pri}}, L\ge0},
		\end{split}
	\end{equation}
	and
	\begin{equation}
		\begin{split}
			&\mathscr{O}({\mathbf{g}_\text{pri}}, H_\text{pert})\\
			&=\text{span}_\mathbb{R}\left\{U_\text{pri}^\dagger(t)H_\text{pert}U_\text{pri}(t)dt\middle|U_\text{pri}(t)\in e^{\mathbf{g}_\text{pri}}\right\}.
		\end{split}
	\end{equation} 
	Then $\mathscr{C}({\mathbf{g}_\text{pri}}, H_\text{pert})=\mathscr{O}({\mathbf{g}_\text{pri}}, H_\text{pert})$.
\end{theorem}
\begin{proof}
	Consider piecewise-continuous $U(t)$. For all $U(t)\in e^{{\mathbf{g}_\text{pri}}}$, there exists $ g_t\in{\mathbf{g}_\text{pri}}$, such that $U(t)=e^{g_t}$. 
	Therefore, $e^{-g_t}H_\text{pert}e^{g_t}=e^{-\text{ad}_{g_t}}(H_\text{pert})\in \mathscr C({\mathbf{g}_\text{pri}}, H_\text{pert})$ and $\mathscr O({\mathbf{g}_\text{pri}}, H_\text{pert})\subset\mathscr C({\mathbf{g}_\text{pri}}, H_\text{pert})$.
	
	Since the total Hilbert space $\mathcal H$ is finite-dimensional, there exists $m<|\mathcal H|^2-1$ such that $\{H_\text{pert},\text{ad}_g(H_\text{pert}),\ldots,\text{ad}_g^{m+1}(H_\text{pert})\}$ is linearly dependent $\forall g\in{\mathbf{g}_\text{pri}}$. Assume that $\{H_\text{pert},\text{ad}_g(H_\text{pert}),\ldots,\text{ad}_g^m(H_\text{pert})\}$ is linearly independent and $e^{gt_i}H_\text{pert}e^{-gt_i}$ is in
	\begin{equation*}
		\text{span}_\mathbb{R}\{H_\text{pert},\text{ad}_g(H_\text{pert}),\ldots,\text{ad}_g^m(H_\text{pert})\}.
	\end{equation*} 
	We can choose $t_i$ for $i=1,\ldots,m$ such that 
	\begin{equation}
		[g,H_\text{pert}]\in\text{span}_\mathbb{R}\left\{e^{gt_i}H_\text{pert}e^{-gt_i}\middle|i=1,\ldots,m\right\},
	\end{equation}
	$\forall g_1,g_2\in{\mathbf{g}_\text{pri}}$, we have
	\begin{equation}
		\begin{split}
			\ls{g_2,\ls{g_1,H_\text{pert}}}&=\sum_{j=1}^sa_j^2e^{g_2t_j'}[g_1,H_\text{pert}]e^{-g_2t_j'}\\
			&=\sum_{j=1}^sa_j^2e^{g_2t_j'}\lr{\sum_{i=1}^ma_i^1e^{g_1t_i}H_\text{pert}e^{-g_1t_i}}e^{-g_2t_j'}\\
			&=\sum_{j=1}^s\sum_{i=1}^ma_j^2a_i^1e^{g_{ij}}H_\text{pert}e^{-g_{ij}},
		\end{split}
	\end{equation}
	for some $g_{ij}\in {\mathbf{g}_\text{pri}}$. The last equality is due to the Baker–Campbell–Hausdorff theorem. Therefore, by induction, $[g_1,\ldots,[g_L,H_\text{pert}]]\in\mathscr{O}({\mathbf{g}_\text{pri}}, H_\text{pert})$ $\forall g_1,\ldots,g_L\in{\mathbf{g}_\text{pri}}$. Thus, $\mathscr C({\mathbf{g}_\text{pri}}, H_\text{pert})\subset \mathscr{O}({\mathbf{g}_\text{pri}}, H_\text{pert})$.
	
	Combining the result with the first part of the proof, we have $\mathscr{C}({\mathbf{g}_\text{pri}}, H_\text{pert})=\mathscr{O}({\mathbf{g}_\text{pri}}, H_\text{pert})$. 
\end{proof}
For time-dependent $H_\text{pert}(t)$, the space can be calculated as
\begin{equation}
	\mathscr C(\mathbf{g}_\text{pri},H_\text{pert}(t))=\bigcup_j \mathscr C(\mathbf{g}_\text{pri},H_j^\text{pert}),
\end{equation} 
where $\lc{H_j^\text{pert}}$ is a basis of the minimal operator subspace containing all $H_\text{pert}(t)$.

It is convenient to choose a representation $\xi$ of ${\mathbf{g}_\text{pri}}$ and $H_\text{pert}$ that is restricted to a subspace of $\mathcal H$ to calculate $\mathscr C({\mathbf{g}_\text{pri}},H_\text{pert})$. $\xi$ needs to preserve the structure of $\mathscr C({\mathbf{g}_\text{pri}}, H_\text{pert})$, namely
\begin{equation}
	[\xi(g),\xi(h_1)]=\xi(h_2)\iff[g,h_1]=h_2,
	\label{rep}
\end{equation}
for all $g\in{\mathbf{g}_\text{pri}}$ and $h_1,h_2\in\mathscr C({\mathbf{g}_\text{pri}},H_\text{pert})$. The representation is bijective and unique up to isomorphism. 

\begin{figure}[htp]
	\centering
	\includegraphics[width=0.4\textwidth]{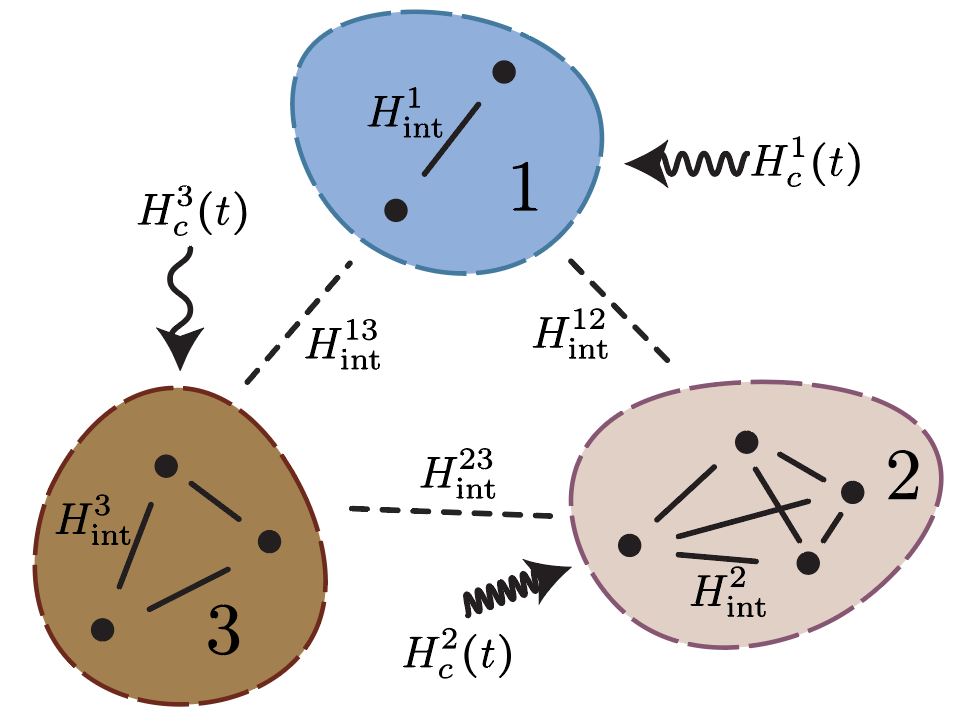}\\
	\caption{A qubit network of three clusters. Different clusters are controlled independently and have their own internal Hamiltonians.}
	\label{cluster}
\end{figure}

When a finite-dimensional representation does not exist, sometimes we can decompose $H_{\text{pert}}$ into components and define a finite-dimensional representation for each component. A representative example is Eq. (3) from the main text where it is convenient to group the bilinear and linear terms into $H_\text{pert}^1$ and $H_\text{pert}^2$, respectively. 
Another example is a qubit network with different clusters that are controlled independently (Fig. \ref{cluster}). In this case, the different interaction Hamiltonians, $H_\text{int}^i$ and $H_\text{int}^{ij}$, are defined as six different $H_\text{pert}^w$.

\section{Finding the Set of Achievable $H_\text{eff}$}\label{appen3}
In this appendix, we first prove that the set of achievable $\bar H^{(0)}$ is the convex hull of $\lc{U_i^\dagger H_\text{pert}U_i}$, $U_i\in e^{\mathbf g_\text{pri}}$. Then, the formal version of Algorithm \ref{L3} in the main text is given. Finally, the result is generalized to characterizing the range of achievable scaling factors when multiple $H_\text{pert}^w$ are being engineered. It is convenient to use a vector representation of a Hamiltonian $|H\rrangle$ in $\mathscr C(\mathbf{g}_\text{pri},H_\text{pert})$ (Eqs. \eqref{cal}), we have the following result.
\begin{theorem}
	Define
	\begin{equation}
		\begin{split}
			&\mathscr{O'}(\mathbf{g}_\text{pri}, H_\text{pert})\\
			&=\left\{\frac{1}{{T_\text{seq}}}\int_0^{T_\text{seq}}  dt|U^\dagger(t)H_\text{pert}U(t)\rrangle\middle | U(t)\in e^{\mathbf{g}_\text{pri}}\right\},
		\end{split}
	\end{equation}
	and
	\begin{equation}
		\mathscr{Q}(\mathbf{g}_\text{pri}, H_\text{pert})=\left\{  |U^\dagger H_\text{pert}U\rrangle |U\in e^{\mathbf{g}_\text{pri}}\right\}.
		\label{achset}
	\end{equation}
	Denote the convex hull of $\mathscr{Q}(\mathbf{g}_\text{pri}, H_\text{pert})$ as $\mathscr{\bar Q}(\mathbf{g}_\text{pri}, H_\text{pert})$. Assume that any $U\in e^{\mathbf g_\text{pri}}$ can be implemented in a time $t_p$ much shorter than $1/\|H_\text{pert}\|$,
	then $\mathscr{O'}(\mathbf{g}_\text{pri}, H_\text{pert})=\mathscr{\bar Q}(\mathbf{g}_\text{pri}, H_\text{pert})$.
	\label{achiev}
\end{theorem}
\begin{proof}
	An element from $\mathscr{\bar Q}(\mathbf{g}_\text{pri}, |H_\text{pert}\rrangle)$ takes the form $\sum_{i=1}^Mp_i  |U_i^\dagger H_\text{pert}U_i\rrangle$ for a positive integer $M$ and probabilities $\{p_i\}$. Let
	\begin{equation}
		g\lr{T_\text{seq}\sum_{j=1}^{i-1}p_j+t}=\begin{cases}
			\frac{g_i}{t_p}&t\in[0,t_p)\\
			0&t\in[t_p,p_i{T_\text{seq}}-t_p)\\
			\frac{-g_i}{t_p}&t\in[p_i{T_\text{seq}}-t_p,p_iT_\text{seq}]
		\end{cases},
	\end{equation}
	for $i=1,2,\ldots,M$. Then 
	\begin{equation}
		\lim_{t_p\rightarrow 0^+}\frac{1}{{T_\text{seq}}}\int_0^{T_\text{seq}} |U^\dagger(t)H_\text{pert}U(t)\rrangle dt=\sum_{i=1}^Mp_i|e^{-g_i}H_\text{pert}e^{g_i}\rrangle.
	\end{equation}
	Therefore, $\mathscr{\bar Q}(\mathbf{g}_\text{pri}, H_\text{pert})\subset\mathscr{O'}(\mathbf{g}_\text{pri}, H_\text{pert})$. For all $U(t)\in e^{{\mathbf{g}_\text{pri}}}$,
	\begin{equation}
		\begin{split}
			&\frac{1}{{T_\text{seq}}}\int_0^{T_\text{seq}} |U^\dagger(t)H_\text{pert}U(t)\rrangle dt\\
			=&\lim_{N\rightarrow\infty}\frac{1}{N}\sum_{i=1}^N \left|U^\dagger\lr{\frac{i{T_\text{seq}}}{N}}H_\text{pert}U\lr{\frac{i{T_\text{seq}}}{N}}\right\rrangle\in\mathscr{\bar Q}(\mathbf{g}_\text{pri}, H_\text{pert}).
		\end{split}
	\end{equation}
	The limit is valid if $t_p\ll 1/\|H_\text{pert}\|$. Thus, $\mathscr{O'}(\mathbf{g}_\text{pri}, H_\text{pert})\subset\mathscr{\bar Q}(\mathbf{g}_\text{pri}, H_\text{pert})$. Therefore, $\mathscr{O'}(\mathbf{g}_\text{pri}, H_\text{pert})=\mathscr{\bar Q}(\mathbf{g}_\text{pri}, H_\text{pert})$.
\end{proof}
If one can implement
\begin{equation}
	\mathcal T\exp\lr{-i\int_0^t H_\text{tot}(t')dt'}=\exp(-iH_\text{pert}t)U_t,
\end{equation}
for all $t>0$ and some $U_t\in e^{\mathbf g_\text{pri}}$, e.g., when  $H_\text{pri}(t)=H_c(t)+pH_\text{int}$ and $H_\text{pert}=(1-p)H_\text{int}$, and $H_\text{pri}(t)=H_c(t)$ and $H_\text{pert}(t)=\epsilon H_c(t)$, the condition $t_p\ll 1/\|H_\text{pert}\|$ in the above proof is unnecessary. Finite $t_p$, however, may limit the efficiency of a sequence. 

$\mathscr{\bar Q}(\mathbf{g}_\text{pri}, H_\text{pert})$ can be approximated using random unitaries $\lc{u_j}$:
\begin{equation}
	\begin{split}
		&\mathscr{\bar Q}(\mathbf{g}_\text{pri}, H_\text{pert})\\
		\approx&\left\{\sum_{j=1}^Jx_j\left| u_j^\dagger H_\text{pert} u_j\right\rrangle\middle| \sum_{j=1}^Jx_j=1,x_j\ge 0, 1\le j\le J\right\}.
	\end{split}
\end{equation}
Choose a matrix $T$ such that $T|H_\text{target}\rrangle=(1,0,0,\ldots,0)$ and define a set of vertices $\lc{\vec v^j=T\left| u_j^\dagger H_\text{pert} u_j\right\rrangle}$, the range of achievable scaling factors, $[s_-,s_+]$, can be found by solving the following linear programs
\begin{equation}
	\begin{array}{lll}
		\text{LP}_\pm:&\text{Find a vector}&\vec x\\
		&\text{that maximizes}&s_\pm=\vec c_\pm\cdot\vec x \text{ with } \vec c_\pm=(\pm 1,0,\ldots,0)\\
		&\text{subject to }&\sum_i x_i=1, -1\le \sum_i v^i_1x_i\le 1\\
		&\text{and}& \sum_iv_k^ix_i=0, k=2,3,\ldots\\
		&\text{and}& \vec x\ge 0
	\end{array}.
	\label{lp}
\end{equation}
$H_\text{target}$ is not achievable if the constraint is infeasible (Fig. \ref{convex}). 
The procedure is summarized as the following algorithm.
\begin{algorithm}[H]
	\caption{Algorithm for finding achievable scaling factors. The function \textsc{FindScale} takes a basis of the primary Lie algebra $\mathcal B({\mathbf{g}_\text{pri}})$, a perturbative Hamiltonian $H_\text{pert}$ and a target effective Hamiltonian $H_\text{target}$ and returns either False when $H_\text{target}$ is not achievable or the range of achievable scaling factors $[s_-,s_+]$.}\label{scalealg}
	\begin{algorithmic}[1]
		\Function{FindScale}{${\mathbf{g}_\text{pri}}$,  $H_\text{pert}$}
		\State Calculate $\mathcal B(\mathscr{C})$ 
		\State find $T$ such that $T|H_\text{target}\rrangle=(1,0,\ldots,0)$
		\For{$j=1,\ldots,J$} 
		\State Sample $u_j\in e^{{\mathbf{g}_\text{pri}}}$
		\State $\vec v^j\gets T|u_j^\dagger H_\text{pert}u_j\rrangle$
		\EndFor
		\State Solve the linear programs LP$_+$ and LP$_-$
		\If {no solution exists}
		\State \Return False
		\Else 
		\State {\Return $[s_-,s_+]$}
		\EndIf
		\EndFunction
	\end{algorithmic}
\end{algorithm}
When
\begin{equation}
	H_\text{pert}=\sum_{w=1}^WH_\text{pert}^w,
\end{equation}
the condition for parallel effective Hamiltonian engineering is
\begin{equation}
	\bigoplus_{w=1}^W|\bar H^{(0)w}\rrangle_{w}=\bigoplus_{w=1}^W|H_\text{target}^w\rrangle_{w}\equiv	|H_\text{target}\rrangle_\text{comp},
	\label{SHEsub}
\end{equation}
where $|\cdot\rrangle_w$ is the representation in $\mathscr{C}_w({\mathbf{g}_\text{pri}},H_\text{pert}^w)$. We have
\begin{theorem}
	Define
	\begin{equation}
		\begin{split}
			&\mathscr{O'}(\mathbf{g}_\text{pri}, \lc{H_\text{pert}^w})\\
			&=\left\{\frac{1}{{T_\text{seq}}}\int_0^{T_\text{seq}}  dt\bigoplus_{w=1}^W|U_\text{pri}^\dagger(t)H_\text{pert}^wU_\text{pri}(t)\rrangle_w\middle | U_\text{pri}(t)\in e^{\mathbf{g}_\text{pri}}\right\},\\
			&\mathscr{Q}(\mathbf{g}_\text{pri}, \lc{H_\text{pert}^w})=\left\{ \bigoplus_{w=1}^W |U_\text{pri}^\dagger H_\text{pert}^wU_\text{pri}\rrangle_w \middle|U_\text{pri}\in e^{\mathbf{g}_\text{pri}}\right\}.
		\end{split}
	\end{equation}
	Then $\mathscr{O'}(\mathbf{g}_\text{pri}, \lc{H_\text{pert}^w})= \mathscr{\bar Q}(\mathbf{g}_\text{pri}, \lc{H_\text{pert}^w})$.
\end{theorem}
For any direction $|H_{\text{target}}\rrangle_{\text{comp}}$ in $\mathscr C_{\text{comp}}$, the achievable range of scaling factors is obtained by solving the linear programs \eqref{lp}.
Find $T$ such that $T|H_\text{target}\rrangle_\text{comp}\propto (1,0,\ldots,0)$, the vertices of the polygon are 
\begin{equation}
	\vec v^j=T\bigoplus_{w=1}^W\left| u_j^\dagger H_\text{pert}^w u_j\right\rrangle_w.
	\label{vert}
\end{equation}
This allows us to use Algorithm \ref{scalealg} to characterize the achievability of $|H_\text{target}\rrangle_\text{comp}$. 

\section{Generating Random Unitaries for Controllability Characterization}\label{appen4}
\begin{figure}[htp]
	\centering
	\includegraphics[width=0.45\textwidth]{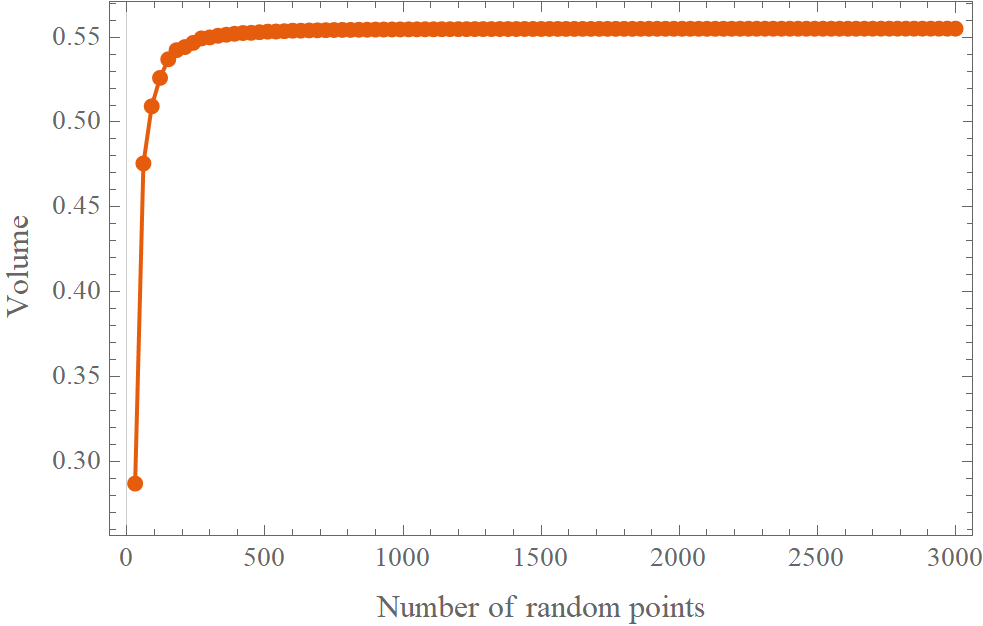}\\
	\caption{Volume of the simulated $\mathscr {\bar Q}$ as a function of the number of samples. It saturates after around 500 samples.}
	\label{volume}
\end{figure}

Here, we present algorithms for sampling random unitaries from $e^{\mathbf{g}_\text{pri}}$. For standard groups such as the unitary groups, Haar random samples can be generated using the QR decomposition \cite{mezzadri2006generate} (Algorithm \ref{algQR}).
\begin{algorithm}[H]
	\caption{Algorithm for generating random unitaries from SU$(n)$ based on QR decomposition.}\label{algQR}
	\begin{algorithmic}[1]
		\Function{RandomUQR}{$n$}
		\State generate two random real $n\times n$ matrices $r_1$ and $r_2$ whose elements are sampled from $\mathcal N(0,1)$
		\State $z\gets (r_1+i r_2)/\sqrt{2}$
		\State use QR decomposition to decompose $z$ into $z=QR$ where $Q$ is unitary and $R$ is upper-triangle
		\State take the signs of the diagonal of $R$ and form a diagonal matrix $\Lambda$
		\State \Return $Q\Lambda$
		\EndFunction
	\end{algorithmic}
\end{algorithm}
For non-standard Lie algebras, we generate random unitaries from $e^{\mathbf g_\text{pri}}$ using a simple random walk algorithm \cite{lawler2010random} (Algorithm \ref{algRW}).
\begin{algorithm}[H]
	\caption{Algorithm for generating random unitaries using random walk.}\label{algRW}
	\begin{algorithmic}[1]
		\Function{RandomULie}{$\mathcal B(\mathbf{g}_\text{pri})$}
		\State $g\gets0$
		\For{$h\in\mathcal B(\mathbf{g}_\text{pri})$}
		\State sample $p$ from $\mathcal N(0,1)$
		\State $g\gets g+ph$
		\EndFor
		\State \Return $e^g$
		\EndFunction
		\Function{RandomwalkU}{$\mathcal B(\mathbf{g}_\text{pri})$,$n_\text{burn}$,$n_\text{thin}$,$n_\text{sample}$}
		\State $x\gets \mathbb{1}$
		\For{$1\le i_\text{burn}\le n_\text{burn}$}
		\State $x\gets \textsc{RandomULie}(B(\mathbf{g}_\text{pri}))\cdot x$
		\EndFor
		\State $l_\text{sample}\gets \lc{}$
		\For{$1\le i_\text{sample}\le n_\text{sample}$}
		\For{$1\le i_\text{thin}\le n_\text{thin}$}
		\State $x\gets \textsc{RandomULie}(B(\mathbf{g}_\text{pri}))\cdot x$
		\EndFor
		\State append $x$ to $l_\text{sample}$ 
		\EndFor
		\State \Return $l_\text{sample}$
		\EndFunction
	\end{algorithmic}
\end{algorithm}
The convergence of Algorithm \ref{algRW} can be checked using, e.g., the coupling method. Alternatively, one can check the volume of the simulated $\mathscr{\bar Q}$. Fig. \ref{volume} shows the volume of the simulated $\mathscr{\bar Q}$ of secular dipoler interaction under the collective control of the qubits. The volume is calculated using the quick hull algorithm \cite{barber1996quickhull,virtanen2020scipy}. 
\section{Qubit Network with Two Clusters}\label{appenexam}
This appendix presents a simple illustrative example to demonstrate how to use Algorithms \ref{L2} and \ref{L3} (Algorithm \ref{scalealg}) to characterize the set of achievable zeroth-order average Hamiltonians. Consider a system with an internal Hamiltonian described by Eq. \eqref{intexp} where
\begin{equation}
	\begin{split}
		\mathbf D=\text{Diag}\lr{
			-1,-1,2},~~~\vec n=(0,0,1)^T,
	\end{split}
\end{equation}
so the internal Hamiltonian includes dipolar interaction between qubits under high field and qubits detuning along $z$ direction:
\begin{equation}
	\begin{split}
		&H_\text{int}=H_\text{int}^1+H_\text{int}^2,\\
		&H_\text{int}^1=\sum_{i<j}^{N}B_{ij}\left(2\z^{i}\otimes\z^{j}-\x^{i}\otimes\x^{j}-\y^{i}\otimes\y^{j}\right),\\
		&H_\text{int}^{2}=\sum_{i=1}^{N}\delta_{i}\z^{i}.
	\end{split}
\end{equation}
Consider collective control over the qubits:
\begin{equation}
	\begin{split}
		H_c(t)&=\omega_1(t)\lr{\cos(\phi(t))\sum_{i=1}^{N}\x^{i}+\sin(\phi(t))\sum_{i=1}^{N}\y^{i}}.
	\end{split}
\end{equation}
Choose the following partitioning:
\begin{equation}
	H_\text{pri}(t)=H_c(t),~~H_\text{pert}^1=H_\text{int}^1,~~H_\text{pert}^2=H_\text{int}^2.
\end{equation} 

\begin{figure}[htp]
	\centering
	\includegraphics[width=0.5\textwidth]{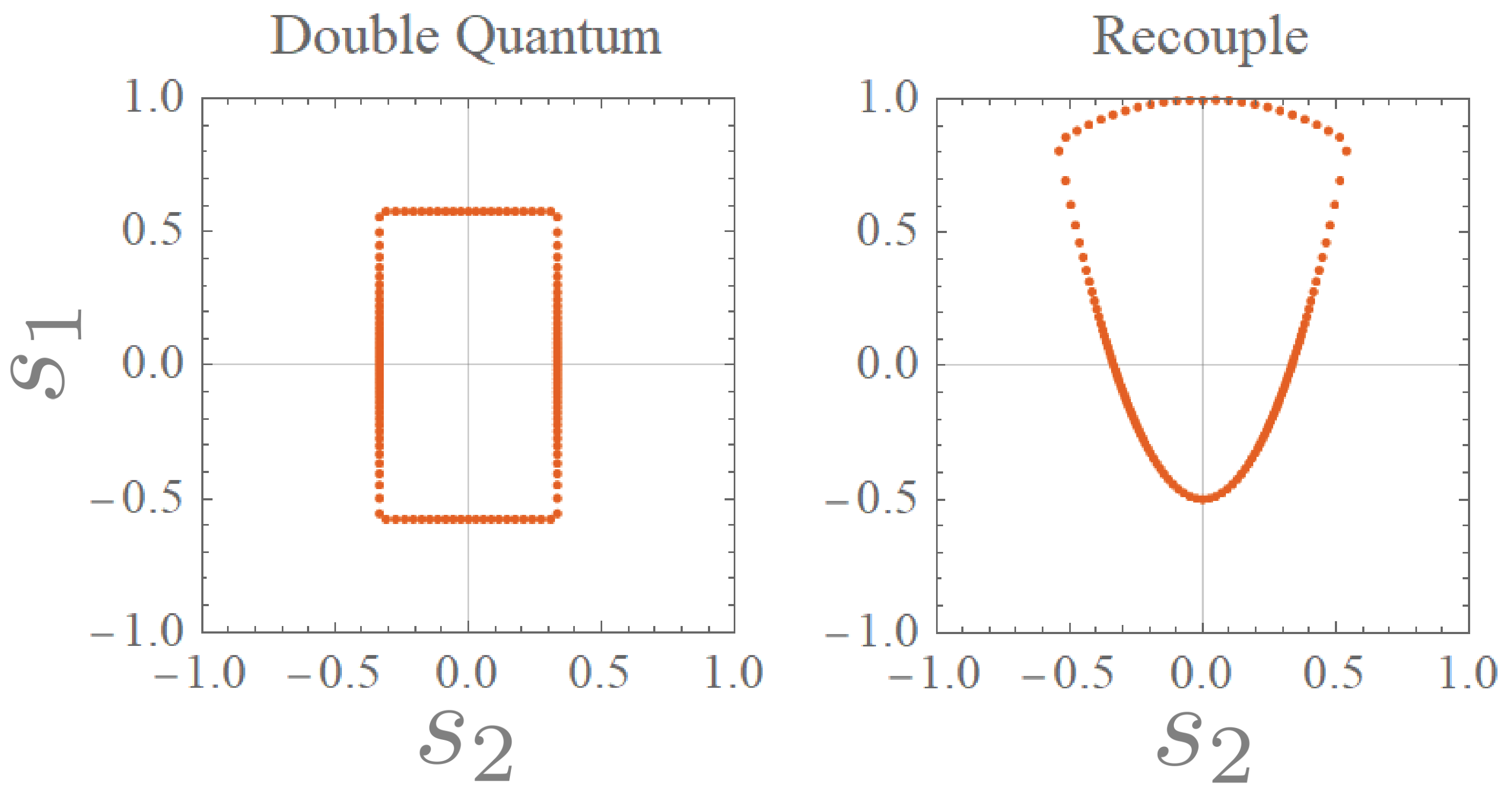}\\
	\caption{Regions of achievable effective Hamiltonians. The left plot corresponds to the targets defined in Eqs. \eqref{tata}, 
		while the right plot shows the case 
		\( H_\text{target}^2 \propto H_\text{pert}^2 \).
		For \( \beta = 0, \frac{\pi}{64}, \ldots, \frac{63\pi}{64} \), each value of 
		\( \beta \) defines a target \( \lvert H_\text{target} \rrangle_\text{comp} \). 
		Algorithm~\ref{scalealg} then determines the corresponding scaling factors 
		\( s_- \) and \( s_+ \), yielding two orange dots with coordinates 
		\( (s_- \cos\beta,\, s_- \sin\beta) \) and 
		\( (s_+ \cos\beta,\, s_+ \sin\beta) \). 
		Sweeping through all values of \( \beta \) produces 128 dots, 
		which trace out the convex set corresponding to all achievable 
		zeroth-order average Hamiltonians.}
	\label{region}
\end{figure}

Typically, no particular representations (Eq. \eqref{rep}) are required when the Hamiltonians act on a finite-dimensional space. However, in order to calculate $\mathbf g_\text{pri}$ and $\mathscr C_{w}(\mathbf g_\text{pri}, H_\text{pert}^w)$ here, we need to choose representations for each $w$ that are independent of $N$. It is easy to verify the following representations satisfy Eq. \eqref{rep}:
\begin{equation}
	\begin{split}
		&\xi_{1}(H_\text{pri})\\
		&=\omega_1(t)\lr{\cos(\phi(t))\sum_{i=1}^2\x^i+\sin(\phi(t))\sum_{i=1}^2\y^{i}},\\
		&\xi_{1}(H_\text{pert})=2\z\otimes\z-\x\otimes\x-\y\otimes\y,\\
		&\xi_2(H_\text{pri})=\omega_1(t)\lr{\cos(\phi(t))\x+\sin(\phi(t))\y},\\
		&\xi_2(H_\text{pert})=\z.
	\end{split}
\end{equation}
Using Algorithm \ref{L2}, an orthonormal basis of $\mathscr C_1(\mathbf g_\text{pri}, H_\text{pert}^1)$ is found to be
\begin{equation}
	\begin{split}
		&\mathcal B(\mathscr C_1)=\\
		&\left\{\frac{2\z\otimes\z-\x\otimes\x-\y\otimes\y}{2\sqrt 6},\frac{\y\otimes\z+\z\otimes\y}{2\sqrt 2},\right.\\
		&~~~~\left.\frac{\x\otimes\z+\z\otimes\x}{2\sqrt 2},\frac{\y\otimes\y-\x\otimes\x}{2\sqrt 2},-\frac{\x\otimes\y+\y\otimes\x}{2\sqrt 2}\right\}.
	\end{split}
\end{equation}
$\mathscr C_2({\mathbf g_\text{pri}},H_\text{pert}^{2})$ is 3-dimensional and spanned by 
\begin{equation}
	\mathcal B(\mathscr C_2)=\left\{\frac{\z}{\sqrt2},\frac{\x}{\sqrt2},\frac{\y}{\sqrt2}\right\}.
\end{equation}
Consider engineering $H_\text{pert}^1$ into a double-quantum Hamiltonian while keeping $H_\text{pert}^2$ in its original form:
\begin{equation}
	\begin{split}
		&s_1H_\text{target}^1=s_1\sqrt{3}\sum_{ij}^{N}B_{ij}\left(\y^{i}\otimes\y^{j}-\x^{i}\otimes\x^{j}\right),\\
		&s_2H_\text{target}^2=s_2\sum_{i=1}^{N}\delta_i\z^{i}.
	\end{split}
	\label{tata}
\end{equation}
The $\sqrt 3$ in the first equation above is due to normalization.
Represented in $\mathscr C_1(\mathbf g_\text{pri},H_\text{pert}^1)$ and $\mathscr C_2(\mathbf g_\text{pri},H_\text{pert}^2)$, the targets can be written as 
\begin{equation}
	\begin{split}
		|s_1H_\text{target}^1\rrangle&=s(0,0,0,\sin\beta,0)^T,\\
		|s_2H_\text{target}^{2}\rrangle&=s(\cos\beta,0,0)^T,\\	
	\end{split}
\end{equation}
where $s_1=s\sin\beta$ and $s_2=s\cos\beta$. Then the composite target Hamiltonian is defined as
\begin{equation}
	|sH_\text{target}\rrangle_\text{comp}=|s_1H_\text{target}^1\rrangle\oplus|s_2H_\text{target}^{2}\rrangle.
\end{equation}
For each value of \( \beta \in [0, 2\pi) \), the range of achievable scaling factors, $s$,
is obtained using Algorithm~\ref{scalealg}. Since $\mathbf g_\text{pri}$ is isomorphic to su(2), we can use Algorithm \ref{algQR} to generate Haar-random unitaries $\lc{u_j}$. The set of random vertices, $\lc{\vec v^j}$, used in Algorithm \ref{scalealg} is given by Eq. \eqref{vert}. Finding the achievable range of $s$ for each $\beta$ traces out the region 
of all achievable zeroth-order average Hamiltonians (Fig.~\ref{region}, left). 
For comparison, the region corresponding to a different target, 
\( s_1H_\text{target}^1 =s_1 H_\text{pert}^1 \) (recoupling), 
is shown on the right side of Fig.~\ref{region}.

\section{Discretization of $\vec B(t)$ for Integration}\label{appen5}
\begin{figure}[htp]
	\centering
	\includegraphics[width=0.5\textwidth]{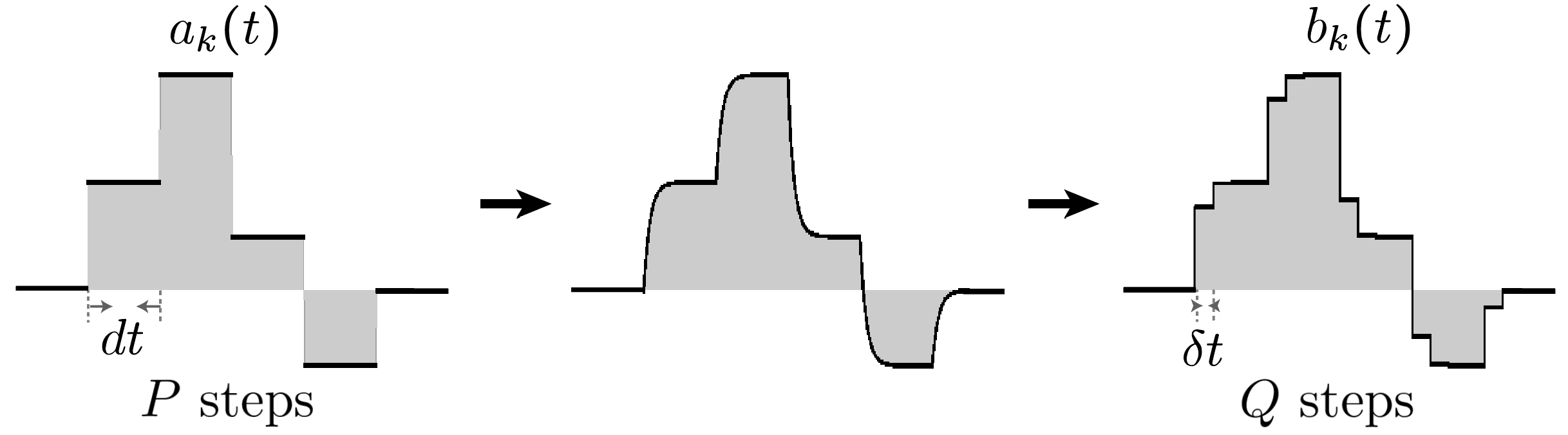}\\
	\caption{Illustration of the discretization of output field parameters. The control parameters $\lc{a_k(t)}$ are mapped to $\vec B(t)$ through the model of the control system. Then the continuous field $\vec B(t)$ is discretized into $Q$ steps for integration.}
	\label{discretize}
\end{figure}

Here, we briefly describe the discretization of $\vec B(t)$ for numerical integration. Note, this is required during pulse finding, but not when calculating the performance of a sequence. The control parameters $a_k(t)$ are typically piecewise-constant:
\begin{equation}
	a_{k,p}=a_k[(p-1/2)dt],\\
\end{equation}
for $p=1,\ldots,P$.
Here, an equal step length $dt$ is assumed for the discretization for simplicity.
Denote components of $\vec B(t)$ in rotation frames of the qubits by $\lc{b_k(t)}$,  they can be discretized into $Q$ steps (Fig. \ref{discretize}) and the value at the $q$th step $b_{k,q}$ is approximated by the value of $b_k(t)$ at the middle of the step
\begin{equation}
	b_{k,q}=b_k[(q-1/2)\delta t],
	\label{dis}
\end{equation}
so that $P dt=Q \delta t=T_\text{seq}$. $Q$ should be large enough in order to accurately represent $\vec B(t)$. The approximation (Eq. \eqref{dis}) is reasonable when the time dependence of $b_k(t)$ is much slower compared to $1/\delta t$. When the assumption does not hold, a better approximation is given by
\begin{equation}
	b_{k,q}=\frac{1}{\delta t}\int_{(q-1)\delta t}^{q\delta t}b_k(t)dt.
	\label{dis2}
\end{equation}
The discretization enables straightforward calculation of the 
$\mathscr C$-integrals (defined below) by first computing them for each step and then composing them using a chain rule.
\section{Calculating $\mathscr C-$integrals for Shaped Pulse}\label{appen6}
In this appendix, we show how to calculate the composite $\mathscr C-$integrals (Eq. \eqref{totr}). Define the vectorized composite $\mathscr C-$integrals as
\begin{widetext}
	\begin{equation}
		\vec c^{(r-1)}_\text{comp}=\lr{\bar c_{11\cdots1}^\text{comp},\ldots,\bar c_{11\cdots|\mathscr C_\text{comp}|}^\text{comp},\ldots,\bar c_{|\mathscr C_\text{comp}||\mathscr C_\text{comp}|\cdots1}^\text{comp},\ldots,\bar c_{|\mathscr C_\text{comp}||\mathscr C_\text{comp}|\cdots|\mathscr C_\text{comp}|}^\text{comp}},
	\end{equation}
	it obeys
	\begin{equation}
		\begin{split}
			\vec c_\text{comp}^{(r-1)}&=\int_0^{T_\text{seq}}dt_1\cdots\int_0^{t_{r-1}}\bigotimes_{r'=1}^r\ls{\bigoplus_{w=1}^W\mathcal D_w(U_\text{pri}^\dagger(t_{r'}))|H_\text{pert}^w\rrangle_w}\\
			&=\int_0^{T_\text{seq}}dt_1\cdots\int_0^{t_{r-1}}\bigotimes_{r'=1}^r\ls{\mathcal D_\text{comp}(U_\text{pri}^\dagger(t_{r'}))|H_\text{pert}\rrangle_\text{comp}},
		\end{split}
	\end{equation}
	where
	\begin{equation}
		\begin{split}
			&\mathcal D_\text{comp}(U_\text{pri}^\dagger(t_{r'}))=\bigoplus_{w=1}^W\mathcal D_w(U_\text{pri}^\dagger(t_{r'})),~~~~
			|H_\text{pert}\rrangle_\text{comp}=\bigoplus_{w=1}^W|H_\text{pert}\rrangle_w.
		\end{split}
	\end{equation}
	For simplicity of notation, we omit the subscript ``comp'' in the following discussion. All representations are understood to be in the composite \(\mathscr{C}\) space.
	
	Under an orthonormal basis (Eq. \eqref{cbasis}) of $\mathscr C(\mathbf g_\text{pri},H_\text{pert})$, the representation of a Hamiltonian $H$, the adjoint representation of a unitary propagator $U=e^g$ and the generator of the adjoint action are denoted by $|H\rrangle$, $\mathcal D(U)$ and $\mathcal D(\text{ad}_g)$ respectively. They obey 
	\begin{equation}
		|UHU^\dagger\rrangle=\mathcal D(U)|H\rrangle,~~~\mathcal D(e^{g})=e^{\mathcal D(\text{ad}_g)},
	\end{equation}
	and can be calculated as
	\begin{equation}
		\begin{split}
			|H\rrangle=\lr{\begin{array}{c}\text{Tr}\lr{Hh_1}\\\vdots\\\text{Tr}\lr{Hh_{|\mathscr C|}}\end{array}},~~\mathcal D(U)=\lr{\begin{array}{ccc}
					\llangle h_1|Uh_1U^\dagger\rrangle&\cdots&\llangle h_1|Uh_{|\mathscr C|}U^\dagger\rrangle\\
					\vdots&\ddots&\vdots\\
					\llangle h_{|\mathscr C|}|Uh_1U^\dagger\rrangle&\cdots&\llangle h_{|\mathscr C|}|Uh_{|\mathscr C|}U^\dagger\rrangle
			\end{array}},~~\mathcal D(\text{ad}_g)=\lr{\begin{array}{ccc}
					\llangle h_1|[g,h_1]\rrangle&\cdots&\llangle h_1|[g,h_{|\mathscr C|}]\rrangle\\
					\vdots&\ddots&\vdots\\
					\llangle h_{|\mathscr C|}|[g,h_1]\rrangle&\cdots&\llangle h_{|\mathscr C|}|[g,h_{|\mathscr C|}]\rrangle
				\end{array}
			},
			\label{cal}
		\end{split}
	\end{equation}
	where $\llangle \cdot|*\rrangle=\text{Tr}(*~\cdot^\dagger)$.
	The discretization of $\lc{b_k(t)}$ is chosen to be fine enough so that during each step both $H_\text{pri}(t)$ and $H_\text{pert}(t)$ can be approximately viewed as constant. Denote their discretized versions corresponding to $\lc{b_{k,q}}$ by
	\begin{equation}
		\begin{split}
			\vec H_\text{pri}&=\lr{H_\text{pri}^1,H_\text{pri}^2,\ldots,H_\text{pri}^Q},\\
			\vec H_\text{pert}&=\lr{H_\text{pert}^{1},H_\text{pert}^{2},\ldots,H_\text{pert}^{Q}}.
		\end{split}
	\end{equation}
	A recipe for calculating the $\mathscr C-$integrals for any sequence requires two ingredients: (1) Their calculation for a square pulse. (2) A chain rule of the $\mathscr C-$integrals that allows to calculate those of a sequence given the $\mathscr C-$integrals of the components of the sequence.
	\subsection{$\mathscr C-$integrals of a square pulse}\label{singleC}
	During the $q$th step, $H_\text{pri}(t)=H_\text{pri}^q$ and $H_\text{pert}(t)=H_\text{pert}^q$, so the $\mathscr C-$integrals become
	\begin{equation}
		\vec c^{(r-1)}=\int_0^{T_\text{seq}}dt_1\cdots\int_0^{t_{r-1}}dt_r\exp\lr{-i\mathcal D(\text{ad}_{H^q_\text{pri}})t_1}\otimes\cdots\otimes\exp\lr{-i\mathcal D(\text{ad}_{H^q_\text{pri}})t_r}|H_\text{pert}^q\rrangle^{\otimes r}.
	\end{equation}
	Here, $\mathcal D(\text{ad}_{H_\text{pri}}^q)=\bigoplus_{w=1}^WD_w(\text{ad}_{H_\text{pri}}^q)$.
	Let $\mathcal D(\text{ad}_{H^q_\text{pri}})=T D T^\dagger$ where $D$ is diagonal and $T$ is a unitary matrix, the integral becomes
	\begin{equation}
		\vec c^{(r-1)}=T^{\otimes r}\text{Diag}\lr{I\lr{\lambda_1,\lambda_1,\ldots,\lambda_1},I\lr{\lambda_1,\lambda_1,\ldots,\lambda_2},\ldots,I\lr{\lambda_{|\mathscr C|},\lambda_{|\mathscr C|},\ldots,\lambda_{|\mathscr C|}}}^T {T^\dagger}^{\otimes r}|H_\text{pert}^q\rrangle^{\otimes r}.
		\label{c1}
	\end{equation}
	where $\lambda_1,\ldots,\lambda_{|\mathscr C|}$ are eigenvalues of $\text{ad}_{H_\text{pri}^q}$ and
	\begin{equation}
		I(\lambda_{i_1},\ldots,\lambda_{i_r})=\int_0^{T_\text{seq}}dt_1\cdots\int_0^{t_{r-1}}dt_re^{-i\lambda_{i_1}t_1}\cdots e^{-i\lambda_{i_r}t_r}.
		\label{c2}
	\end{equation}
	The above integrals admit closed-form analytic expressions, which can be precomputed and stored to avoid repeated numerical integration.
	The diagonalization of $\text{ad}_{H_\text{pri}^q}$ can be found using the numerical Lanczos method. 
	\subsection{$\mathscr C-$integrals of a shaped pulse} 
	Denote the $\mathscr C-$integral of the $q$th step by $\vec c^{(r-1)}(q)$. Define
	\begin{equation}
		\vec c^{(r-1)}_\text{tog}(q)=\ls{\mathcal D(U_1^\dagger)\cdots\mathcal D( U_{q-1}^\dagger)}^{\otimes r}\cdot\vec c^{(r-1)}(q),
		\label{togc}
	\end{equation}
	where $U_q$ is the primary unitary of the $q$th step.
	Then 
	\begin{equation}
		\vec c^{(r-1)}=\sum_{r_1+\cdots+r_v=r}\sum_{Q\ge q_1>\cdots>q_v\ge 1}\vec c^{(r_1-1)}_\text{tog}(q_1)\otimes\vec c^{(r_2-1)}_\text{tog}(q_2)\otimes\cdots\otimes \vec c^{(r_v-1)}_\text{tog}(q_v).
		\label{chainrule}
	\end{equation}

	For $r=1,2$ and 3, we have
	\begin{equation}
		\begin{split}
			\vec c^{(0)}&=\sum_{q=1}^Q \vec c_\text{tog}^{(0)}(q),\\
			\vec c^{(1)}&=\sum_{q=1}^Q\vec c_\text{tog}^{(1)}(q)+\sum_{q_1=2}^Q\sum_{q_2=1}^{q_1-1}\vec c_\text{tog}^{(0)}(q_1)\otimes\vec c_\text{tog}^{(0)}(q_2),\\
			\vec c^{(2)}&=\sum_{q=1}^Q\vec  c_\text{tog}^{(2)}(q)+\sum_{q_1=2}^Q\sum_{q_2=1}^{q_1-1}\ls{\vec c_\text{tog}^{(0)}(q_1)\otimes\vec c_\text{tog}^{(1)}(q_2)+\vec c_\text{tog}^{(1)}(q_1)\otimes\vec c_\text{tog}^{(0)}(q_2)}+\sum_{q_1=3}^Q\sum_{q_2=2}^{q_1-1}\sum_{q_3=1}^{q_2-1}c_\text{tog}^{(0)}(q_1)\otimes c_\text{tog}^{(0)}(q_2)\otimes c_\text{tog}^{(0)}(q_3).
		\end{split}
	\end{equation}
	\section{Calculating the Modified $\mathscr C-$integrals}\label{appen7}
	For simplicity, here we only consider $r=2$. The vectorized modified $\mathscr C-$integral is
	\begin{equation}
		\vec c^{'(1)}=\int_0^{T_\text{seq}}dt_1\int_0^{t_1}dt_2e^{-\theta(t_1-t_2)}\mathcal D( U_\text{pri}^\dagger)(t_1)\otimes \mathcal D(U_\text{pri}^\dagger)(t_2)|H_\text{pert}(t_1)\rrangle\otimes|H_\text{pert}(t_2)\rrangle
	\end{equation}
	
	Following the same procedure introduced in the last appendix, during the $q$th step, the modified $\mathscr C-$integral is
	\begin{equation}
		\vec c^{'(1)}=\int_0^{T_\text{seq}}dt_1\int_0^{t_{1}}dt_2e^{-\theta(t_1-t_2)}\exp\lr{-i\mathcal D(\text{ad}_{H^q_\text{pri}})t_1}\otimes\exp\lr{-i\mathcal D(\text{ad}_{H^q_\text{pri}})t_2}|H_\text{pert}^q\rrangle^{\otimes 2}.
	\end{equation}
	Let $\mathcal D(\text{ad}_{H^q_\text{pri}})=T D T^\dagger$ where $D$ is diagonal and $T$ is a unitary matrix, the integral becomes
	\begin{equation}
		\vec c^{'(1)}=T^{\otimes 2}\text{Diag}\lr{I'\lr{\lambda_1,\lambda_1},I'\lr{\lambda_1,\lambda_2},\ldots,I'\lr{\lambda_{|\mathscr C|},\lambda_{|\mathscr C|}}}^T {T^\dagger}^{\otimes 2}|H_\text{pert}^q\rrangle^{\otimes 2}.
	\end{equation}
	where $\lambda_1,\ldots,\lambda_{|\mathscr C|}$ are eigenvalues of $\text{ad}_{H_\text{pri}^q}$ and
	\begin{equation}
		I'(\lambda_{i_1},\lambda_{i_2})=\int_0^{T_\text{seq}}dt_1\int_0^{t_{1}}dt_2e^{-i\lambda_{i_1}t_1} e^{-i\lambda_{i_2}t_2}e^{-\theta(t_1-t_2)}.
	\end{equation} 
	
	Denote the modified $\mathscr C-$integral of the $q$th step by $\vec c^{'(r-1)}(q)$. Define
	\begin{equation}
		\vec c^{'(r-1)}_\text{tog}(q)=\ls{\mathcal D(U_1^\dagger)\cdots\mathcal D( U_{q-1}^\dagger)}^{\otimes r}\cdot\vec c^{'(r-1)}(q).
	\end{equation}
	Then 
	\begin{equation}
		\vec c^{'(1)}=\sum_{q=1}^Q\vec c_\text{tog}^{'(1)}(q)+\sum_{q_1=2}^Q\sum_{q_2=1}^{q_1-1}e^{-\theta\delta t(q_1-q_2)}\vec c_\text{tog}^{'(0)}(q_1)\otimes\vec c_\text{tog}^{'(0)}(q_2).
	\end{equation} 
	Here, the modified $\vec c^{'(0)}$ are defined as 
	\begin{equation}
		\begin{split}
			&\vec c^{'(0)}(q_1)=\int_0^{\delta t}dt_1e^{-\theta t_1}\exp\lr{-i\mathcal D(\text{ad}_{H^{q_1}_\text{pri}})t_1}|H_\text{pert}^{q_1}\rrangle,\\
			&\vec c^{'(0)}(q_2)=\int_0^{\delta t}dt_2e^{\theta t_2}\exp\lr{-i\mathcal D(\text{ad}_{H^{q_2}_\text{pri}})t_2}|H_\text{pert}^{q_2}\rrangle.
		\end{split}
	\end{equation}
\end{widetext}
\section{Achievability of Zeroth-order Control Robustness}\label{appen8}
In the presence of an error Hamiltonian $\Delta H(t,\mu+\epsilon)$,
assume the following: (1) there exists an implementation of the identity operator that is robust to $\epsilon$ to at least zeroth order (e.g., a waiting period where $a_k(t)\equiv 0$); (2) $\Delta H$ continuously depends on $a_k(t)$ and $\mu$; and (3) the correlation time of $\Delta H$ is finite so that we can compose gates without interference, then the following result holds when single-qubit universal control is available.
\begin{theorem}
	Zeroth-order robustness to $\epsilon$ is always achievable to any systematic error Hamiltonian $\Delta H$.
\end{theorem}
\begin{proof}
	The total Hamiltonian is
	\begin{equation}
		H_\text{tot}(t)=H_c(t)+\epsilon\frac{\partial }{\partial \epsilon}\Delta H(t,0)+O(\epsilon^2).
	\end{equation}
	The goal is to design a unitary $U_\text{target}$ so that 
	\begin{equation}
		\begin{split}
			&U(T_\text{seq},0)=U_\text{target},\\
			&U(T_\text{seq},\epsilon)=U_\text{target}+O(\epsilon^2).
		\end{split}
		\label{goal}
	\end{equation}
	Suppose that we have found a sequence $U_0(T_\text{seq},\epsilon)$ that achieves the first condition in Eqs. \eqref{goal}. In general, 
	\begin{equation}
		U_0(T_\text{seq},\epsilon)=U_\text{target}e^{-i\epsilon \sigma_0\theta}+O(\epsilon^2).
	\end{equation}
	If we could implement a sequence $U^*$ so that
	\begin{equation}
		U^*(T_\text{seq},\epsilon)=e^{i\epsilon \sigma_0\theta}+O(\epsilon^2),
	\end{equation}
	then $U_0U^*$ achieves a robust implementation of $U_\text{target}$ in zeroth order. 
	
	Let $U_1=\mathbb 1+O(\epsilon^2)$ and $U_2=\mathbb 1+O(\epsilon)$.
	Since we can continuously modify $U_1$ into $U_2$, there exists $U_r=\exp(-i r \epsilon\sigma')+O(\epsilon^2)$, $\forall r>0$ and some $\sigma'$. Therefore, if we choose $r$ small enough, we can achieve $U_r^N\approx \exp(-i s \epsilon\sigma')+O(\epsilon^2)$ to arbitrary accuracy, $\forall s>0$ and some positive integer $N$.
	
	Due to universal control, we can rotate $\sigma'$ to any direction. Thus, we can implement $U_\alpha(s)=\exp(-i (s \epsilon\sigma_\alpha+\epsilon\sigma_{r_\alpha}))+O(\epsilon^2)$, for $\alpha=\pm x, \pm y$ and $\pm z$. Here, $\sigma_{r_\alpha}$ corresponds to the error introduced by the imperfect rotation. Since we can choose any $s>0$ and $\sigma_{r_\alpha}$ does not depend on $s$, therefore, $\forall \sigma_0$, $\exists s_\alpha>0$ so that $\prod_\alpha U_\alpha (s_\alpha)=e^{i\epsilon \sigma_0\theta}+O(\epsilon^2)$.
\end{proof}
The result is a generalization of the result from \cite{brown2004arbitrarily}. All the errors considered in this work (Rabi field variation, variations in linear or nonlinear models) satisfy the assumptions and therefore can be corrected to at least zeroth order. Additionally, since $U_\text{target}$ can be any unitary from SU(2), requiring control robustness does not affect the controllability, so when characterizing the achievable sets of $U_\text{target}$ and $H_\text{target}$, we do not need to consider $\Delta H$.
\section{Achievable Set of $\rho_\text{target}$}\label{appen9}
To characterize the set of achievable $\rho_\text{target}$, we consider a more general setting where the expectation-valued measurement allows an average over different sequences, so
\begin{equation}
	\rho_\text{target}=\frac{1}{M}\sum_{m=1}^MU_m(T_\text{seq})\rho_0U_m^\dagger(T_\text{seq}),
\end{equation}
where $U_m\in e^{\mathbf g_\text{pri}}$. This has been used in, for example, in preparing pseudo-pure states \cite{cory1997ensemble,knill1998effective}. 
Then the set of achievable $\rho_\text{target}$ is $\mathscr{Q}(\mathbf{g}_\text{pri}, \rho_0)$ (Eq. \eqref{achset}) and therefore can be characterized using Algorithm \ref{L3}. $M$ is at most the dimension of $\mathscr C(\mathbf g_\text{pri},\rho_0)$. If $\rho_\text{target}$ is achievable and has the same spectrum as $\rho_0$, then $M=1$.
\section{Robust Control in Nonlinear Systems}\label{appen10}
In this appendix, we show how to evaluate the derivatives of $\Delta H$ (Eq. \eqref{derive}) in a system governed by the following state-space differential equation:
\begin{equation}
	\frac{d\vec B}{dt}=h(\vec B, \vec a,t;\vec \mu_0).
\end{equation}
A variation $\vec\epsilon$ in the model parameter $\vec\mu$ results in a different output field $\vec B(t,\vec\mu_0+\vec\epsilon)=\vec B(t,\vec \mu_0)+\sum_i\epsilon_i\frac{\partial\vec B}{\partial\mu_i}(t,\vec \mu_0)+\cdots$. The differential equation becomes
\begin{widetext}
	\begin{equation}
		\frac{d}{dt}\lr{\vec B+\sum_i\epsilon_i\frac{\partial\vec B}{\partial\mu_i}+\cdots}=h(\vec B(\vec \mu_0), \vec a,t;\vec \mu_0)+\sum_i\epsilon_i\mathbf J\lr{h,\vec B}|_{\vec \mu=\vec \mu_0}\frac{\partial\vec B}{\partial\epsilon_i}(\vec \mu_0)+\cdots,
		\label{consysdiff}
	\end{equation}
	where $\mathbf J$ is the Jacobian matrix. By equating terms of the same order in $\epsilon_i$, we can obtain coupled differential equations that can be solved for $\frac{\partial \vec B}{\partial \mu_i},\ldots$, enabling evaluation of the error Hamiltonian $\Delta H$. 
	For the resonant circuit in Fig. \ref{concirc}. In a rotating frame of frequency $\omega_r$, it obeys the following differential equation given by Eq. \eqref{e74}. Using Eq. \eqref{consysdiff}, the equations for $x(\alpha_L=0)$ and $\frac{\partial x}{\partial \alpha_L}(\alpha_L=0)$ are
\begin{equation}
	\left.
	\frac{d}{dt}\lr{\begin{array}{c}
			x\\\frac{\partial x}{\partial \alpha_L}
	\end{array}}=\lr{\begin{array}{cc}
			A(x)&0\\\frac{\partial A}{\partial \alpha_L}(x)&A(x)
	\end{array}}\lr{\begin{array}{c}
			x\\\frac{\partial x}{\partial \alpha_L}
	\end{array}}+\alpha(t)\lr{\begin{array}{c}
			u\\\frac{\partial u}{\partial \alpha_L}
	\end{array}}\right|_{\alpha_L=0},
\end{equation}
\end{widetext}
where
\begin{equation}
	\frac{\partial A}{\partial\alpha_L}(x,\alpha_L=0)=\lr{\begin{array}{ccc}
			\frac{R|\tilde I_L|^2}{L_0}&0&-\frac{|\tilde I_L|^2}{L_0}\\
			0&0&0\\
			0&0&0
	\end{array}},~~~~\frac{\partial u}{\partial\alpha_L}=0.
\end{equation}
The distortion of the input pulse and the corresponding error Hamiltonian are thus calculated by solving the system of differential equations.

	\bibliography{ref1}
	
\end{document}